\documentclass[twocolumn, twocolappendix, trackchanges]{aastex631}% arxiv

\usepackage{amsmath}

\usepackage{placeholder}
\usepackage{comment}
\usepackage{todonotes}
\usepackage{placeins} 

\begin{document}

%\todo{modify linenumbers}
%\title{Search for Diffuse Supernova Neutrino Background in Super-Kamiokande Gadolinium experiment}
\title{Search for Diffuse Supernova Neutrino Background\\ with 956.2~days of Super-Kamiokande Gadolinium Dataset}

\newcommand{\AFFicrr}{\affiliation{Kamioka Observatory, Institute for Cosmic Ray Research, University of Tokyo, Kamioka, Gifu 506-1205, Japan}}
\newcommand{\AFFkashiwa}{\affiliation{Research Center for Cosmic Neutrinos, Institute for Cosmic Ray Research, University of Tokyo, Kashiwa, Chiba 277-8582, Japan}}
\newcommand{\AFFipmu}{\affiliation{Kavli Institute for the Physics and
Mathematics of the Universe (WPI), The University of Tokyo Institutes for Advanced Study,
University of Tokyo, Kashiwa, Chiba 277-8583, Japan }}
\newcommand{\AFFmad}{\affiliation{Department of Theoretical Physics, University Autonoma Madrid, 28049 Madrid, Spain}}
\newcommand{\AFFubc}{\affiliation{Department of Physics and Astronomy, University of British Columbia, Vancouver, BC, V6T1Z4, Canada}}
\newcommand{\AFFbu}{\affiliation{Department of Physics, Boston University, Boston, MA 02215, USA}}
\newcommand{\AFFuci}{\affiliation{Department of Physics and Astronomy, University of California, Irvine, Irvine, CA 92697-4575, USA }}
\newcommand{\AFFcsu}{\affiliation{Department of Physics, California State University, Dominguez Hills, Carson, CA 90747, USA}}
\newcommand{\AFFcnm}{\affiliation{Institute for Universe and Elementary Particles, Chonnam National University, Gwangju 61186, Korea}}
\newcommand{\AFFduke}{\affiliation{Department of Physics, Duke University, Durham NC 27708, USA}}
\newcommand{\AFFgifu}{\affiliation{Department of Physics, Gifu University, Gifu, Gifu 501-1193, Japan}}
\newcommand{\AFFgist}{\affiliation{GIST College, Gwangju Institute of Science and Technology, Gwangju 500-712, Korea}}
\newcommand{\AFFuh}{\affiliation{Department of Physics and Astronomy, University of Hawaii, Honolulu, HI 96822, USA}}
\newcommand{\AFFicl}{\affiliation{Department of Physics, Imperial College London , London, SW7 2AZ, United Kingdom }}
\newcommand{\AFFkek}{\affiliation{High Energy Accelerator Research Organization (KEK), Tsukuba, Ibaraki 305-0801, Japan }}
\newcommand{\AFFkobe}{\affiliation{Department of Physics, Kobe University, Kobe, Hyogo 657-8501, Japan}}
\newcommand{\AFFkyoto}{\affiliation{Department of Physics, Kyoto University, Kyoto, Kyoto 606-8502, Japan}}
\newcommand{\AFFliv}{\affiliation{Department of Physics, University of Liverpool, Liverpool, L69 7ZE, United Kingdom}}
\newcommand{\AFFmiyagi}{\affiliation{Department of Physics, Miyagi University of Education, Sendai, Miyagi 980-0845, Japan}}
\newcommand{\AFFnagoya}{\affiliation{Institute for Space-Earth Environmental Research, Nagoya University, Nagoya, Aichi 464-8602, Japan}}
\newcommand{\AFFkmi}{\affiliation{Kobayashi-Maskawa Institute for the Origin of Particles and the Universe, Nagoya University, Nagoya, Aichi 464-8602, Japan}}
\newcommand{\AFFpol}{\affiliation{National Centre For Nuclear Research, 02-093 Warsaw, Poland}}
\newcommand{\AFFsuny}{\affiliation{Department of Physics and Astronomy, State University of New York at Stony Brook, NY 11794-3800, USA}}
\newcommand{\AFFokayama}{\affiliation{Department of Physics, Okayama University, Okayama, Okayama 700-8530, Japan }}
\newcommand{\AFFosaka}{\affiliation{Department of Physics, Osaka University, Toyonaka, Osaka 560-0043, Japan}}
\newcommand{\AFFox}{\affiliation{Department of Physics, Oxford University, Oxford, OX1 3PU, United Kingdom}}
\newcommand{\AFFqmul}{\affiliation{School of Physics and Astronomy, Queen Mary University of London, London, E1 4NS, United Kingdom}}
\newcommand{\AFFregina}{\affiliation{Department of Physics, University of Regina, 3737 Wascana Parkway, Regina, SK, S4SOA2, Canada}}
\newcommand{\AFFseoul}{\affiliation{Department of Physics and Astronomy, Seoul National University, Seoul 151-742, Korea}}
\newcommand{\AFFsheff}{\affiliation{School of Mathematical and Physical Sciences, University of Sheffield, S3 7RH, Sheffield, United Kingdom}}
\newcommand{\AFFshizuokasc}{\affiliation{Department of Informatics in
Social Welfare, Shizuoka University of Welfare, Yaizu, Shizuoka, 425-8611, Japan}}
\newcommand{\AFFstfc}{\affiliation{STFC, Rutherford Appleton Laboratory, Harwell Oxford, and Daresbury Laboratory, Warrington, OX11 0QX, United Kingdom}}
\newcommand{\AFFskk}{\affiliation{Department of Physics, Sungkyunkwan University, Suwon 440-746, Korea}}
\newcommand{\AFFtodai}{\affiliation{Department of Physics, University of Tokyo, Bunkyo, Tokyo 113-0033, Japan }}
\newcommand{\AFFtit}{\affiliation{Department of Physics, Institute of Science Tokyo, Meguro, Tokyo 152-8551, Japan }}
\newcommand{\AFFtus}{\affiliation{Department of Physics and Astronomy, Faculty of Science and Technology, Tokyo University of Science, Noda, Chiba 278-8510, Japan }}
\newcommand{\AFFtriumf}{\affiliation{TRIUMF, 4004 Wesbrook Mall, Vancouver, BC, V6T2A3, Canada }}
\newcommand{\AFFtokai}{\affiliation{Department of Physics, Tokai University, Hiratsuka, Kanagawa 259-1292, Japan}}
\newcommand{\AFFtsinghua}{\affiliation{Department of Engineering Physics, Tsinghua University, Beijing, 100084, China}}
\newcommand{\AFFynu}{\affiliation{Department of Physics, Yokohama National University, Yokohama, Kanagawa, 240-8501, Japan}}
\newcommand{\AFFllr}{\affiliation{Ecole Polytechnique, IN2P3-CNRS, Laboratoire Leprince-Ringuet, F-91120 Palaiseau, France }}
\newcommand{\AFFbari}{\affiliation{ Dipartimento Interuniversitario di Fisica, INFN Sezione di Bari and Universit\`a e Politecnico di Bari, I-70125, Bari, Italy}}
\newcommand{\AFFnapoli}{\affiliation{Dipartimento di Fisica, INFN Sezione di Napoli and Universit\`a di Napoli, I-80126, Napoli, Italy}}
\newcommand{\AFFroma}{\affiliation{INFN Sezione di Roma and Universit\`a di Roma ``La Sapienza'', I-00185, Roma, Italy}}
\newcommand{\AFFpadova}{\affiliation{Dipartimento di Fisica, INFN Sezione di Padova and Universit\`a di Padova, I-35131, Padova, Italy}}
\newcommand{\AFFkeio}{\affiliation{Department of Physics, Keio University, Yokohama, Kanagawa, 223-8522, Japan}}
\newcommand{\AFFwinnipeg}{\affiliation{Department of Physics, University of Winnipeg, MB R3J 3L8, Canada }}
\newcommand{\AFFkcl}{\affiliation{Department of Physics, King's College London, London, WC2R 2LS, UK }}
\newcommand{\AFFwarwick}{\affiliation{Department of Physics, University of Warwick, Coventry, CV4 7AL, UK }}
\newcommand{\AFFral}{\affiliation{Rutherford Appleton Laboratory, Harwell, Oxford, OX11 0QX, UK }}
\newcommand{\AFFwu}{\affiliation{Faculty of Physics, University of Warsaw, Warsaw, 02-093, Poland }}
\newcommand{\AFFbcit}{\affiliation{Department of Physics, British Columbia Institute of Technology, Burnaby, BC, V5G 3H2, Canada }}
\newcommand{\AFFtohoku}{\affiliation{Department of Physics, Faculty of Science, Tohoku University, Sendai, Miyagi, 980-8578, Japan }}
\newcommand{\AFFicise}{\affiliation{Institute For Interdisciplinary Research in Science and Education, ICISE, Quy Nhon, 55121, Vietnam }}
\newcommand{\AFFilance}{\affiliation{ILANCE, CNRS - University of Tokyo International Research Laboratory, Kashiwa, Chiba 277-8582, Japan}}
\newcommand{\AFFibs}{\affiliation{Center for Underground Physics, Institute for Basic Science (IBS), Daejeon, 34126, Korea}}
\newcommand{\AFFglasgow}{\affiliation{School of Physics and Astronomy, University of Glasgow, Glasgow, Scotland, G12 8QQ, United Kingdom}}
\newcommand{\AFFoecu}{\affiliation{Media Communication Center, Osaka Electro-Communication University, Neyagawa, Osaka, 572-8530, Japan}}
\newcommand{\AFFminn}{\affiliation{School of Physics and Astronomy, University of Minnesota, Minneapolis, MN  55455, USA}}
\newcommand{\AFFsilesia}{\affiliation{August Che\l{}kowski Institute of Physics, University of Silesia in Katowice, 75 Pu\l{}ku Piechoty 1, 41-500 Chorz\'{o}w, Poland}}
\newcommand{\AFFtoyama}{\affiliation{Faculty of Science, University of Toyama, Toyama City, Toyama 930-8555, Japan}}
\newcommand{\AFFbmcc}{\affiliation{Science Department, Borough of Manhattan Community College / City University of New York, New York, New York, 1007, USA.}}
\newcommand{\AFFnumazu}{\affiliation{National Institute of Technology, Numazu College, Numazu, Shizuoka 410-8501, Japan}}

\AFFicrr
\AFFkashiwa
\AFFmad
\AFFbmcc
\AFFbu
\AFFbcit
\AFFuci
\AFFcsu
\AFFcnm
\AFFduke
\AFFllr
\AFFgifu
\AFFgist
\AFFglasgow
\AFFuh
\AFFibs
\AFFicise
\AFFicl
\AFFbari
\AFFnapoli
\AFFpadova
\AFFroma
\AFFilance
\AFFkeio
\AFFkek
\AFFkcl
\AFFkobe
\AFFkyoto
\AFFliv
\AFFminn
\AFFmiyagi
\AFFnagoya
\AFFkmi
\AFFpol
\AFFnumazu
\AFFsuny
\AFFokayama
\AFFoecu
\AFFox
\AFFral
\AFFseoul
\AFFsheff
\AFFshizuokasc
\AFFsilesia
\AFFstfc
\AFFskk
\AFFtohoku
%\AFFtokai
%\AFFtokyo
\AFFtodai
\AFFipmu
\AFFtit
\AFFtus
\AFFtoyama
\AFFtriumf
\AFFtsinghua
\AFFwu
\AFFwarwick
\AFFwinnipeg
\AFFynu

%%%%%%%%%%%%%%%%%%%%%%%%%%%%%%%%%%%%%%%%%%%%%%%%%%%%%%%%%%%%%%%%%%%%
%ICRR
\author{K.~Abe}
\AFFicrr
\AFFipmu
\author[0000-0002-2110-5130]{S.~Abe}
\AFFicrr
\author[0000-0001-6440-933X]{Y.~Asaoka}
\AFFicrr
\AFFipmu
\author[0000-0003-3273-946X]{M.~Harada}
\AFFicrr
\author[0000-0002-8683-5038]{Y.~Hayato}
\AFFicrr
\AFFipmu
\author[0000-0003-1229-9452]{K.~Hiraide}
\AFFicrr
\AFFipmu
\author[0000-0002-8766-3629]{K.~Hosokawa}
\author{T.~H.~Hung}
\AFFicrr
\author[0000-0002-7791-5044]{K.~Ieki}
\author[0000-0002-4177-5828]{M.~Ikeda}
\AFFicrr
\AFFipmu
% ...

%%%%%%%%%%%%%%%%%%%%%%%%%%%%%%%%%%%%%%%%%%%%%%%%%%%%%%%%%%%%%%%%%%%%%

\correspondingauthor{M.~Harada}
\email{mharada@km.icrr.u-tokyo.jp}
\correspondingauthor{A.~D.~Santos}
\email{andrew.santos@ipmu.jp}
\correspondingauthor{R.~Rogly}
\email{rrogly@llr.in2p3.fr}
\correspondingauthor{A.~Beauchêne}
\email{antoine.beauchene@llr.in2p3.fr}

\collaboration{248}{The Super-Kamiokande Collaboration}
\noaffiliation

% \todo[inline]{Update AAS tex v7.0?}
%\textcolor{red}{Change later!!!}
%\collaboration{20}{(AAS Journals Data Editors)}
%\collaboration{20}{(Super-Kamiokande Collaboration)}



%\todo[inline]{Abstract should be less than 250 words}
\begin{abstract}
We report the search result for the Diffuse Supernova Neutrino Background (DSNB) in neutrino energies beyond 9.3~MeV in the gadolinium-loaded Super-Kamiokande (SK) detector with $22,500\times956.2$$~\rm m^3\cdot day$ exposure.
%$22.5{\rm k}\times956.2$$~\rm m^3\cdot day$ exposure. 
Starting in the summer of 2020, SK introduced 0.01\% gadolinium (Gd) by mass into its ultra-pure water to enhance the neutron capture signal, termed the SK-VI phase. 
This was followed by a 0.03\% Gd-loading in 2022, a phase referred to as SK-VII.
We then conducted a DSNB search using 552.2~days of SK-VI data and 404.0~days of SK-VII data through September 2023. 
This analysis includes several new features, such as two new machine-learning neutron detection algorithms with Gd, an improved atmospheric background reduction technique, and two parallel statistical approaches.
No significant excess over background predictions was found in a DSNB spectrum-independent analysis, and 90\% C.L. upper limits on the astrophysical electron anti-neutrino flux were set.  
Additionally, a spectral fitting result exhibited a $\sim1.2\sigma$ disagreement with a null DSNB hypothesis, comparable to a previous result from 5823~days of all SK pure water phases. 
\end{abstract}

%\todo[inline]{Put keywords}


%theoretical
%\input{section/dsnb}
\section{Diffuse supernova neutrino background} \label{sec:dsnb}
Core-collapse supernovae (CCSNe) are known as some of the most dynamic phenomena in the Universe. 
To understand the CCSN mechanism, knowledge of the deep core of the exploding star is essential.
Neutrinos are one of the few ways to access the core of a star. Since they are not sensitive to electromagnetic interactions, the information encoded within a neutrino flux is largely unaltered. 
Owing to this, observing a time-dependent neutrino flux from a CCSN burst could provide important information about the CCSN explosion mechanism~\citep{totani_future_1998, kachelries_exploiting_2005, Janka2012, scholberg_supernova_2012, Takiwaki2014, mirizzi_supernova_2016, horiuchi_what_2018, Burrows2023}.
%Owing to this, detecting neutrinos from a CCSN burst could provide important information about the CCSN explosion mechanism from the neutrino energy spectrum and luminosity with emission time~\citep{totani_future_1998, kachelries_exploiting_2005, scholberg_supernova_2012, mirizzi_supernova_2016, horiuchi_what_2018}.
Despite the growing focus on detecting neutrinos from CCSNe, 
neutrino detectors are primarily sensitive to those occurring within our own galaxy, which are rare events~\citep{Tammann1994}.
%neutrino detectors worldwide are primarily sensitive only to the rare events occurring within our own galaxy~\citep{Tammann1994}.

Another avenue for studying CCSNe is through the observation of the cumulative neutrino fluxes from all past supernovae in the Universe.
This is termed the Diffuse Supernova Neutrino Background~(DSNB), or Supernova Relic Neutrinos~(SRNs).
%Current neutrino detectors around the world aim to observe the DSNB for the first time. 
For most detectors, the target signal channel is the inverse beta decay (IBD) of protons induced by electron antineutrinos due to the large cross-section within the MeV signal range.%, even though the DSNB flux comprises nearly equal amounts of all neutrino flavors due to flavor conversion via neutrino oscillations. 

The DSNB flux is affected by the cosmological expansion of the Universe, such that it is redshifted before reaching the Earth, and the amount of redshift depends on when each supernova occurred in the history of the Universe.  
The magnitude of the flux depends heavily on the supernova rate, which can be predicted using astrophysical measurements of the star formation rate (SFR). %of massive stars. 
%This modeling can predict the stellar star formation rate, referred to as the star formation rate (SFR).
Therefore, the magnitude and shape of the DSNB flux provide unique information about the cosmic history of massive star formation. Compared to the neutrino emission from CCSNe, the impact from Type-Ia SNe neutrino emission is estimated to be $10^{-6}$ smaller, and the effect is negligible~\citep{Anandagoda2023}.
%The DSNB flux shape is sensitive to various redshift contributions.
%Then, by combining flux shape and magnitude, we can learn about the star-formation rate depending on the redshift. 
The shape of the DSNB flux also results from the combined effect of various factors, such as the equation of state of neutron stars, the shockwave revival time of CCSNe, neutrino propagation in dense matter, and the stellar initial mass function~\citep{beacom_diffuse_2010, lunardini_diffuse_2016, suliga_towards_2022, ando_diffuse_2023}. %, volpe_neutrinos_2024}.
In addition, the neutrino mass ordering affects the DSNB spectral shape for each neutrino flavor. 
Furthermore, potential exotic physics, such as neutrino decay~\citep{Tabrizi2021, Ivanez2023}, general neutrino interactions with dark matter~\citep{farzan_dips_2014}, and non-trivial sterile-active neutrino state mixings~\citep{DeGouvea2020}, could impact the spectrum.

\begin{figure*}[htb!]
\epsscale{0.8}
%\plotone{figure/relic_flux_model_pub_169_2}
%\plotone{figure/relic_flux_model_pub_v3}
\plotone{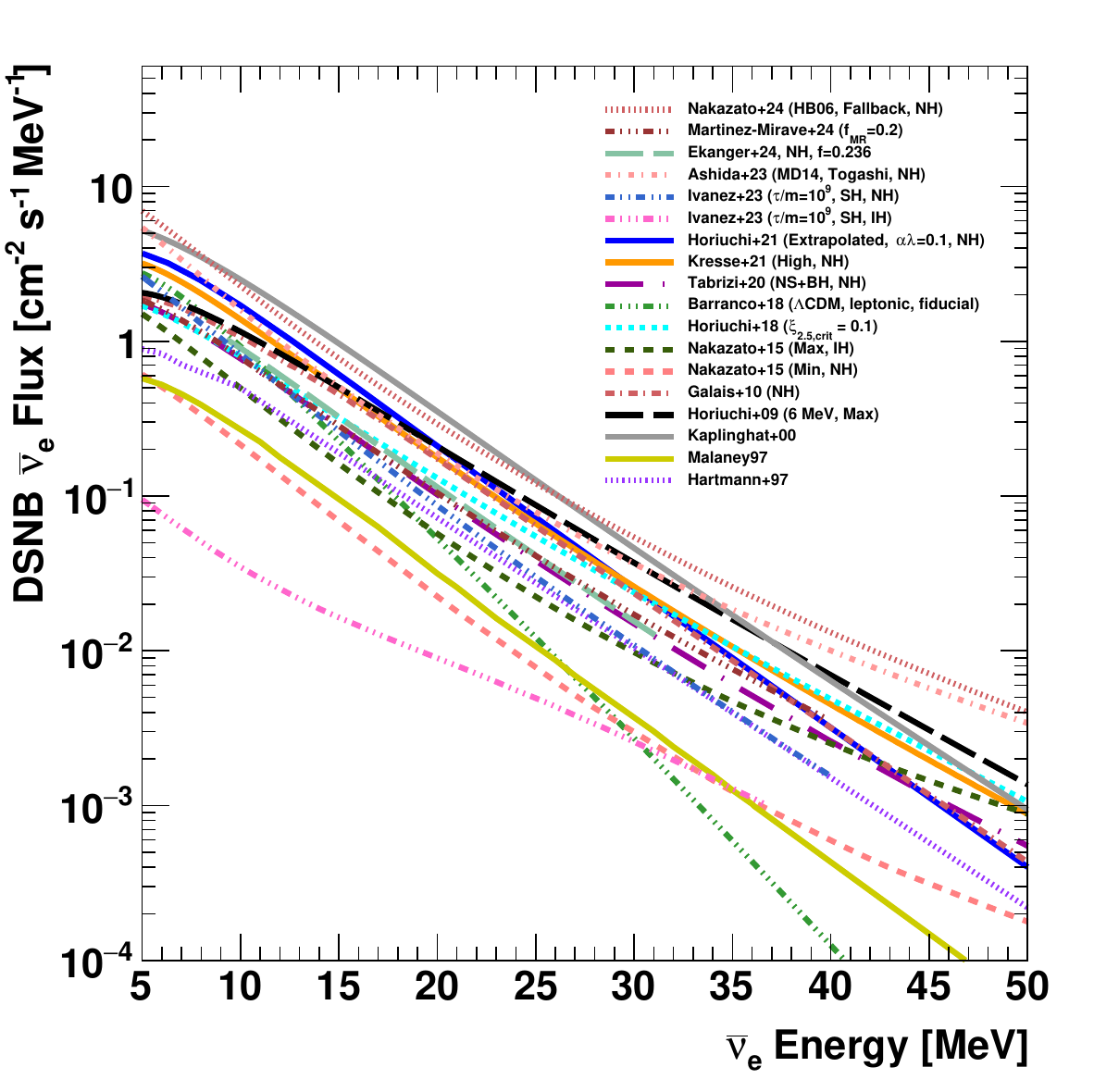}
\caption{\label{fig:dsnbfluxpred}
DSNB $\bar{\nu}_e$ flux predictions from various predictions~\citep{Nakazato2024, Martinez2024, Nick2024, Ashida2023, Ivanez2023, Horiuchi2021, Kresse2021, Tabrizi2021, Barranco2018, Horiuchi2018, Nakazato2015, Galais2010, Horiuchi2009, Kaplinghat2000, Malaney1997, Hartmann1997}. 
Representative parameter sets are chosen for some of the flux models:  
``NH'' and ``IH'' represent the neutrino normal and inverted mass hierarchy assumptions in the calculation, respectively. 
``HB06'' and ``MD14'' refer to the calculation of  \citet{Hopkins_2006} and \citet{Madau_2014} for the SFR, respectively. 
``SH'' represents the ``strongly hierarchical'' defined in \citet{Ivanez2023}.
For the \citet{Nakazato2024} model, only the fallback supernova contribution to the DSNB is shown. We only draw the maximum flux model of~\citet{Horiuchi2009} model with a neutrino temperature of 6~MeV.
Refer to each publication for further detailed descriptions.
%\todo[inline]{Add citation}
}
\end{figure*}

In recent years, advances in DSNB theoretical predictions have grown significantly. 
Figure~\ref{fig:dsnbfluxpred} summarizes modern DSNB $\bar{\nu}_e$ flux predictions.
The current upper bound of predictions, which is not experimentally excluded, is marked by the highest-flux assumptions for the astrophysical parameters of Kaplinghat+00~\citep{Kaplinghat2000}. 
%Those with higher flux predictions, such as \citet{Totani1995}, have been experimentally excluded at more than $3\sigma$ by \citet{Abe2021}.
A systematic investigation of combined factors contributing to the DSNB flux is performed by \citet{Nakazato2015}. 
The minimum and maximum fluxes of these combinations are shown in Figure~\ref{fig:dsnbfluxpred}. 

%Within the range of fluxes predicted by the models described above, there is an abundance of new predictions resulting from the latest observational astrophysics findings and related assumptions since the start of the 2020s.
%Recently, the black-hole-forming supernovae effect on the DSNB flux often manifests itself in a variety of ways in addition to the ordinary supernova explosion, as can be seen in \citet{Horiuchi2018}, which estimates the DSNB flux with a black hole fraction based on the compactness parameter of stars. 
In modern predictions, the impact on the DSNB flux of failed SNe (those forming black holes before the shockwave reaches the surface) alongside ordinary CCSNe is incorporated in various approaches, as seen in \citet{Horiuchi2018, Ashida2022, Ashida2023}.
%CCSNe introducing black hole formation before shockwave reaching surface, called failed SN, alongside ordinary core-collapse supernovae, on the DSNB flux is incorporated in various ways and argued, as seen in \citet{Horiuchi2018, Ashida2022, Ashida2023}. 
Moreover, the impact of binary star systems, including their mergers and mass transfer dynamics, is incorporated into the DSNB flux calculation, as argued in \citet{Horiuchi2021}, and then further updated in \citet{lunardini2025arxiv} based on modeling from \citet{Burrows2023}. %summarizes the update of astrophysical contribution, such as the binary star evolution and failed SN, by connecting between steller evolution simulation and CCSN simulation, by C-O core mass and compactness parameter. 
%The modeling of neutrino with C-O core mass and compactness, are done by fitting various CCSN burst simulations given by \citet{Burrows2023}. 
Another illustrative example is the work of \citet{Nick2022}, which considers the late-phase neutrino emission originating from the proto-neutron star (PNS) cooling in flux calculations, which is revisited in \citet{Nick2024} with an up-to-date 3D explosion model and SFR.

Although the existence of the DSNB is theoretically sound, the event rate on the Earth is quite low, $\sim~0.1$~event~kton$^{-1}$~yr$^{-1}$ for water Cherenkov detectors, and this signal is overwhelmed by backgrounds. 
Thus, despite the ensemble of dedicated background-reduction techniques, prior searches have only placed upper limits on the flux. 

Super-Kamiokande published a search result for the DSNB using 20 years of pure-water data~\citep{Abe2021} and placed the most stringent upper limit for the astrophysical electron antineutrino flux above the 15.3~MeV region.
In contrast, below 15.3~MeV, the DSNB searches conducted by liquid scintillator experiments such as KamLAND~\citep{KamLAND2022} and Borexino~\citep{Borexino2021} can set tighter upper limits. %These results are reviewed below.

Recently, the Super-Kamiokande experiment started a new detector phase using dissolved gadolinium sulfate, termed the `Super-Kamiokande Gadolinium project', or `SK-Gd', to further reduce backgrounds and enhance the signal generated by neutron captures~\citep{Beacom2004, Abe2022a, Abe2024}. 
Thanks to the increased signal efficiency from Gd-loading, the first result of SK-Gd~\citep{Harada2023} showed comparable DSNB sensitivity to the pure-water Super-Kamiokande result~\citep{Abe2021}, which had five times the livetime of this SK-Gd search. 

Here, we present the results of the DSNB search using 956.2~days of SK-Gd data, which include updates to neutron detection techniques for SK-Gd, a new background reduction strategy, and two statistical analysis methods.
%This article describes the updates in neutron detection techniques for SK-Gd, background reduction, statistical analysis methods, and the results of the DSNB search using 956.2~days of SK-Gd data. 
This article is organized into the following sections: 
In Section~\ref{sec:superk}, we describe the Super-Kamiokande detector, specifically its configuration and data acquisition.
%---namely, the configuration of the detector and the data acquisition.
%In Section~\ref{sec:superk}, we describe the Super-Kamiokande detector---namely, the configuration of the detector, the data acquisition system, and the characteristics of each detector phase. 
In Section~\ref{sec:sigbkg}, we introduce the DSNB signal and backgrounds in the $\mathcal{O}$(1--10)~MeV region.
Section~\ref{sec:eventselection} details the event selection scheme to isolate the DSNB signal while removing background events. 
%Section~\ref{sec:eventselection} shows the event selection scheme to isolate DSNB $\bar{\nu}_e$ while removing background events in the $\mathcal{O}$(1--10)~MeV region. 
In Section~\ref{sec:binnedanalysis}, we divide the data into energy bins to compare the predicted and observed events after background reduction. 
With this, we search for an astrophysical $\bar{\nu}_e$ flux by testing a background-only hypothesis. 
Next, in Section~\ref{sec:spectralfitting}, we introduce an energy spectrum analysis with unbinned probability density functions (PDFs), providing details on the fitting procedure and subsequent results.
In the final two sections, we present the results obtained, draw conclusions for this study, and discuss future prospects.
%we discuss the results obtained, the conclusions for this study, and the prospects.

%experiment
%\input{section/detector}
\section{Super-Kamiokande Gadolinium project} \label{sec:superk}
Super-Kamiokande (SK)~\citep{Fukuda2003} is a large underground water Cherenkov detector experiment, consisting of a volume of $50,000~\rm m^3$ of water. 
%Its current configuration has Gadolinium introduced to the tank, which denotes the SK-Gd phase. 
The detector is located 1000~m underground (2700~m.w.e.) in the Kamioka mine in Japan.
This overburden enables the reduction of muons originating from cosmic-ray interactions in the atmosphere, known as `cosmic-ray muons', by a factor of $10^5$, limiting their crossing rate to approximately 2 Hz throughout the entire detector. 
The detector is cylindrical in shape with a diameter of 39.3~m and a height of 41.4~m. 
The tank is optically separated into an inner detector (ID) to observe physics events and an outer detector (OD), which surrounds the ID for vetoing cosmic-ray muon events. 

The ID is 33.8~m in diameter and 36.2~m in height. 
On the surface of the ID tank, 11,129 20-inch photomultiplier tubes (PMTs) are mounted facing inward, corresponding to approximately 40\% photocathode coverage. 
%The gaps between PMTs are covered by a black sheet to reduce light reflection.
A black sheet covers the gaps between PMTs to reduce light reflection.
There is a buffer of 2~m between the ID and OD support structure and the outer walls of the tank, which defines the OD. 
It is equipped with 1185 8-inch PMTs mounted on the outside of the PMT support structure facing outward and contains a total volume of 17,500$~\rm m^3$.
%It is equipped with 1185 8-inch PMTs facing outwards from the outer surface of ID PMT support structure and contains 17.5~kton of the total volume. 
The outer walls of the tank and the space between OD PMTs are lined with a reflective layer made from Tyvek to enhance the detection efficiency of Cherenkov photons produced by cosmic-ray muons.
%The outer walls of the tank and the space among OD PMTs are covered by a white Tyvek sheet with high reflectivity due to collecting the photons to tag the cosmic-ray muons.

The ID PMTs have a 3-ns timing resolution with about 21\% quantum efficiency at a peak wavelength of $\sim$~$380~\rm nm$. 
The water quality in the SK detector tank is tightly controlled, circulated, and purified~\citep{Abe2022a}. 
Due to the energy resolution achieved by the large number of high-performance PMTs and well-controlled water quality, SK is sensitive to a wide energy range of neutrino events, spanning from a few MeV to a few TeV.

Data acquisition in SK is achieved using online triggers based on the number of PMT hits within specific time windows.
The trigger process employs multiple thresholds based on the number of PMT hits within a 200-ns window, which is the approximate time it takes for a Cherenkov photon to traverse the diagonal of the ID. 
These triggers classify event types and store PMT hits associated with each event in computer disk storage.  
We also have a trigger using OD PMTs, called the `OD trigger,' to veto cosmic-ray muon events.
Notably, we can identify events with energies above approximately 6~MeV using the Special High Energy (SHE) trigger, which collects the hits in a $[-5, 35]~\rm \mu s$ around the main hit timing peak, after the upgrade of electronics in 2008~\citep{Yamada2010}. 
Furthermore, this trigger creates a subsequent $500$-$\mu$s wide timing window, termed the ``after  (AFT) trigger'' window, 
to collect all hits that occur within this time interval, 
allowing for the offline search of delayed neutron captures.
%with which delayed neutron captures can be searched offline. 
Thus, we can search for neutrons within a total of $535~\rm \mu s$ from the main trigger timing. 

A novel event selection technique utilizing the detection of accompanying neutrons, termed ``neutron tagging,'' became available with this electronics upgrade, and demonstrated by \citet{Zhang2015} and \citet{Watanabe2009}.
The maximum AFT trigger rate was approximately once every 21~ms until March 3rd, 2022. 
After that, this trigger rate was increased to three times per 21~ms.
In general, trigger thresholds are controlled based on the trigger rate and dark hits rate, which can change with time. 
However, for the SK-Gd period considered in this DSNB search, the SHE trigger threshold is stably set to 60 hits, and the OD trigger threshold remains fixed at 22 hits for almost the entire period until 2023.

From July 2020, gadolinium (Gd) has been introduced into the pure water of SK, marking the start of SK-Gd. 
In SK-Gd, thermal neutron capture on Gd (mainly ${}^{157}\rm Gd$) enables a brighter signal than conventional pure water data in SK, resulting in about a total of 8~MeV.
We confirmed that detecting Gd-signals improves neutron identification~\citep{Harada2022}. 
Until June 2022, SK-Gd operated with a Gd concentration of 0.011\%~\citep{Abe2022a}, a period referred to as `SK-VI.'
In SK-VI, the neutron captures on Gd account for about half of all captures. 
In 2022, the concentration of Gd was increased to 0.033\%~\citep{Abe2024} to start the period `SK-VII.' 
This improvement increases neutron captures on Gd from around 50\% to 75\% of all captures. 
Table~\ref{tab:runsummary} summarizes trigger conditions, Gd concentration, and operational live time.
\begin{deluxetable}{c c c c}[htb!]
\tablecaption{ Summary of the Gd concentration, AFT trigger rate limit, and live time for each phase. \label{tab:runsummary}} 
\tablecolumns{4}
\tablewidth{0pt}
\tablehead{
    \colhead{ SK phase } &
    \colhead{ Gd conc. } &
    \colhead{ AFT limit } & 
    \colhead{ Live time [days]}
}
\startdata
SK-VI & 0.011\% & 1/21~ms & 474.1 \\
SK-VI & 0.011\% & 3/21~ms & 78.1 \\
SK-VII & 0.033\% & 3/21~ms & 404.0 
\enddata
\end{deluxetable}

Events passing the SHE trigger requirement are further classified based on the presence of a coincident OD trigger, which is caused by high-energy events in the OD, such as high-energy electron-like events and incoming cosmic-ray muons. 
The SHE event is regarded as the ``prompt'' event, while the neutron capture in the AFT trigger is called the ``delayed" event.
We will continue to use this terminology throughout the rest of this article.
The vertex, direction, energy, and other basic characteristics of the prompt event are reconstructed using the same algorithm as the SK solar neutrino analysis~\citep{Hosaka2006, Cravens2008, Abe2011a, Abe2016a, Abe2024a}. 
In what follows, we employ the conventional expression of the event energy introduced in~\citet{Abe2016a, Abe2024a}, and use electron or positron equivalent kinetic energy $E_{\rm rec}$ by subtracting the electron mass 0.511~MeV from the total reconstructed energy.

%signal and background
%\input{section/signalbackground}

\section{Signal and background} \label{sec:sigbkg}
This analysis targets inverse beta decay (IBD) events from electron antineutrinos, whose resulting positrons have energies in the range of $\mathcal{O}(1-10)$~MeV.
The IBD process has the largest cross section in this signal energy region and is accompanied by a neutron.
By requiring the coincidence of one neutron with the prompt positron, most background events without subsequent neutron capture---such as solar neutrinos and radioimpurity decays---are rejected. 
Major background events in this energy region after neutron tagging include reactor neutrinos, decays of radioactive isotopes from muon spallation on oxygen nuclei, and atmospheric neutrinos. 
This section provides a detailed description of the modeling of the signal flux and each background source. 
Signal and background estimations are done using SKG4, which is a Geant4~\citep{Agostinelli2003, Allison2006, Allison2016}-based detector Monte-Carlo (MC) simulation software for SK~\citep{Harada2020}.

\subsection{DSNB Signal Modeling}
The prompt signal events in this search are positrons from IBD, and the delayed signal is a neutron capture.
The kinematics of each positron, neutron, and initial electron antineutrino---such as directional correlation among initial and final state particles and energies---are computed by SKSNSim~\citep{Nakanishi2024} based on the \citet{Strumia2003} calculation. 
The interaction vertex is sampled uniformly in the ID tank, and the incoming direction of neutrinos is assumed to be isotropic.
To produce a wide variety of DSNB theoretical models, we generate IBD events with total positron energies ranging from 1 to 90 MeV uniformly and then apply weighting factors to the MC events according to various DSNB predictions afterwards. 
Some of the background sources, such as reactor neutrinos and spallation $^{9}\rm Li$, which will be introduced later, are also modeled in this way.

%with statistics proportional to the live time of each run operation. 

\subsection{Atmospheric Neutrinos} \label{subsec:atmbg}

Events originating from atmospheric neutrino interactions form a significant background, despite the fact that atmospheric neutrinos are more energetic than the DSNB search region, ranging from a few hundred MeV to GeV. 
This is because the prompt events generated by atmospheric neutrinos do not always carry the majority of their initial energy.
The first of these backgrounds is neutral current quasi-elastic (NCQE) interactions, which are significant below $E_{\rm rec} = 20~\rm MeV$. 
For any flavor, atmospheric neutrino NCQE scattering off oxygen yields 
\begin{eqnarray}
&&\nu(\bar{\nu}) + {}^{16}\text{O} \to \nu(\bar{\nu}) + {}^{15}\text{O}^*+n, \nonumber \\
&&\nu(\bar{\nu}) + {}^{16}\text{O} \to \nu(\bar{\nu}) + {}^{15}\text{N}^*+p.
\end{eqnarray} 

\noindent
For these interactions, a nucleon is ejected, and the remaining daughter nucleus promptly emits de-excitation gamma rays~\citep{Langanke1995, Ankowski2013}. 
De-excitation through $\gamma$-emission is determined by the oxygen shell from which the nucleon is ejected. 
The energy of the de-excitation gamma ray is mostly below 10~MeV.
However, de-excitation by gamma rays sometimes occurs above 10~MeV when the $s_{1/2}$ state is involved~\citep{Ejili1993, Ankowski2012}. 
A more detailed picture of the de-excitation processes in oxygen nuclei during NCQE interactions is provided by the T2K experiment~\citep{Abe2014c, Abe2019b, Abe2025}, and by SK analyses using atmospheric neutrinos~\citep{Wan2019, Sakai2024}.

At higher energy regions above 16~MeV in the DSNB search energy window, charged current quasi-elastic (CCQE) interactions and pion-producing events make a notable contribution. 
A representative event type contributing to these backgrounds is an electron from muon decay, including those from muons originating from the decay of charged pions. 
When the muons or pions are below their Cherenkov thresholds, only the electron signal will be visible. 
At these energies, the muons come to rest such that the electron-reconstructed energies form a Michel spectrum, which is below 50.8~MeV. 
These Michel electrons form the dominant background just above the DSNB energy region of interest. 

Also, the CCQE scattering off hydrogen and oxygen of electron-type neutrinos can directly produce electrons. 
For example, in the case where an atmospheric electron antineutrino interacts through IBD, this is exactly the same as the DSNB IBD signal.
Given that the energy of this electron reflects the parent neutrino energy, the expected event rate increases with energy, unlike the invisible muon decay events that peak around 50~MeV. 
Thus, in the DSNB signal region, this type of event is secondary to those caused by invisible muon decay.
\begin{comment}
We should also note that, as a secondary effect, visible but very low-energy muons and pions behave as DSNB. 
These can be easily rejected because the typical Cherenkov light shape originating from muon and pion is quite different from electron's. 
\end{comment}

To simulate atmospheric neutrino events, we utilize the HHKM2011~\citep{Honda2007,Honda2011} flux model as input to the neutrino event generator NEUT, version~5.6.4~\citep{Hayato2021}. 
Since atmospheric neutrinos are largely at $\mathcal{O}$(100)~MeV to GeV-scale, the momentum imparted on nucleons sometimes allows for secondary interactions with other nuclei in water.
After the neutrino interaction, the propagation of the produced particles is simulated by SKG4. 
In contrast to the conventional SK simulation conducted with GEANT3, SKG4 allows for the selection of hadronic interaction models, including those that account for the behavior of fast neutrons, by changing the physics list for each particle.
This time we selected the Li\`{e}ge intranuclear cascade (INCL) model~\citep{Boudard2013}. 
INCL adopts a $G4PreCompound$ model for nuclear deexcitation~\citep{quesada2011}, based on \citet{Gudima1983}, and it affects the neutron and gamma-ray emission as a final state of atmospheric neutrino events.
From the discussion by \citet{Hino2025arxiv} based on measurements of de-excitation gamma-rays from oxygen after interaction with a fast neutron~\citep{Ashida2024, Tano2024}, this model agrees more precisely with the experimental data than the conventional nuclear de-excitation model named Bertini (BERT) model~\citep{wright2015}.
Additionally, \citet{Sakai2024, Han2025, Abe2025} support the INCL model, showing better agreement for the number of neutrons and gamma-ray emission between SK measurement and MC simulation than the BERT model.

\begin{comment}
These can produce secondary $\gamma$-emission on the timescale of the initial knock-out nucleon thermalization. Since this thermalization can be fast enough to be contained within the 200~ns of the SHE prompt trigger window, PMT hits from the initial NCQE interaction and from secondary $\gamma$-emission can be collected together. The multiple $\gamma$-emission then leads to multiple Cherenkov cones in the prompt event, and the total prompt energy can easily exceed 10~MeV. This puts them in the DSNB signal energy region. In addition, a varying number of neutrons can be produced in the final state due to the secondary interactions of the initial knock-out nucleon. More details about NCQE interactions and their associated systematic uncertainties in SK were studied by~\citet{Sakai2024}.
\end{comment}

\subsection{Cosmic Ray Muon Spallation}\label{subsec:spabg}

The SK detector is exposed to cosmic ray muons at a rate of about 2~Hz. 
%Although the 1000~m overburden shields SK from cosmic ray muons by reducing their flux by several orders of magnitude, the detector is still exposed to the remaining rate of about 2~Hz. 
%These muons may break up oxygen nuclei through spallation with the production of hadronic showers.
These muons create electromagnetic and hadronic showers, and these may break up oxygen nuclei through spallation.
These showers finally result in the creation of radioisotopes, of which the subsequent $\beta$ decays with a time scale of $\mathcal{O}(0.01)$ to $\mathcal{O}(10)$~s mimic the signal of a DSNB prompt event. 
%This radioisotopes decay on a time scale of $\mathcal{O}(0.01)$ to $\mathcal{O}(10)$~s. 
Given the weak timing correlation between the muon event and the spallation event compared to the muon crossing rate, removing the spallation background using correlation with the muon is difficult. 
Also, the event rate of this type of background below 20~MeV in SK is about $10^6$ times higher than DSNB-predicted event rates, making it the most harmful background at energies below 16~MeV. 

Most of the spallation events produce a single $\beta$ particle with an energy below 20~MeV, which can be largely removed by neutron tagging. 
However, some of the radioisotopes, such as $\rm {}^8He$ and $\rm {}^9Li$, produce neutrons in coincidence with their $\beta$ decay. 
%Although the typical neutron energy from these nuclei is a few MeV higher than that of IBD, we cannot directly measure the neutron's energy or separate the short distances between the interaction and capture vertex through our event reconstruction.
This striking similarity to the topology of IBD events mimics the DSNB signal. 
In addition, accidental coincidences between $\beta$ signals and PMT noise-hit clusters, or signals from radioactive decay of radon~\citep{Nakano2020}, inevitably remain even after requiring the detection of one neutron capture. 
Thus, it is necessary to employ a dedicated reduction technique exploiting various correlations between muons, hadronic showers, and spallation isotopes.

\begin{figure}[htb!]
\epsscale{1.}
\plotone{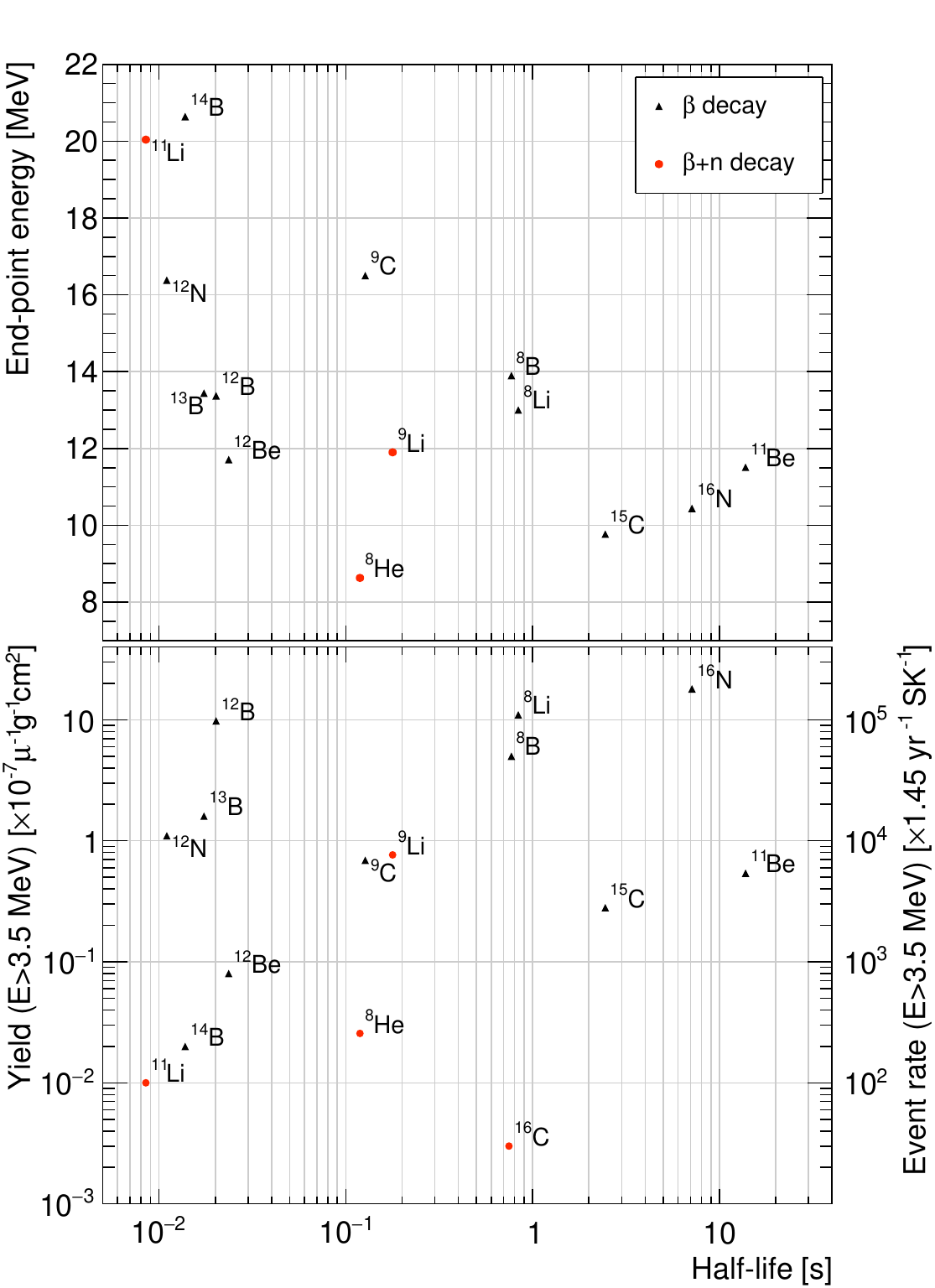}
%\plotone{figure/spatime_merge.pdf}
%\plotone{figure/spatimeene.pdf}
%\plotone{figure/spatimeyield2.pdf}
\caption{\label{fig:spatimeyield}
The half-lives, endpoint energies, and yields of $\beta$ decays of expected radioisotopes produced by oxygen nucleus spallation from cosmic ray muons. 
The yields are taken from \citet{Li2014}, and the yields for these isotopes account for the fraction of neutrons accompanying the channel.
Event rates are calculated considering the average muon rate ($\sim2~\rm Hz$) and average track length ($\sim2300~\rm cm$).
}
\end{figure}

Muon spallation characteristics in SK are studied using simulations based on the FLUKA toolkit~\citep{Battistoni2007} by~\citet{Li2014, Li2015a, Li2015b, Nairat2024}, and demonstrated by \citet{Scott2024}. 
Figure~\ref{fig:spatimeyield} summarizes the lifetimes, endpoint energies, and yields of spallation radioisotopes above 3.5~MeV.%, which is close to the SK energy lower limit. 
%The spallation isotope event rate is significantly larger than the expected event rate of DSNB, $\mathcal{O}(0.1)$ events per year in SK, without any event selection.

Radioisotopes shown with red markers in Figure~\ref{fig:spatimeyield} represent those that have a $\beta+n$ decay branch, such as $\rm ^{11}Li$, $\rm ^{16}C$, $\rm ^{8}He$, and $\rm ^{9}Li$ in this case. 
$\rm ^{11}Li$ has a short half-life, which can be easily removed using time correlation with the parent muon. 
In addition, the yield of $\rm ^{11}Li$ is expected to be rather small compared to other spallation isotopes; thus, the contribution of $\rm ^{11}Li$ can be neglected. 
Similarly, the $\rm ^{16}C$ yield is quite small and the endpoint energy of the $\rm ^{16}C$ $\beta+n$ decay channel is approximately 5.5~MeV, which falls outside the range of Figure~\ref{fig:spatimeyield}; therefore, this is negligible in this analysis.
Furthermore, $\rm ^{8}He$ is a subdominant component due to the low yield compared to $\rm ^{9}Li$. 

In contrast, $\rm ^{9}Li$ has a relatively long half-life~(0.178~s) and has a high yield compared to other isotopes. 
The total yield of $\rm {}^{9}Li$ is $0.76\times10^{-7}~\mu^{-1}~{\rm g }^{-1}~{\rm cm}^2$ above 3.5~MeV, with a sufficiently high end-point energy to contaminate the DSNB signal energy window.
Therefore, $\rm ^{9}Li$ forms a non-negligible background even after neutron tagging in this analysis.

Although liquid-scintillator experiment measurements have demonstrated better agreement with theory~\citep{SAbe2023,An2024}, the predicted yields of spallation isotopes from oxygen are still inconsistent with measurements; as \citet{Zhang2016} showed, the measured yield in SK is smaller than the expectation in \citet{Li2014} by a factor of 3.1–4.7. 
This indicates that there is still a limited understanding of the composition of radioisotope production. 
Thus, we employ yield measurement results in our analysis.

\subsection{Reactor Neutrinos}\label{subsec:reactor}
\begin{figure}[htb!]
\centering
\epsscale{0.9}
\plotone{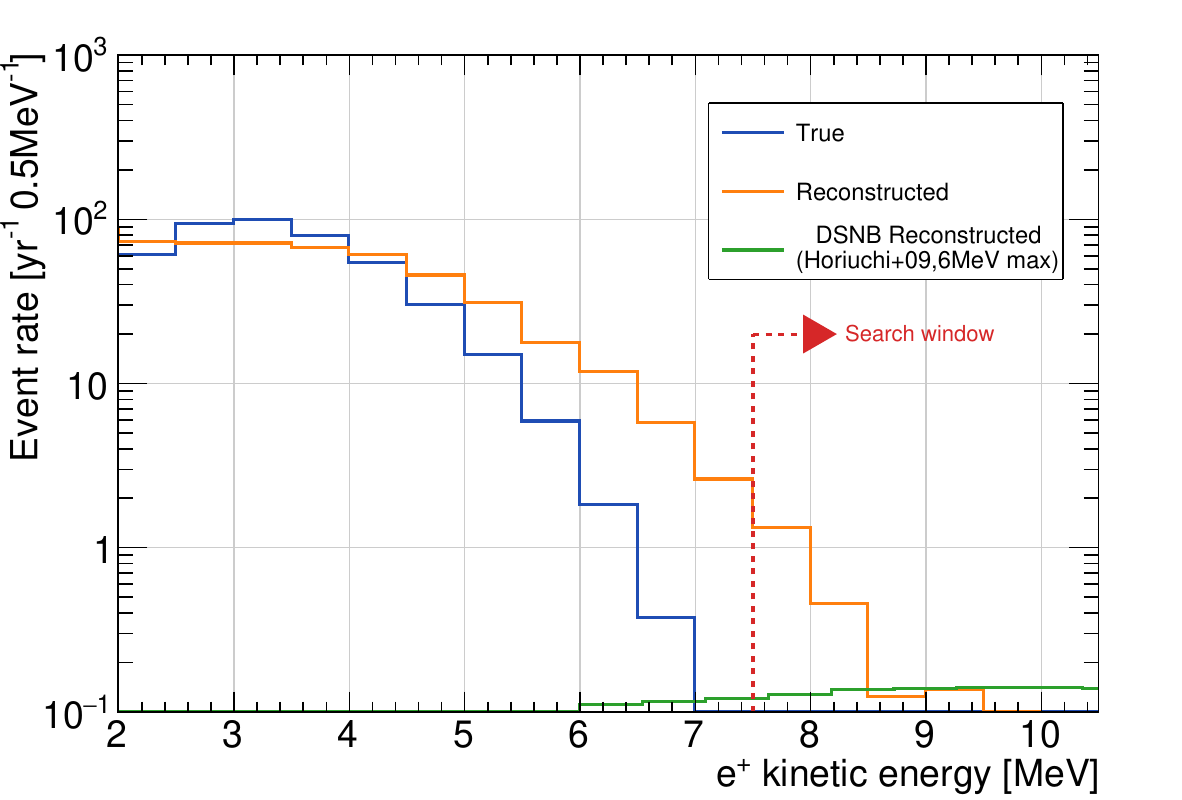}
%\plotone{figure/reactorflux_sk6_v2.pdf}
\caption{\label{fig:reactor}
Positron kinetic energy of predicted reactor neutrino IBD events. An example of the DSNB flux is also shown as well.
The vertical axis shows the average event rate per year. 
The variation of the operating period of Japanese reactors is averaged into the IBD event creation rate.
}
\end{figure}

Electron antineutrinos created in nearby reactors irradiate SK. 
Then, these neutrinos undergo IBD interaction and mimic DSNB signals because they have the same signature.
While we know the precise locations of these reactors, the directional information carried by neutrinos is mostly lost through IBD~\citep{vogel_angular_1999}.
The flux estimations of these reactor neutrino events are performed by~\citet{SKReact} based on the reactor neutrino model of \citet{Baldoncini2015}.
This calculation takes into account Japanese reactor activities during the SK-Gd observation period, along with neutrino oscillations due to the distance from reactors. 
%This flux at SK is predicted up to $E_\nu = 9.3~\rm MeV$. 
Figure~\ref{fig:reactor} shows the expected reactor neutrino event spectra during the SK-VI period as functions of both true and reconstructed kinetic energy, along with the DSNB flux example. 
%in the true and reconstructed kinetic energy spectra of expected reactor events along with the DSNB flux example. 
We can see that reactor neutrinos contribute significantly to the backgrounds covering the DSNB signal at MeV-scale energies. %{\bf below search window}. 
In addition, energy resolution effects induce more observed contributions of these backgrounds at slightly higher energies. 
%are what primarily determine the contribution of reactor neutrino events{\bf in the signal window.
Thus, in this search, we consider the reactor neutrino background up to  $E_\nu = 9.3~\rm MeV$, equivalent to 7.5~MeV in the positron kinetic energy.

\subsection{Solar Neutrinos}
The production chain of heavier elements in the Sun's core leads to the decay of ${}^8\text{B} \to {}^8\text{Be}^*+e^++\nu_e$. % for which the $\nu_e$ energy enters into the DSNB search region. 
For any neutrino flavor, electron elastic scattering through $\nu(\bar{\nu})+e^- \to \nu(\bar{\nu})+e^-$ off free electrons in SK will generate a prompt event, and the energy enters the DSNB search region.
Luckily, the following two features of this scattering make solar neutrinos an easily reducible background: First, since no neutrons are produced, neutron tagging will largely remove these events.
Second, for samples without a neutron tagging requirement,  we can still exploit the strong correlation between the reconstructed electron direction and the direction toward the Sun due to the forward-scattering nature of electrons~\citep{Abe2021,Abe2024a}.
%For samples without a neutron tagging requirement, though, we can still exploit the directionality of the forward-scattered electrons because the angle between the reconstructed electron direction and the direction toward the Sun provides a strong discriminator~\citep{Abe2021,Abe2024a}.

%\input{section/reduction}
\section{Event selection} \label{sec:eventselection}
This analysis searches for IBD events, characterized by the temporal and spatial coincidence of a prompt positron event and a delayed neutron event, resulting from thermal neutron capture on Gd or hydrogen nuclei. 
We apply a series of event selection criteria to observed data from the SK-VI and SK-VII periods associated with an SHE trigger and a subsequent AFT trigger when available. 
The lower energy threshold of the analysis region is set to $E_{\rm rec} = 7.5$~MeV due to the sufficient SHE trigger efficiency at this energy and the negligible amount of reactor neutrino background events (see Section~\ref{subsec:reactor}). 
The upper energy bound of the DSNB signal region of interest depends on the analysis method described later. 

The following sections describe four stages of event selections: primary noise reduction (Section~\ref{subsec:noisered}), spallation event reduction~(Section~\ref{subsec:spared}), atmospheric neutrino event reduction~(Section~\ref{subsec:atomred}), and delayed neutron identification~(Section~\ref{subsec:ntagred}).

\subsection{Basic Noise and Low-Quality Event Reduction} \label{subsec:noisered}
We first select the SHE-triggered events without OD triggers, and pair them with a subsequent AFT window when available. 
As noted in Section~\ref{sec:superk} and Table~\ref{tab:runsummary}, the AFT trigger rate varies depending on the detector phase.
%As noted in Section~\ref{sec:superk} and Table~\ref{tab:runsummary}, the AFT trigger rate was limited to one AFT per 21~ms until the middle of SK-VI, when it was increased to three AFT triggers per 21~ms. 
%Thus, the fraction of SHE triggers with a subsequent AFT trigger was equivalent to $\sim90\%$ on average before trigger modification and then above 99\% after trigger modification. 
Next, the collected events with $E_{\rm rec}$ below 79.5~MeV undergo a set of cleaning cuts to remove events from PMT noise,  
radioactivity from the detector wall, and cosmic-ray muon activity.
In particular, the candidate events are required to have a reconstructed vertex 2~m or more inside the ID wall to avoid radioactive backgrounds and poorer reconstruction performance. 
This defines a fiducial volume of 22,500~$\rm m^3$. 
To further remove backgrounds originating from the detector walls without shrinking the fiducial volume, we impose an energy-dependent ``effective" distance criterion. 
This distance is calculated from the wall to the reconstructed vertex along the axis defined by the reconstructed event direction.
In addition, we exclude events that occur within $50~\rm\mu s$ of high-charge events, defined as those with a total charge deposited on PMTs exceeding 500~p.e. equivalent. 
This cut rejects events activated by cosmic-ray muons, including decay electrons and nuclear events that occur rapidly following muon interactions. 
Finally, we apply a vertex reconstruction quality cut to remove non-electron-like noise events based on the PMT hit timing distributions per event. 
The inefficiency associated with this quality cut is below 1\%, as validated by the IBD signal MC simulation. 
Events that pass these reduction criteria are hereafter referred to as DSNB candidates.

\subsection{Spallation Reduction} \label{subsec:spared}
We utilize timing and spatial correlations between DSNB candidates and cosmic ray muons to remove the spallation background, called ``spallation cuts.''
%As described in Section~\ref{subsec:spabg}, no reliable simulation model for muon spallation is available for this analysis. 
In this analysis, a data-driven study is conducted to reduce the spallation background. 
Given that the maximum endpoint energies of the electrons or positrons are about 20.5~MeV, spallation event reduction algorithms are applied up to $E_{\rm rec} = 23.5$~MeV, taking into consideration energy resolution effects. 
The overall concept of this reduction is the same as that of previous SK analyses~\citep{Abe2021, Harada2023}, which consists of some pre-treatment cuts, a detailed likelihood approach, and a robust cut for high-energy spallation. 

One notable improvement from previous work~\citep{Abe2021} is that the shower neutrons from muon interactions are now identified by the Gd capture signal, as theoretically validated by \citet{Nairat2024}, and measured by \citet{Shinoki2023}. 
It has become possible to efficiently identify muons that are likely to cause hadronic showers, i.e., spallation.
The timing between DSNB candidates and muons causing a neutron shower, along with spatial correlations between DSNB candidates and the neutron shower, is employed to remove such background events.
More details of these ``neutron cloud cut'' criteria and other pre-treatments are described in Appendix~\ref{appendix:ncloudcut}.
Other steps for reducing the spallation events exactly follow those of previous searches~\citep{Abe2021, Harada2023}.

\subsection{Atmospheric Neutrino Reduction} \label{subsec:atomred}
To reduce atmospheric neutrino backgrounds, we employ the same event selection steps as in previous searches~\citep{Abe2021, Harada2023}. 
These make use of the reconstructed Cherenkov angle ($\theta_c$), the PMT activity before the main PMT hit peak from the prompt signal, the reconstructed particle decays after this peak, the clearness of the PMT hit pattern, 

\begin{eqnarray}
    L_{\rm clear} = \frac{N_{\rm t riplets}(\theta_c \pm 3^\circ)}{N_{\rm triplets}(\theta_c \pm 10^\circ)},
\end{eqnarray}

\noindent where the number of hit PMT triplets $N_{\rm triplets}$ are counted that give a Cherenkov angle within a given difference from the overall Cherenkov angle $\theta_c$, and the average charge deposited per PMT hit.
More details are provided in Appendix~\ref{appendix:thirdred}.

In addition, a new atmospheric neutrino background reduction step is introduced in this analysis to target NCQE events. 
These and certain CCQE processes can produce secondary $\gamma$-emission on the timescale of the initial knock-out nucleon thermalization. 
Since this thermalization is fast enough to be contained within the SHE prompt trigger window, PMT hits from the initial NCQE interaction and secondary $\gamma$-emission are collected together. 
The multiple $\gamma$-emission then leads to multiple Cherenkov cones in the prompt event, and the total prompt energy can easily exceed 10~MeV. 
Furthermore, a varying number of neutrons can be produced in the final state due to the secondary interactions of the initial knock-out nucleon.

\begin{figure}[ht!]
\centering
\epsscale{0.7}
\plotone{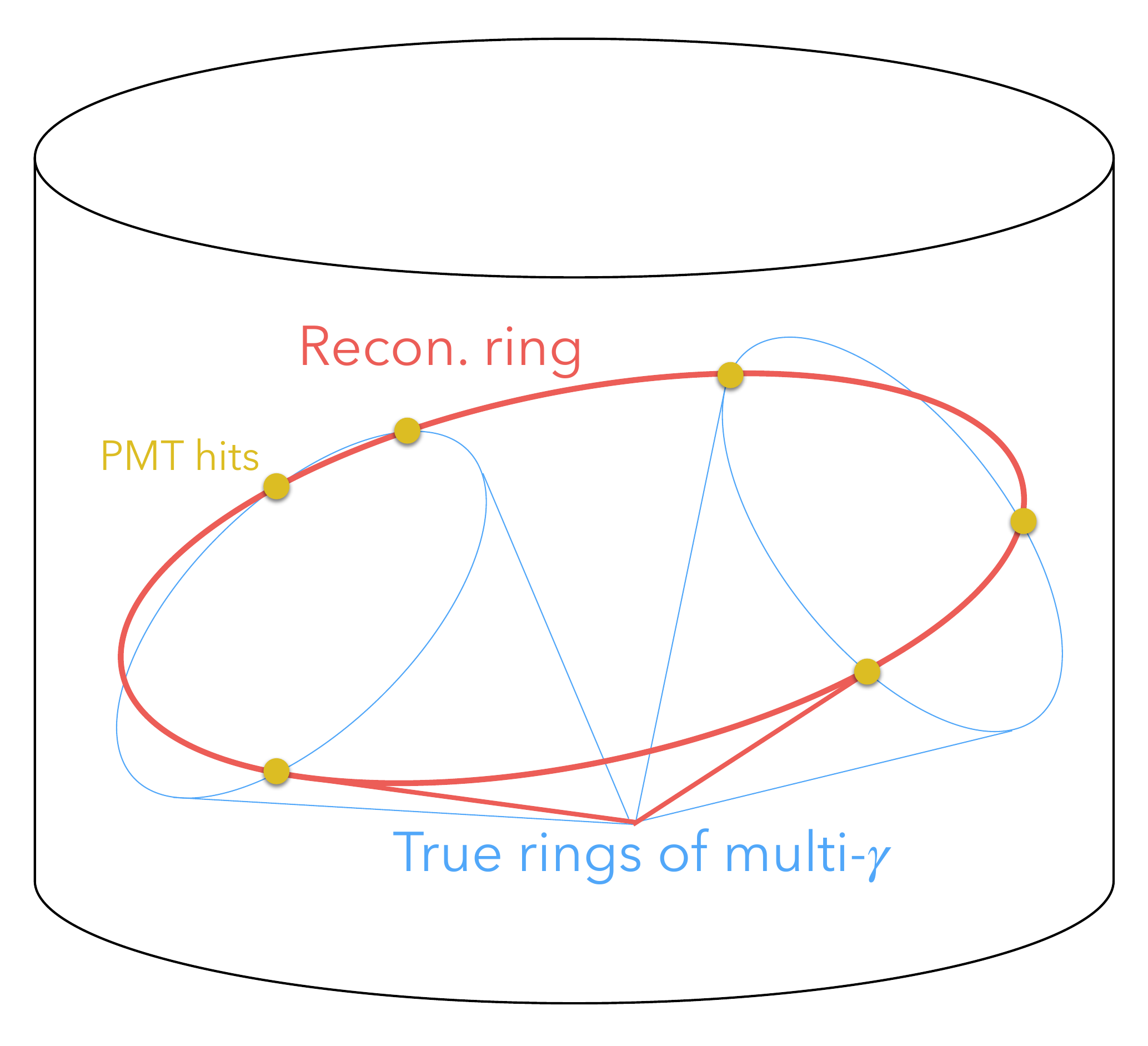}
%\plotone{figure/illust_multiring.pdf}
\caption{\label{fig:illustmultiring}
Illustration of a single event with multiple Cherenkov cones. The reconstructed $\theta_c$ appears to have a larger angle.
}
\end{figure}

In past analyses~\citep{Abe2021,Harada2023}, NCQE backgrounds were targeted in one of two main ways. 
The first was through the reconstructed Cherenkov angle ($\theta_c$) selection, and the second was the number of neutrons observed after the candidate prompt event. 
%in which the opening angle for a single cone is calculated for the SHE trigger. 
As first introduced by~\citet{Malek2003}, the reconstructed Cherenkov angle of multi-cone prompt events tends to have a large $\theta_c$ compared to the single electron-like event due to the hit pattern, as illustrated in Figure~\ref{fig:illustmultiring}.
Thus, events whose $\theta_c$ value is significantly larger than $42^\circ$ can be rejected as NCQE events. 
%The second method 
Next, the requirement of identifying a single neutron capture in the final state removes many NCQE events because NCQE interactions can have neutron multiplicities different from one. 
With these two methods, the NCQE remaining rate was reduced to below 10\%, but more NCQE events were still present compared to nominal DSNB predictions. 
%Precisely, the $\theta_c$ reconstruction is not always capable of detecting multiple Cherenkov cones that point in similar directions. 
If multiple Cherenkov cones in a prompt event point in similar directions, the reconstructed Cherenkov angle cannot distinguish the hit pattern from that of a single Cherenkov cone.

\begin{figure*}[hbt!]
\centering
\epsscale{0.8}
%\plotone{figure/msg_def_all_updated.PNG}
\plotone{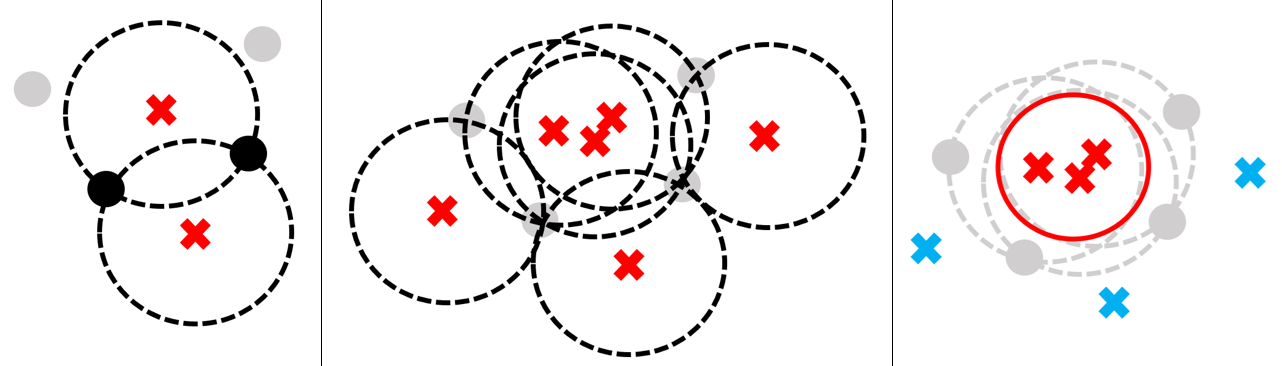}
\caption{Main steps for calculating MSG. (Left) Start with one pair of PMT hits (black dots). With origin at the reconstructed prompt vertex, define any cones with axis (red cross) and $42^\circ$ opening angle whose edges (black, dashed) coincide with both PMT hits. (Middle) Iterate through every possible pair of PMT hits. After this, each PMT hit pair will have zero, one, or two candidate cones. (Right) Identify the largest cluster of candidate cones, defined as those whose axes fit within $28.4^\circ$ of a single direction (red, solid). Axes in the cluster are kept in red, while those outside the cluster are changed to blue. Clustering is done in multiple iterations to maximize the magnitude of the sum of the largest cluster's axis vectors. MSG is defined as this magnitude divided by the total number of candidate cones.\label{fig:msg_def}}
\end{figure*}

A new reduction variable is introduced for further reducing NCQE backgrounds, termed the ``multiple scattering goodness'' (MSG) variable. 
In the context of the DSNB search, this variable was first introduced by~\citet{Bays2012} to quantify the multiple Coulomb scattering of electrons, thereby reducing solar neutrino backgrounds.
Since multiple Coulomb scattering limits the directional resolution of non-showering electrons, MSG provides a measure of angular resolution.
%This variable was originally developed to classify the reliability of electron scattering events from solar neutrinos and their directional reconstruction~\citep{Abe2024a}.
%This is an analogous situation to separating NCQE from IBD. 
%It has previously quantified the classification of the reliability of electron scattering events from solar neutrinos~\citep{Abe2024a}. 
It is also capable of distinguishing multi-$\gamma$ events from single-$e^\pm$ events more explicitly. 
%It has previously quantified the reliability of multiply-scattered electrons by solar neutrinos to reconstruct the initial neutrino direction~\citep{Abe2024a}.
%This is an analogous situation to separating NCQE from IBD. 
Instead of being sensitive to the overall PMT pattern for $\theta_c$, the MSG variable is sensitive to the substructure of the PMT hits. 
The main steps for calculating MSG are shown in Figure~\ref{fig:msg_def}.

For each event, this algorithm identifies cones with $42^\circ$ opening angle originating from the reconstructed prompt vertex that could explain the PMT hit pattern.
The axis of each candidate cone defines a unit vector $\vec{u}_{i}$ that points in the direction of the cone.
The value of MSG is the magnitude of the sum of the axis unit vectors in the largest cone cluster divided by the total number $N_{\rm cones}$ of candidate cones, or
\begin{eqnarray}
    {\rm MSG} = \frac{1}{N_{\rm cones}}\left | \sum_{i\in{\rm cluster}}\vec{u}_i\right|.
\end{eqnarray}
The largest cluster is taken as the most candidate cones whose edges fit within a broader cone of $50^\circ$ opening angle.

%\onecolumngrid
\twocolumngrid

\begin{figure}[htb!]
\centering
\epsscale{1.1}
\plotone{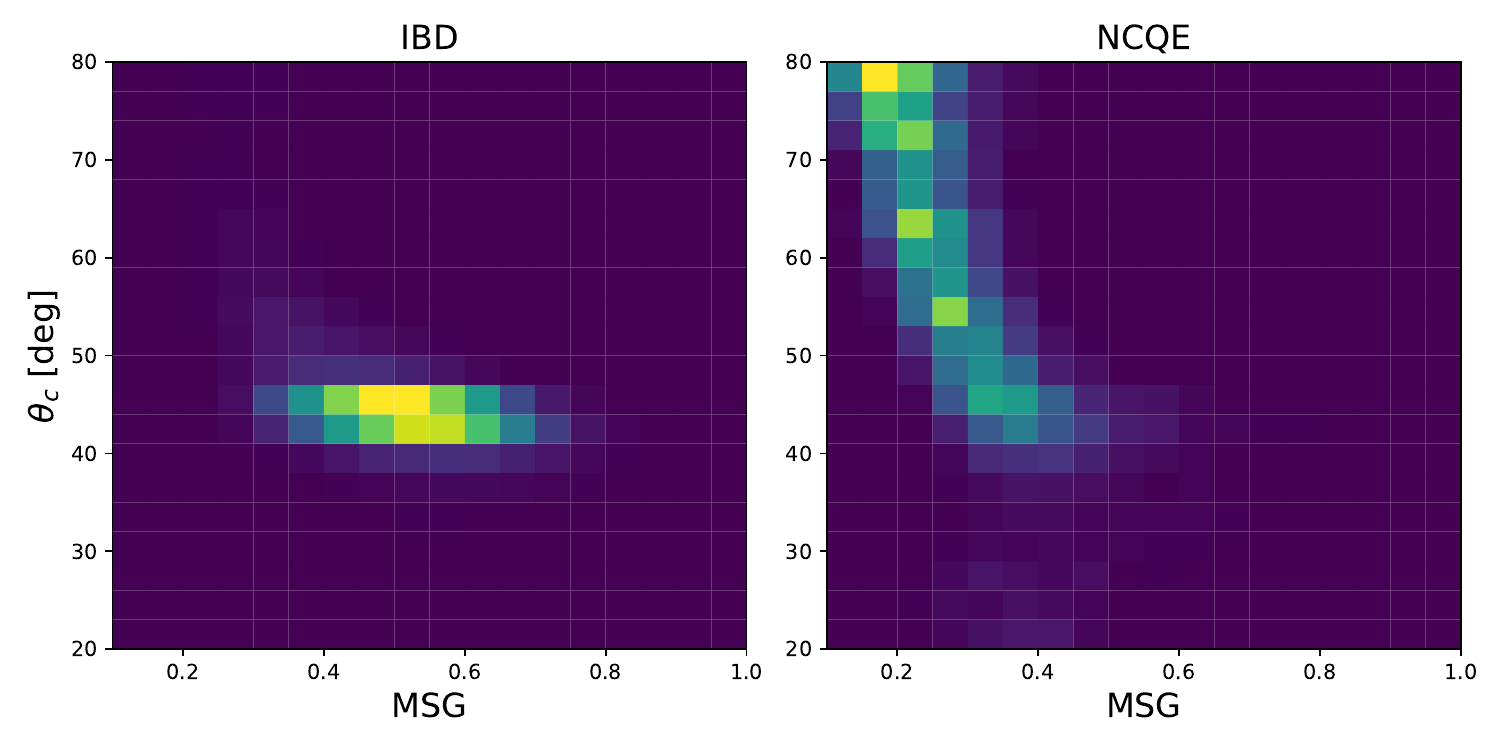}
%\plotone{figure/sk6_thetac_vs_msg_nothirdred_wgt.pdf}
\caption{Cherenkov angle versus MSG for IBD and NCQE events for $E_{\text{rec}}\in[11.5, 23.5]$~MeV. As the angle between multiple $\gamma$'s decreases for NCQE events, the MSG value increases. In the characteristic $\theta_c$ region for IBD signals, the NCQE MSG values are, on average, lower than those of IBD. \label{fig:thetac_vs_msg}}
\end{figure}

\begin{figure}[htb!]
\centering
\epsscale{0.8}
\plotone{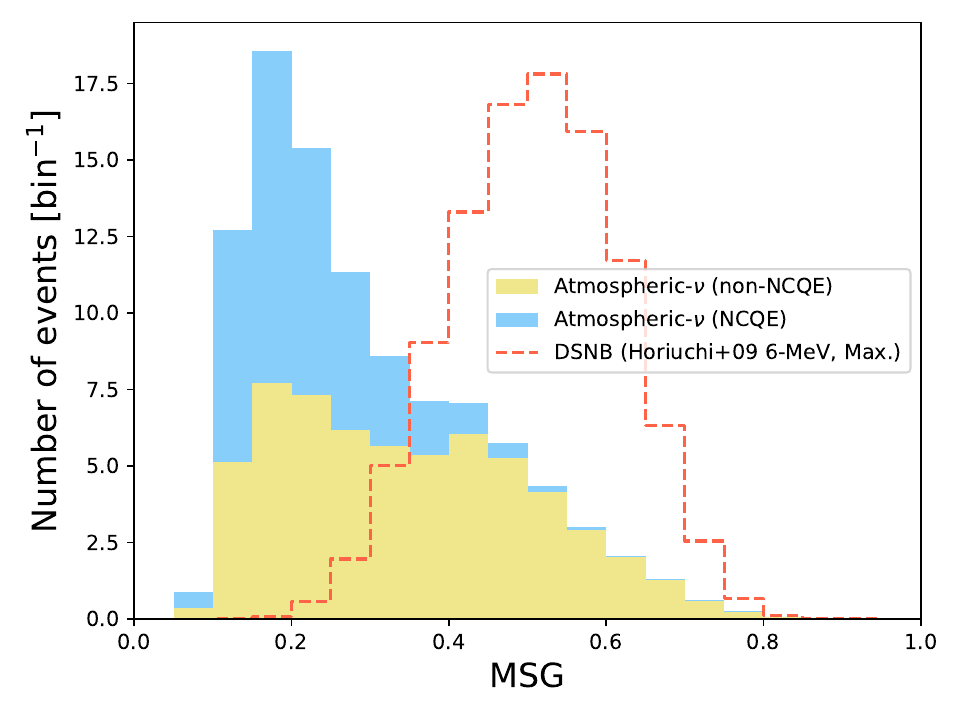}
\plotone{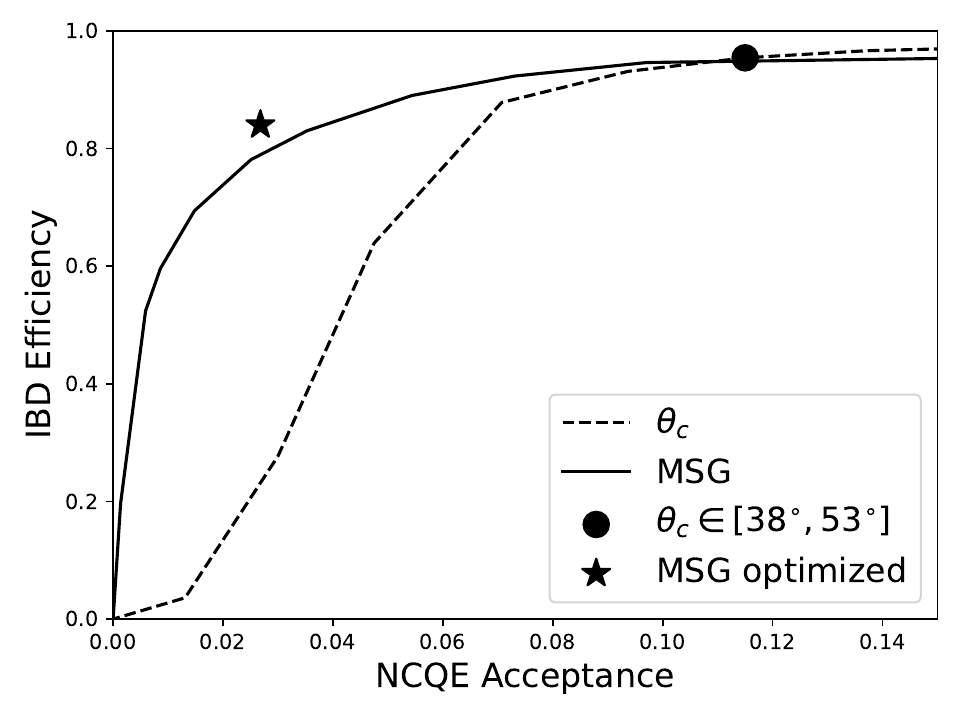}
%\plotone{figure/sk6_msg_dist_paper.pdf}
%\plotone{figure/final_roc_ncqe_paper.pdf}
\caption{\label{fig:msg}
MSG distributions for atmospheric neutrino backgrounds and IBD signal events (amplified by a factor of 100) for $E_{\text{rec}}\in[11.5, 23.5]$~MeV (top).  
ROC curves for MSG-only and $\theta_c$-only reduction with an old $\theta_c$ working point and MSG working point for $E_{\text{rec}}\in[11.5, 23.5]$~MeV as optimized in Section~\ref{sec:binnedanalysis} in 2~MeV bins for $E_{\text{rec}}\in[11.5, 23.5]$~MeV (bottom). 
Note that the MSG ROC curve shown is defined with a global MSG threshold over the whole energy range, which leads to an offset in the MSG optimized point.
}
\end{figure}

As shown in Figure~\ref{fig:thetac_vs_msg}, there is a population of NCQE events in the signal-like $\theta_c$ region around $42^\circ$, whereas these can be reduced by introducing MSG cut criteria.
Smaller MSG values indicate that multiple cones are more likely, while larger values suggest the presence of a single cone, as shown in the top panel of Figure~\ref{fig:msg}.
The event selection using MSG further distinguishes DSNB signals from NCQE background events beyond the conventional $\theta_c$ event selection, as shown by the Receiver Operating Characteristic (ROC) curve in the bottom panel of Figure~\ref{fig:msg}. In addition, part of the non-NCQE background, such as pion-producing interactions, produces multiple cones and can be reduced, as shown in Figure~\ref{fig:msg}.

\subsection{Neutron Tagging} \label{subsec:ntagred}
As described in Section~\ref{sec:superk}, SHE and subsequent AFT triggers record PMT hits within $[-5,~535]~\mu$s from the SHE trigger time. 
We can search offline for a hit cluster originating from a neutron capture and classify the prompt event based on the number of tagged neutrons.
%The delayed neutron captures that occur within this time window are then identified and paired with the prompt events. 
%The classification method for prompt events based on the following neutrons is termed ``neutron tagging.'' 
In this analysis, we require exactly one tagged neutron to identify the event as an IBD event.
Previous DSNB searches in SK used a Boosted Decision Tree (BDT) for neutron detection in pure water~\citep{Abe2021}, as well as a box cut-based neutron capture selection in the first Gd-loaded phase~\citep{Harada2023}. 
In this study, we retrained the BDT to include neutron captures on Gd and independently developed a Neural Network (NN) for neutron identification. 
DSNB search results using both approaches are discussed in the following sections, including the cross-validation of their performances and physics inferences.
Both neutron detection approaches include a pre-selection that requires hit clusters above a certain threshold in a given timing window, where the time-of-flight from the reconstructed prompt vertex to each PMT is subtracted.
These threshold criteria are defined as seven or more hits in 14~ns for the NN and as five or more hits in 10~ns for the BDT. 
Feature variables and an output score for both the NN and BDT are calculated for each hit cluster to judge whether the hit cluster is a neutron signal. 
The neutron detection efficiency $\epsilon_n$ is determined by $\epsilon_{\rm pre}\times\epsilon_{\rm score}$, where the $\epsilon_{\rm pre}$ represents the efficiency that the pre-selection picks a neutron,
%neutron picked by the pre-selection,
and $\epsilon_{\rm score}$ is the selection efficiency by the output score of both neutron identification tools.

The NN employs 12 variables representing the number of PMT hits, spatial features of PMT hits, the RMS of the hit timing peak, and the distance from the ID wall. These are calculated for each candidate searched from the window $[4, 535]~\rm\mu s$.
%The timing scan starts at $4~\rm\mu s$, and $\epsilon_{\rm pre}$ is 75.0\% for SK-VI and 85.1\% for SK-VII.
%The number of noise candidates per event after pre-selection, averaged over the entire periods, is 47 for SK-VI and 58 for SK-VII, respectively. These values vary by 5\% depending on the noise rate and water circulation for both periods.
%The timing scan starts at $4~\mu s$ from the prompt event, while the termination of the neutron search is defined $535~\rm\mu s$ for SK-VI and $270~\rm\mu s$ for SK-VII.
Details about variables and optimization are described in Appendix~\ref{appendix:nn}.

The BDT takes in 22 variables related to the spatial topology of the PMT hits, their timing distributions, and the charge deposited. 
In SK-IV, the neutron search window for the BDT sample began at 14~$\mu$s after the SHE-triggered timing, whereas it is placed at 2~$\mu$s in SK-Gd, as neutron captures happen faster due to Gd-loading. 
Further information on BDT neutron identification inputs can be found in Appendix~\ref{appendix:bdt}, and details about the training of the algorithm are provided by~\citet{Giampaolo2023}.

\begin{figure}[htb!]
\centering
%\PlaceHolder[0.5\linewidth]
%\plotone{figure/ntag_roc_sum.pdf}
\plotone{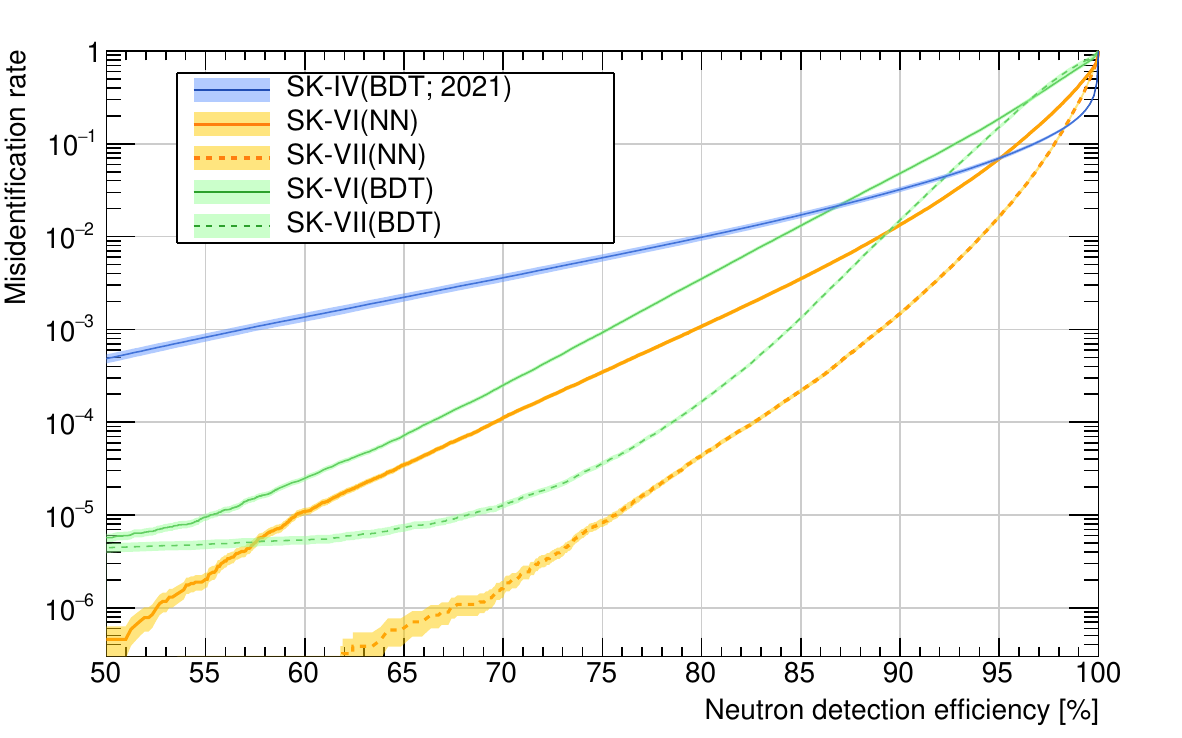}
\caption{
\label{fig:ntagroc}
Neutron detection efficiency $\epsilon_{\rm score}$ and misidentification probability of noise candidate. 
The $\epsilon_{\rm pre}$ for NN (BDT) are 75.0\% (77.2\%) for SK-VI and 85.1\% (87.3\%) for SK-VII.
The number of noise candidates per event after pre-selection, averaged over the entire periods, is 47 for SK-VI and 58 for SK-VII, respectively. 
%Signal efficiency of neutron tagging and background misidentification rate for the NN (Orange, solid line), and BDT (Green, short-dashed line), respectively. 
%BDT working points are energy-dependent below 29.5~MeV and are chosen such that the misidentification rate is below $10^{-3}$ per event. Here, the working points corresponding to the lowest and highest BDT misidentification rates used are shown.
%The error band representing the uncertainty in signal efficiency is evaluated by the agreement of the MC sample with the calibration data (See Appendix~\ref{appendix:ambe}).
%The red long-dashed line shows the working point of the NN referring to the misidentification rate.
}
\end{figure}

Both neutron identification approaches explore neutron-like clusters based on the NN/BDT output and count up the number of neutrons ($N_{\rm n}$) for each prompt DSNB candidate.
Figure~\ref{fig:ntagroc} shows the averaged neutron detection efficiency and misidentification probability as the output score threshold is varied for both NN/BDT. 
A comparison to the SK-IV pure-water BDT performance is included.
We can see the curve is significantly improved compared to the pure-water case, owing to the enhanced neutron signal by Gd.

%The number of noise candidates per event after pre-selection, averaged over the entire periods, is 47 for SK-VI and 58 for SK-VII, respectively. These values vary by 5\% depending on the noise rate and water circulation for both periods.
%\todo[inline]{Move plot to Section5.3.3}
%Figure~\ref{fig:ntagroc} shows the neutron tagging performance in the form of signal efficiency against the background misidentification rate for NN and BDT.  
%In the plot, adopt two performance metrics to be directly related to final samples requiring $N_{\rm n}=1$.
%First, we should note that the signal efficiency in Figure~\ref{fig:ntagroc} represents the probability of an IBD event to satisfy the $N_{\rm n}=1$ condition. 
%The misidentification rate indicates how often the approach will confuse PMT noise or low-energy radioactivity as a neutron capture per event.
%The misidentification rate is employed to estimate the residual background after neutron tagging, referred to later in Section~\ref{subsec:accidental}. 
%We can see that loosening the cut criteria will increase signal efficiency but also lead to misidentification. 
%Adding to this comprehensible behavior, with sufficiently loose criteria will more frequently mistake backgrounds as neutrons, which causes $N_{\rm n}>1$ and, therefore, decreases IBD efficiency by our definition.

%\textcolor{red}{We can see that the BDT achieves the highest signal efficiency, while the NN method can choose a criterion with a lower misidentification rate, thanks to the strict capture time selection, as shown in Appendix~\ref{appendix:nn}. }

\subsection{Validation of Event Selection} \label{subsec:validation}
\subsubsection{Calibration Samples}

Before applying the full event selection to all data (i.e., ``unblinding''), we define validation steps to verify our event reduction for both signal- and background-like events. 
Some of these are entirely new procedures to the SK DSNB search. 
We begin with data from the LINear ACcelerator (LINAC) calibration~\citep{nakahata_calibration_1999} and then focus on the calibration using the radioactive source Americium-Beryllium (Am/Be)~\citep{Abe2022a}.
We focus on validating the overall agreement between the observable distributions of data and MC simulations, while also defining systematic uncertainties arising from event reduction steps. 
The LINAC monochromatic electron events behave similarly to the IBD prompt signal from the DSNB. 
The Am/Be source emits neutrons, and these events are analogous to those of IBD because the neutron behavior in source energies is similar to that of IBD neutrons.
%and the Am/Be neutron captures are analogous to those of IBD. 
%With these calibration data, we can then generate samples to validate the distributions of all observables for signal-like events in the atmospheric neutrino background reduction steps and subsequent neutron tagging.
Comparing these calibration data and MC, we can validate the effects of atmospheric neutrino background reduction (Section~\ref{subsec:atomred}) and subsequent neutron tagging (Section~\ref{subsec:ntagred}) on the IBD signal.

After verifying that the LINAC data distributions closely match those of the MC, the associated uncertainties on the IBD signal efficiency for each reduction are estimated. % by established approaches~\citep{Abe2021, Harada2023} using these calibration samples. 
We compare the selection efficiency of data ($\epsilon_{\rm data}$) and MC ($\epsilon_{\rm MC}$) and define the relative 1$\sigma$ uncertainty of each reduction step as $(\epsilon_{\rm MC}-\epsilon_{\rm data})/\epsilon_{\rm data}$.

In a similar procedure with Am/Be calibration data for both NN and BDT algorithms, we estimate the systematic uncertainty on the neutron detection efficiency. 
As a function of either algorithm's cut point and the various calibration configurations, we take the difference in predicted and observed tagging efficiency as the 1$\sigma$ systematic uncertainty. 
%Since we insert real detector noise into the MC samples, the uncertainty on the predicted misidentification rate is dictated by the statistical uncertainty at a given algorithm working point.
Appendix~\ref{appendix:ambe} details the comparison of Am/Be calibration data with MC samples.

\subsubsection{Background Samples}

The remaining steps before unblinding include exploiting background-dominated samples. 
First, we consider the NCQE background behavior.
A dedicated SK-Gd study on atmospheric NCQE interaction modeling followed the event selection used in SK DSNB searches~\citep{Sakai2024}. 
With a Cherenkov angle selection of $\theta_c$ above $50^\circ$, it was demonstrated that the predicted $\theta_c$, energy, and neutron multiplicity distributions agree well with data. 
%First, we apply all reductions to the data below 12~MeV. This sample is dominated by spallation backgrounds, which serves as a reliable and representative sample on remaining spallation predictions.

Next, for data with $E_{\rm rec} > 29.5~\rm MeV$, we apply all reduction criteria since we assume a comparatively negligible DSNB contribution based on a wide range of theoretical DSNB models.
These events are dominated by atmospheric neutrino CC backgrounds, notably the decays of invisible muons and pions, categorized as part of the ``non-NCQE'' backgrounds (Section~\ref{subsec:atmbg}).
Since this background contributes significantly to the DSNB signal region, this sample helps validate the scaling of remaining atmospheric background predictions in the adjacent signal region. 

%Finally, we focus on the whole energy range of the analysis. 
%We apply all reduction criteria except for atmospheric background reduction to compare our overall atmospheric background predictions with data and then apply all reduction criteria except for spallation cut to compare our overall spallation background predictions with data. 
%These two samples have negligible contributions from DSNB events.\\

Table~\ref{tab:unblinding} presents a summary of these validation steps.
%A summary of these validation steps is included in Table~\ref{tab:unblinding}. 
Once all of them are performed while demonstrating good data/MC agreement, we proceed to unblind the full dataset with all reduction criteria applied. 
This includes preparing the final samples for both the spectrum-independent and spectrum-dependent searches. 
These two statistical approaches are done in parallel, and we detail their procedures and results in the following sections.

\begin{deluxetable}{l l}[htb!]
\tablecaption{Validation samples ahead of data unblinding. \label{tab:unblinding}}%, as described in Section~\ref{subsec:validation}. \label{tab:unblinding}} 
\tablecolumns{2}
\tablewidth{0pt}
\tablehead{
%    Sample & Validation
Sample & Item Checked
}
\startdata
LINAC & ~IBD prompt signal \\
Am/Be & ~IBD delayed signal (Appendix~\ref{appendix:ambe})\\
%\begin{tabular}{l}
%IBD delayed signal \\ (Appendix~\ref{appendix:ambe})\\
%\end{tabular}\\
$\theta_c \in [50^\circ,~90^\circ]$ sample& 
\begin{tabular}{l}
Atm. NCQE\\(See~\citet{Sakai2024})\\
\end{tabular}\\
%$[7.5,11.5]$~MeV & remaining spallation backgrounds \\
%$[7.5,79.5]$~MeV & overall atmospheric backgrounds \\
%& overall spallation backgrounds \\
$[29.5,79.5]$~MeV sample& 
\begin{tabular}{l}
Scaling of Atm. non-NCQE \\(Section~\ref{subsec:ccnorm})\\
\end{tabular}\\
\enddata
\end{deluxetable}

% ----- binned search -----
\section{DSNB Spectrum-independent electron antineutrino search} \label{sec:binnedanalysis}
In this section, we describe the search for electron antineutrino IBD events over the expected backgrounds on a bin-by-bin basis. 
This search makes no explicit assumptions about the theoretical model of the IBD signal, ensuring that the result can be applied to any astrophysical electron antineutrino flux. 
The upper energy bound of the signal region is set to $E_{\rm rec} < 29.5$~MeV due to a low expected DSNB flux and increasing atmospheric neutrino background at higher energies.
First, we discuss the backgrounds that should be taken into account in this analysis, cut optimization, and the resulting signal efficiency. Then, after estimating the systematic uncertainties of the backgrounds, we introduce the final search result.

\subsection{Scaling of Atmospheric Non-NCQE Background} \label{subsec:ccnorm}
Events with $E_{\rm rec} = 29.5-79.5~\rm MeV$ are utilized as the validation sample for the atmospheric non-NCQE background, as mentioned in Section~\ref{subsec:validation}. 
We start with the same procedure as \citet{Abe2021, Harada2023}, which determines non-NCQE normalization by comparing $E_{\rm rec}$ distributions between data and MC in this high-energy sideband region. 
To get accurate estimations, we compared the $E_{\rm rec}$ distribution without the neutron tagging step for the SK-VI sample. 
In SK-VII, we get more neutron detection efficiency, so a loosened neutron selection is applied to obtain statistics before comparison. 

\subsection{Accidental Coincidences} \label{subsec:accidental}
%By applying all reduction steps to simulated events, we can predict the expected remaining events of each event category. 
With our MC samples, we can predict the remaining events of nearly all event categories after neutron tagging.
Additionally, we should estimate instances where a coincidence between a prompt event and a misidentified delayed signal in our neutron search algorithm occurs,  referred to as ``accidental coincidences.''
A significant contribution to accidental coincidences comes from spallation isotopes, since we do not fully simulate the spallation background yield due to the large uncertainties in isotope production.
%Although a significant contribution to accidental coincidences comes from spallation isotopes, we do not fully simulate all spallation backgrounds due to the large uncertainties in isotope production, as mentioned above.
%As mentioned in Section~\ref{subsec:spabg}, the only exception is an estimate of $\rm^9Li$ decays, which have neutron emission.
Furthermore, we do not simulate solar neutrino events. 
Thus, we evaluate the accidental background events in a data-driven way. %from these non-simulated backgrounds in the following manner.

We first apply all reduction steps except for neutron tagging to  the unblinded full dataset, which is divided into 2~MeV energy bins.
%Next, we define the difference $\Delta N_i$ between the resulting data per bin $i$ and the sum of MC predictions as the ``missing" MC contributions from remaining spallation and solar backgrounds.
From the misidentification rate $f_{\text{mis}}$ shown in Figure~\ref{fig:ntagroc}, we estimate the number of accidental coincidence background events $B_{\text{acc}}$ as

\begin{eqnarray} \label{eq:accidentals}
    B_{\text{acc}} = f_{\text{mis}} \times N_i,
\end{eqnarray}
where $N_i$ is the number of events before neutron tagging in each energy bin $i$.
%This estimate assumes that the misidentification rate applies to the backgrounds in the remaining data.

%This type of background is an important source to consider when optimizing spallation cuts and neutron identification, especially for energies below 16~MeV, as most events in this region are likely spallation events.

\subsection{Cut Optimization}
%The spectrum-independent analysis is performed separately using the BDT and the NN because the cut optimization policy for both neutron tagging algorithms differs.
%However, all reduction steps are chosen to be the same until the neutron identification step to be comparable.
This section describes the optimization for the spallation cut, atmospheric neutrino event reduction, and neutron identification.
The energy binning is selected to 2~MeV, chosen to match the SK energy resolution at $\mathcal{O}(10)$~MeV.

% another wording (Andrew, original)
% The spectrum-independent analysis is done separately with the BDT and NN neutron tagging to generate parallel results to compare. For this analysis, the cut criteria for both were chosen such that the two approaches have highly similar mis-ID rates and tagging efficiencies. For the NN, we adopt the cut criteria of NN output score ($a$) to 0.98 so that the $f_{\rm mis}$ satisfy the 0.02 with an assumption of $f_{\rm mis}$ being independent of the prompt event energy. For the BDT, the cut criteria slightly vary per energy region, but the mis-ID rate is forced to sit near the NN's $f_{\rm mis}$ level of 0.02. Final tagging efficiencies are similar (show Figure?).

\subsubsection{Spallation Cut}
%We optimized the spallation likelihood cut criteria based on the ROC curve as shown in Figure~\ref{fig:sparoc8MeV} for each 2-MeV bin. 
The optimization scheme is identical to that of \citet{Abe2021}, which computes the working point of the spallation likelihood cut threshold specified in Section~\ref{subsec:spared}, to maximize sensitivities using the Rolke method~\citep{Rolke2005} under the null signal assumption. 
The working points are determined by taking into account all backgrounds, including $\rm ^9Li$, accidental background, and all other types of backgrounds that remain after applying the optimal NN neutron tagging cut.
%Shifting spallation cut criteria affects $\rm ^9Li$ and accidental background because most of the accidental background consists of remaining spallation events. 
Above 16~MeV, there are insufficient event statistics to fine-tune the cut criteria. 
Therefore, the cut point is chosen to maximize signal efficiency as we expect minimal contribution from the remaining spallation background in this energy region.
%since we strictly removed the spallation background in this energy region.

\subsubsection{Positron Event Selection}
The reduction criteria targeting atmospheric neutrinos are determined by comparing atmospheric and IBD signal MC predictions in each 2~MeV $E_{\rm rec}$ bin because the distributions of these values for signal and background events vary with energy. 
%This is because the distributions of these values for signal and background events vary with energy, and the resolution of our energy reconstruction is about this size. 
The figure of merit used in the optimization steps is taken to be $S/\sqrt{(S+B)}$. 
%$S (S+B)^{-\frac{1}{2}}$. 

Given the strong correlation between the Cherenkov angle and MSG observables, these two reduction steps are optimized together. 
%The Cherenkov angle lower and upper values are varied by $1^\circ$, while the minimum MSG value is varied by 0.01. 
The Cherenkov angle selection is optimized to be a tight interval that rejects visible $\mu$/$\pi$ events at low values and NCQE backgrounds at high values, while
the MSG selection is set to a minimum threshold to reject multi-cone events. 
More comments about final NCQE background contamination levels from these cuts are discussed in Appendix~\ref{appendix:msg}.

Finally, high ring clearness and high charge-per-hit indicate Cherenkov-visible $\mu/\pi$-like events. An upper bound for these parameters is optimized to remove such events from the final sample.
%The cuts on ring clearness and charge-per-hit variables are optimized separately, whereby events with high ring clearness and high charge-per-hit are more likely to be mu/pi-like and therefore are rejected.

%For the ring clearness and charge-per-hit variables, we treat the optimization of each separately. 
%Event selection for these sets a maximum value: High ring clearness and high charge-per-hit are more $\mu$/$\pi$-like and should be rejected. 

\subsubsection{Neutron Identification}
The spectrum-independent DSNB search is performed separately using both the NN-based and BDT-based neutron identification to allow cross-comparison.
%In both analyses, we require the number of neutron signals to be one for identifying IBD events. 
For this analysis, neutron identification criteria for both methods were chosen to ensure that the two methods have similar misidentification rates. % despite different approaches.

%NN
For the NN-based approach, we select the cut value of the NN score such that the $f_{\rm mis}$ satisfies an expected misidentification rate of 0.02\% per prompt event with the assumption of $f_{\rm mis}$ being independent of the prompt event energy. %, to avoid complicating the cut.
The search time window is optimized to maximize the signal-to-noise ratio, resulting in the $[4,535]~\rm \mu s$ for SK-VI and $[4,270]~\rm \mu s$ for SK-VII, respectively.
More details are given in Appendix~\ref{appendix:nn}.
%The optimal cut for the NN output score is determined to be 0.99 for SK-VI and 0.98 for SK-VII.

% too much detail?
%We explored the working neutron-search window, which was 535~$\mu$s in SK-IV, because increasing 
%The cross-section of neutron capture makes the neutron capture timing shorter. 
%We applied NN cut to signal MC with the various combinations of NN cut point so that $f_{\rm mis}$ keeps around 0.02. As a result, we succeeded in shortening the search time window range to $270~\mu$s and lowering the NN cut threshold in SK-VII. 
%The optimal cut for the NN output score is determined to be 0.99 for SK-VI and 0.98 for SK-VII. 

%BDT
For the BDT-based approach, the neutron selection criteria are determined in bins of 2~MeV, considering the rates of atmospheric neutrino MC, accidental coincidences, and IBD MC predictions. 
%\todo[inline]{This should be in section 4.4?}
%In SK-Gd, accidental coincidences become much smaller at a given neutron identification rate compared to pure water conditions, due to much lower misidentification rates rendered possible by Gd-captures. 
%We also apply a stricter pre-selection to BDT events using $N_{10}$, \texttt{bse}, and $N_c$ (see more details in Appendix~\ref{appendix:bdt}). 
We also apply a stricter selection to each candidate using three characteristic variables: the reconstructed Cherenkov photon count, the number of PMT hits in clusters of three, and total number of PMT hits for the neutron candidate (see Appendix~\ref{appendix:bdt} for more details). 
%These quantities are the number of hits in 10~ns TOF-corrected, the reconstructed number of Cherenkov photons generated by the neutron capture candidate, and the number of PMTs found in at least one cluster (defined as a group of at least 3 PMT hits within $14.1^\circ$ when comparing their directions from the reconstructed prompt vertex). 
%%In the end, the BDT misidentification rate across the analysis energy range stays below 0.1\% per prompt event (as in Figure~\ref{fig:ntagroc_opt}), which approaches that of the NN, for a similar signal tagging efficiency.

\begin{figure*}[htb!]
\centering
%\plottwo{figure/ntag_roc2.pdf}{figure/ntag_roc2.pdf}
\plottwo{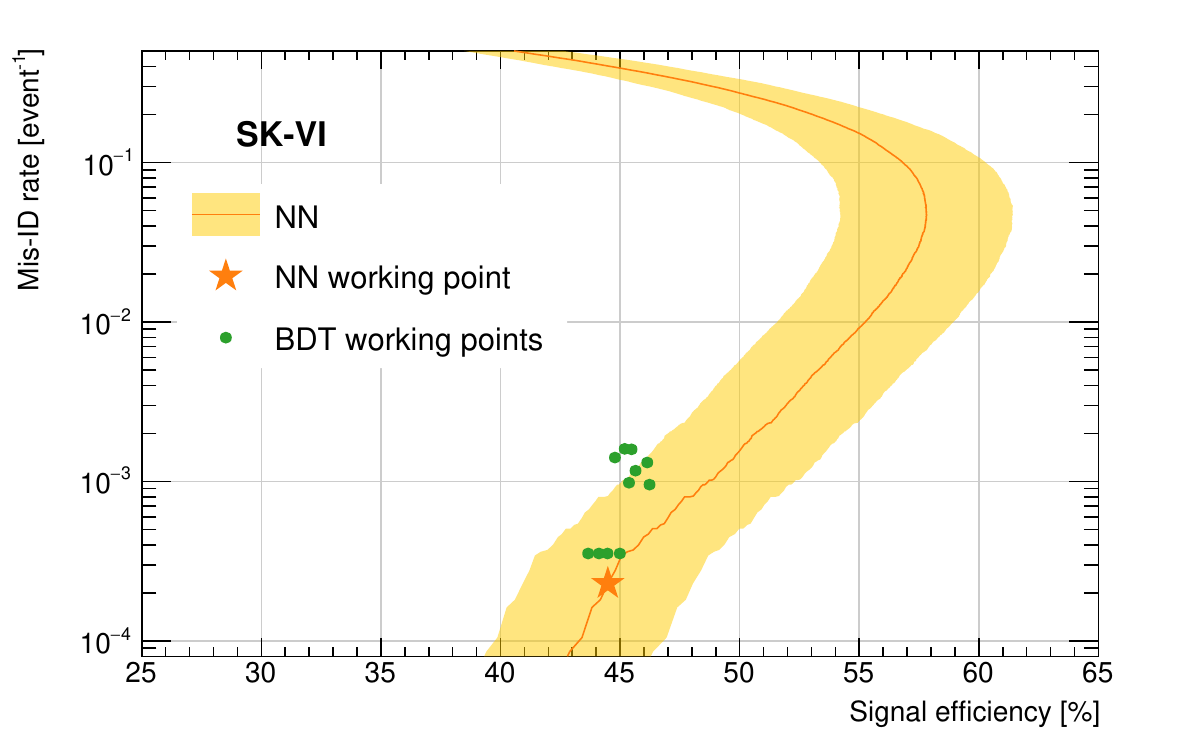}{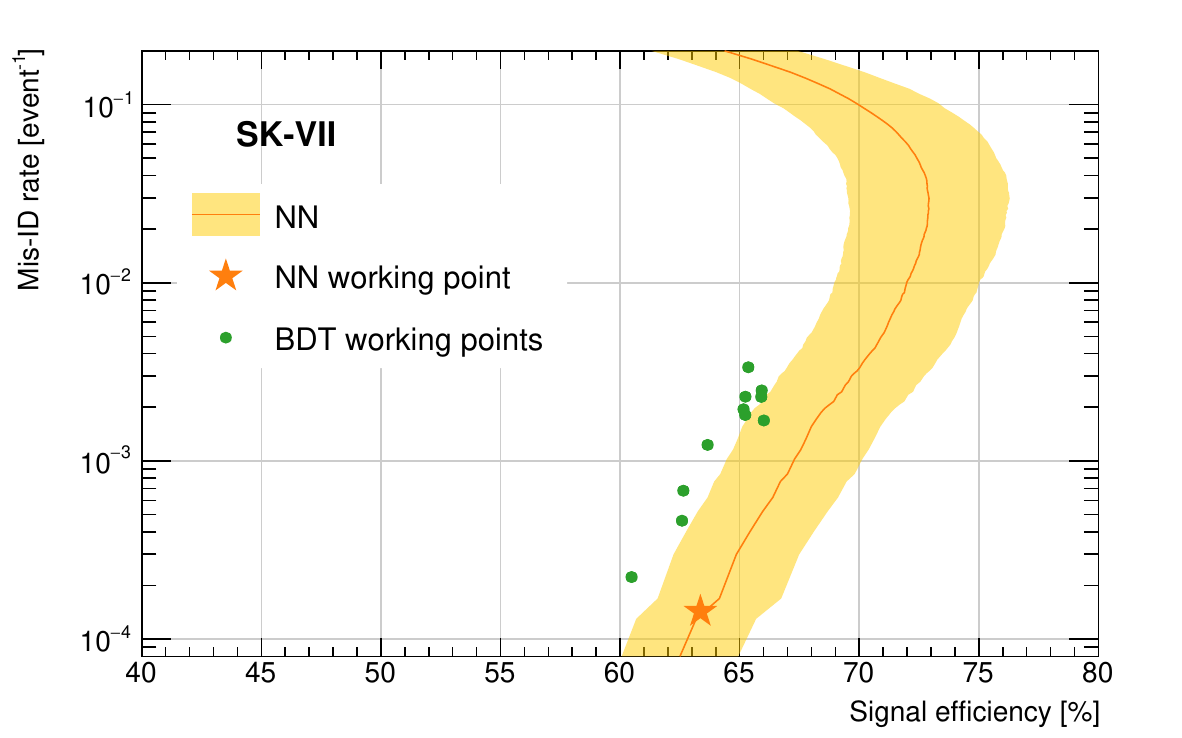}
%\plottwo{figure/ntag_opt_sk6.pdf}{figure/ntag_opt_sk7.pdf}
\caption{
\label{fig:ntagroc_opt}
Signal efficiency of neutron tagging and misidentification rate for the NN (Orange, solid line) and BDT (Green, circles) with strict pre-cut, respectively. 
The signal efficiency represents the probability of an IBD event to satisfy the $N_{\rm n}=1$ condition. 
%BDT working points are energy-dependent below 29.5~MeV and are chosen such that the misidentification rate is below $10^{-3}$ per event. 
%Here, the working points corresponding to the lowest and highest BDT misidentification rates used are shown.
The error band representing the uncertainty in signal efficiency is evaluated by the agreement of the MC sample with the calibration data (See Appendix~\ref{appendix:ambe}).
%The red long-dashed line shows the working point of the NN referring to the misidentification rate.
}
\end{figure*}

Figure~\ref{fig:ntagroc_opt} shows the neutron tagging performance in the form of signal efficiency against the background misidentification rate for NN and BDT.  
In the plot, we adopt two performance metrics that are directly related to the final samples requiring $N_{\rm n}=1$.
First, the signal efficiency in Figure~\ref{fig:ntagroc_opt} represents the probability of an IBD event to satisfy the $N_{\rm n}=1$ condition. 
%The misidentification rate indicates how often the approach will confuse PMT noise or low-energy radioactivity as a neutron capture per event.
The misidentification rate $f_{\rm mis}$ is explained in Section~\ref{subsec:accidental}. 
%The misidentification rate is the same definition as $f_{\rm mis}$, defined in Section~\ref{subsec:accidental}. 
%The misidentification rate is employed to estimate the residual background after neutron tagging, referred to later in Section~\ref{subsec:accidental}. 
Loosening the cut criteria will increase signal efficiency but also lead to more misidentification. 
Eventually, with sufficiently loose criteria, we will more frequently mistake backgrounds as neutrons, which causes $N_{\rm n}>1$ and, therefore, decreases IBD efficiency by our definition.

%We can see that the BDT achieves the highest signal efficiency, while the NN method can choose a criterion with a lower misidentification rate, thanks to the strict capture time selection, as shown in Appendix~\ref{appendix:nn}. 

\subsection{Signal Efficiency of Final Sample}
\begin{figure}[htb!]
\centering
\epsscale{0.9}
%\PlaceHolder[0.45\textwidth]
%\plotone{figure/srn_sig_eff_sk6.pdf}
%\plotone{figure/srn_sig_eff_sk7.pdf}
%\plotone{figure/srn_sig_eff_sk6_v2.pdf}
%\plotone{figure/srn_sig_eff_sk7_v2.pdf}
%\plotone{figure/srn_sig_eff_sk6_v3.pdf}
%\plotone{figure/srn_sig_eff_sk7_v3.pdf}
\plotone{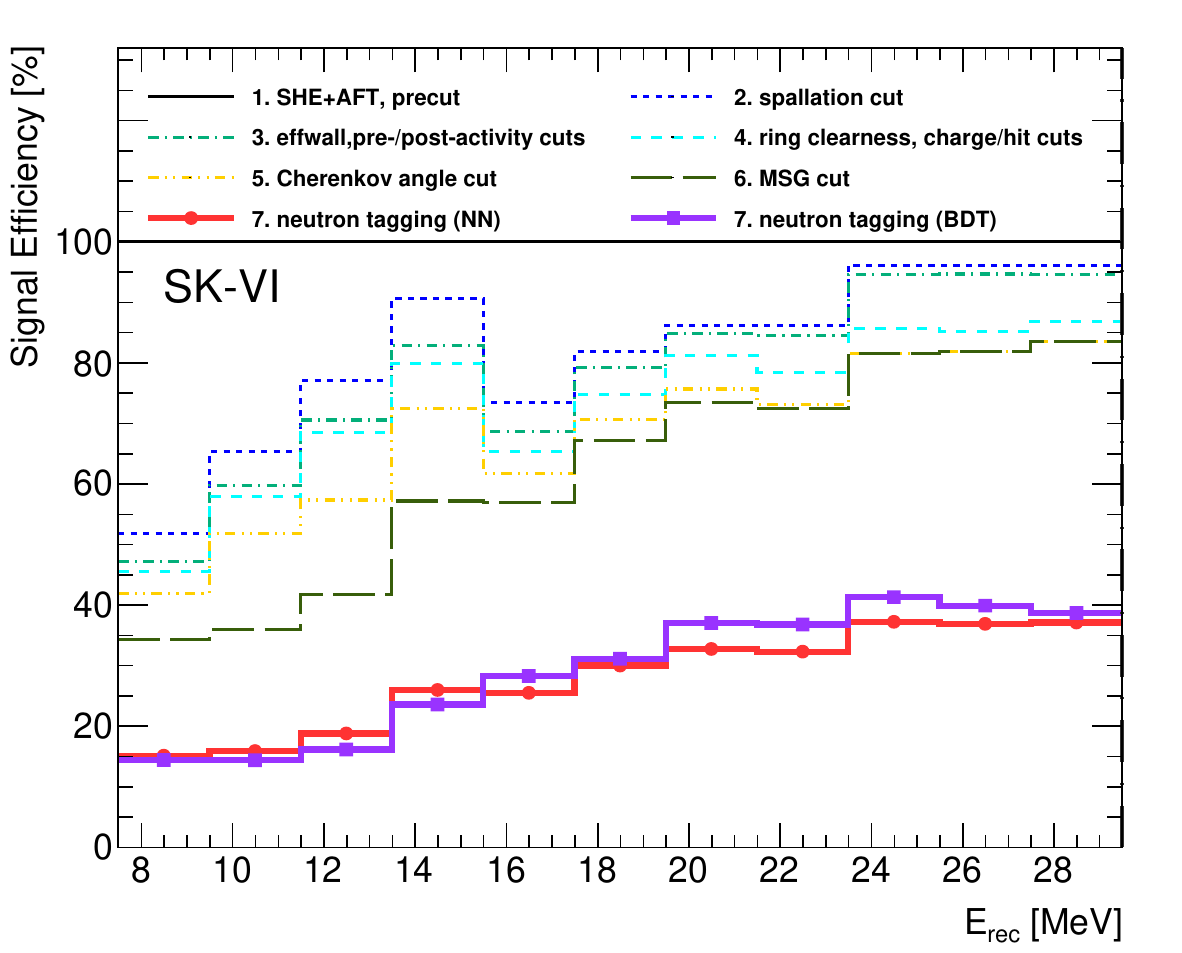}
\plotone{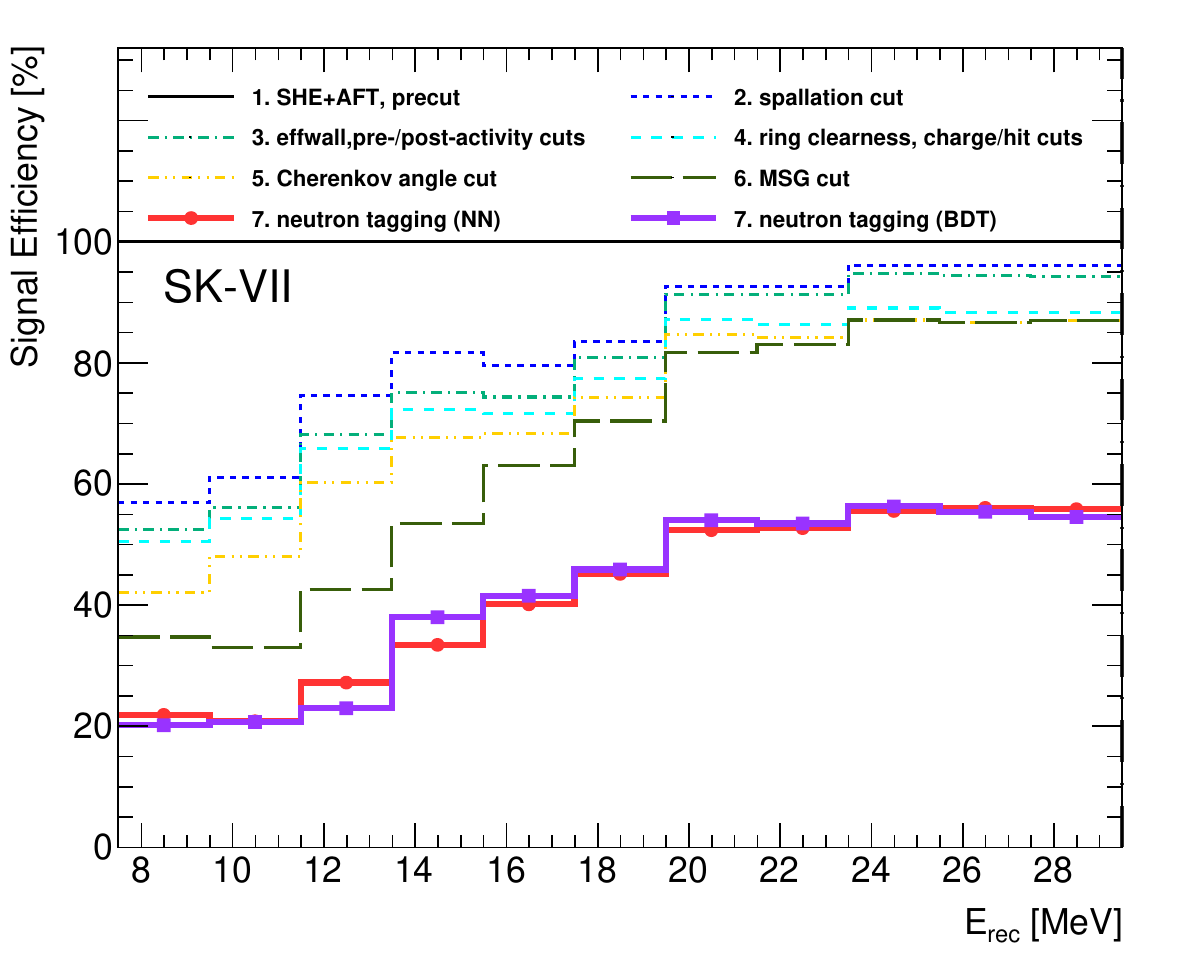}
\caption{\label{fig:signaleff}
DSNB signal efficiencies for the signal energy region, for SK-VI (top) and SK-VII (bottom),
normalized after the trigger requirement and basic noise reductions.
Each line shows the cumulative efficiency at each stage, performed in the order shown in the legend. Red (purple) with circle (square) points show the final efficiency after NN (BDT)-based neutron tagging. 
}
\end{figure}

Figure~\ref{fig:signaleff} shows the IBD signal efficiency for each 2~MeV $E_{\rm rec}$ bin after applying each optimized signal selection criterion for SK-VI and SK-VII, respectively. 
%A series of atmospheric neutrino reduction criteria, other than the MSG cut, are detailed in \citet{Abe2021}. 
The final IBD signal efficiency is shown for both the NN-based and BDT-based methods for comparison. 
These final efficiencies are also summarized in Table~\ref{tab:sigeff}.
The efficiencies before neutron tagging are comparable between SK-VI and SK-VII, whereas the neutron tagging effect on the final efficiencies increases markedly in SK-VII compared to SK-VI.
%The efficiencies before neutron tagging for both SK-VI and SK-VII are comparable, while the neutron tagging effect increases markedly in SK-VII.
Systematic uncertainties on the signal efficiency for each reduction are summarized in Table~\ref{tab:signalsys}. 
 
\begin{deluxetable}{c c c c c}[htb!]
\tablecaption{Signal efficiency after applying all reductions, including both NN and BDT-based neutron tagging, for each 2-MeV $E_{\rm rec}$ bin in the signal energy region. \label{tab:sigeff}} 
\tablecolumns{5}
\tablewidth{0pt}
\tablehead{
    \colhead{$E_{\rm rec}$~[MeV]} & 
    \multicolumn{4}{c}{Signal efficiency[\%]} \\
    & \multicolumn{2}{c}{SK-VI} & \multicolumn{2}{c}{SK-VII}\\
    & NN & BDT & NN & BDT
}
\startdata
7.49 - 9.49 & 15.1\% & 14.4\% & 21.9\% & 20.2\% \\
9.49 - 11.5 & 15.9\% & 14.4\% & 20.8\% & 20.7\% \\
11.5 - 13.5 & 18.8\% & 16.2\% & 27.2\% & 23.0\% \\
13.5 - 15.5 & 26.0\% & 23.6\% & 33.4\% & 38.0\% \\
15.5 - 17.5 & 25.5\% & 28.3\% & 40.1\% & 41.6\% \\
17.5 - 19.5 & 30.0\% & 31.1\% & 45.2\% & 45.9\% \\
19.5 - 21.5 & 32.8\% & 37.1\% & 52.4\% & 54.0\% \\
21.5 - 23.5 & 32.3\% & 36.8\% & 52.7\% & 53.5\% \\
23.5 - 25.5 & 37.2\% & 41.3\% & 55.5\% & 56.3\% \\
25.5 - 27.5 & 36.9\% & 39.9\% & 56.1\% & 55.4\% \\
27.5 - 29.5 & 37.1\% & 38.7\% & 55.8\% & 54.6\% \\
\enddata
\end{deluxetable}

\begin{deluxetable}{c c c}[htb!]
\tablecaption{ Systematic uncertainties on the signal efficiency from background reduction, calculated as described in Section~\ref{subsec:validation}. \label{tab:signalsys}} %Section~\ref{subsec:spared} and \ref{subsec:validation}. \label{tab:signalsys}} 
\tablecolumns{3}
\tablewidth{0pt}
\tablehead{
    \colhead{} & 
    \multicolumn{2}{c}{Relative systematic error} \\
    Cut & 
    SK-VI &
    SK-VII
}
\startdata
%Spallation cut &  &  \\
%$d_{\rm eff}$ cut &  &  \\
%$L_{\rm clear}$ cut & 0.79\% & 0.58\% \\
$q_{50}/n_{50}$ cut & 0.20\% & 0.25\% \\
$\theta_{\rm C}$ cut & 1.3\% & 0.94\% \\
MSG cut & 1.7\% & 1.4\% \\
Neutron tagging (NN/BDT) & 8.4\%/5.0\% & 3.4\%/6.0\% \\
\enddata
\end{deluxetable}

\subsection{Uncertainties on Background Estimation}
This section presents the systematic uncertainties corresponding to each background component. 
In the present analysis, we evaluated uncertainties related to the MSG cut and updated the uncertainties for atmospheric non-NCQE events and neutron tagging. 
Table~\ref{tab:bk_sys_unc} summarizes the relative systematic uncertainties assigned to each background category.

\subsubsection{Atmospheric NCQE Background}
%\subsubsection{Atmospheric Neutrinos}
%We retrieve the scaling factor of the NCQE events from the cross-section measurement done by T2K~\citep{Abe2019b} as $f_{\rm \nu-NCQE}$ and $f_{\rm \bar{\nu}-NCQE}$. NCQE-like events, including 2p2h interaction, are normalized by this factor. The uncertainty of this measurement result should be added to other systematic uncertainties.
%For the systematic uncertainty relying on the NCQE scaling, we take the same concept of uncertainties on spallation ${}^9$Li and reactor neutrinos as in~\citet{Abe2021}, for which the values are estimated to be 60\% and 100\% on the total predictions, respectively. 
In light of the new MSG cut, we update the uncertainty from the previous SK analyses~\citep{Abe2021, Harada2023} on the remaining NCQE level. 
This is accomplished by an MC-driven estimate.
%\todo[inline]{modify after test with T2K} 
In particular, we examine the difference in the reduction efficiency between two distinct MC models.
As discussed in Section~\ref{subsec:atmbg}, knock-out nucleons from NCQE interactions are energetic enough to partake in secondary interactions with other nuclei to produce secondary $\gamma$-emission. 
The way in which these ``nuclear cascades" occur impacts the multi-cone behavior of NCQE events and, therefore, MSG reduction. A discussion of nuclear cascade modeling and NCQE events was performed by~\citet{Abe2025}. 
For these reasons, we generate one sample with the INCL model and another with the BERT model.
%For these reasons, we generate one sample with a Liège Intranuclear cascade~\citep[INCL;][]{Boudard2013} model and another with the Bertini cascade model~\citep[BERT;][]{wright2015}.
The BERT model shows the most discrepant results from INCL in some validation results~\cite{Hino2025arxiv, Sakai2024}, such that it should provide a reliable estimate of the maximal difference in MSG cut efficiency between all possible models.
Based on the discrepancy of the NCQE predicted remaining rate using the BERT and INCL models, we conservatively estimate 20\% as the additional uncertainty in the level of remaining NCQE backgrounds, independent of energy. 
%This results in an estimated 19\% event rate difference between the two models throughout the signal energy range. 
%To be cautious, we take 20\% as the additional uncertainty on NCQE magnitude independent of energy. 
Other uncertainties on the atmospheric neutrino flux and NCQE cross-section are assumed to be the same as previous SK analyses~\citep{Abe2021, Harada2023}, which were estimated to be $\pm68\%$.
In total, the new uncertainty for the remaining atmospheric neutrino NCQE background is estimated as $\pm71\%$ by combination with the additional $\pm20\%$ uncertainty in quadrature.
%, including those for neutron multiplicity and reconstructed energy spectrum shape. 
%, calculated using the same methodology as the previous analyses~\citep{Abe2021, Harada2023}. 
%such as the 30\% for neutron multiplicity uncertainty and approximately 37\% for the spectrum of 
%Summing these uncertainties along with those on neutron multiplicity and spectral shape in quadrature gives 71\%.
%To be cautious, we take 20\% as the additional NCQE uncertainty and add in quadrature to the previous uncertainty of 68\% given by~\citet{Abe2021} to obtain 71\%.

\subsubsection{Atmospheric Non-NCQE Background}
%To estimate the uncertainty on the other atmospheric neutrino backgrounds, categorized as ``non-NCQE'' backgrounds, we start with the same procedure as \citet{Abe2021, Harada2023}, which determine non-NCQE normalization by comparing $E_{\rm rec}$ distributions between data and MC in the high-energy sideband region, described in Section~\ref{subsec:validation}. 
The overall systematic uncertainty on the flux of atmospheric neutrinos and the cross-section for non-NCQE interactions are determined by the same procedure as in Section~\ref{subsec:ccnorm}.
%the uncertainty of the sideband fit done in Section~\ref{subsec:ccnorm}.
This is an energy-binned fit of sideband MC to the data for which we extract a $\pm1\sigma$ uncertainty.
%To make up for the limited statistics in SK-VI, we compared the events between data and MC before applying neutron tagging, adding in quadrature of neutron multiplicity uncertainty, same value as the NCQE case described above.
In SK-VI, since this fit is performed before neutron tagging, we add a 30\% systematic uncertainty on the neutron multiplicity of atmospheric neutrino interactions in quadrature, as in \citet{Harada2023}. 
Then, the resulting systematic uncertainties are $\pm36\%$ for SK-VI and $\pm41\%$ for SK-VII. 
%The result is... 
%In addition, we cross-check these estimates by adding in quadrature the difference of the cut efficiencies between data/MC samples in [29.5, 59.5]~MeV for each reduction step independently. 
%Using this approach, we find that the non-NCQE uncertainty is similar to the fitting result---39\% for SK-VI and 51\% for SK-VII.

\subsubsection{Lithium-9 Background}
Below 16~MeV, most of the background from spallation after neutron tagging consists of $\rm^9 Li$ ($\beta+n$) decays. 
The normalization is taken from \citet{Zhang2016}, which measured a yield of $0.86 \pm 0.12 \, (\mathrm{stat.}) \pm 0.15 \, (\mathrm{sys.})$~${\rm kton^{-1}\cdot day^{-1}}$ in SK. 
The systematic uncertainty on the scaling is 22\% in yield uncertainty, taken from \citet{Zhang2016}. 
Additionally, according to \citet{Abe2021}, there is approximately a 50\%  uncertainty in our data-driven estimation of the $\rm^9 Li$ remaining rate after spallation cuts. %estimation detailed in \citet{Abe2021}, is 50\%. 
Uncertainties related to the reduction steps other than the spallation cut, primarily from neutron tagging, are summed in quadrature to the systematic uncertainty estimate, resulting in a total of $\pm55\%$.

%For the accidental coincidence event, the uncertainty on the $f_{\rm mis}$ only contributes to the scale uncertainty for each bin. Thus, the 5\% due to the statistical uncertainty in $f_{\rm mis}$ evaluation is assigned to this type of event.

\subsubsection{Reactor Neutrinos}
The reactor neutrino background is estimated by scaling the IBD simulation following the reactor neutrino flux introduced in Section~\ref{subsec:reactor}. 
These events populate only the lowest energy bin, ranging from $7.5-9.5$~MeV, as shown in Figure~\ref{fig:binned_spectrum}. 
%The flux strongly depends on each reactor's activity, which has large uncertainties. 
The flux strongly depends on the activity of each reactor. 
In this analysis, we conservatively assign a $\pm100\%$ systematic uncertainty on the reactor neutrino events.

\subsubsection{Accidental Coincidence Background}
For the accidental coincidence background, the uncertainty on $f_{\rm mis}$ from Equation~\ref{eq:accidentals} should be considered. Since we evaluate $f_{\rm mis}$ by real detector noise, the statistical uncertainty of $f_{\rm mis}$ at a given algorithm's working point over the entire period is assigned as the total uncertainty, which is approximately $\pm5\%$ for both SK-VI and SK-VII.
%Since we insert real detector noise into the MC samples, the uncertainty on the predicted misidentification rate is dictated by the statistical uncertainty at a given algorithm working point, as approximately 5\% for both SK-VI and SK-VII.

\begin{deluxetable}{c c c}[htb!]
\tablecaption{ Systematic uncertainties on background predictions for the spectrum-independent electron antineutrino search.\label{tab:bk_sys_unc}} 
\tablecolumns{3}
\tablewidth{0pt}
\tablehead{
    \colhead{} & 
    \multicolumn{2}{c}{Relative systematic error} \\
    Event category & SK-VI & SK-VII
}
\startdata
Atmospheric-$\nu$ (NCQE) & $\pm71\%$ & $\pm71\%$ \\
Atmospheric-$\nu$ (non-NCQE) & $\pm36\%$ & $\pm41\%$ \\
Spallation ${}^9$Li & $\pm55\%$ & $\pm55\%$ \\
Reactor-$\nu$ & $\pm100\%$ & $\pm100\%$ \\
Accidental coincidence & $\pm5\%$ & $\pm5\%$ \\
\enddata
\end{deluxetable}

\subsection{Results}
\subsubsection{Final Data Samples}
\begin{figure}[htbp!]
\centering
%\plottwo{figure/final_spectrum_comb67_linear.pdf}{figure/sk6_7_binned_BDT_final_linear.pdf}
%\plotone{figure/final_spectrum_comb67_linear.pdf}
%\plotone{figure/sk6_7_binned_NN_final_linear.pdf}
%\plotone{figure/sk6_7_binned_BDT_final_linear.pdf}
\plotone{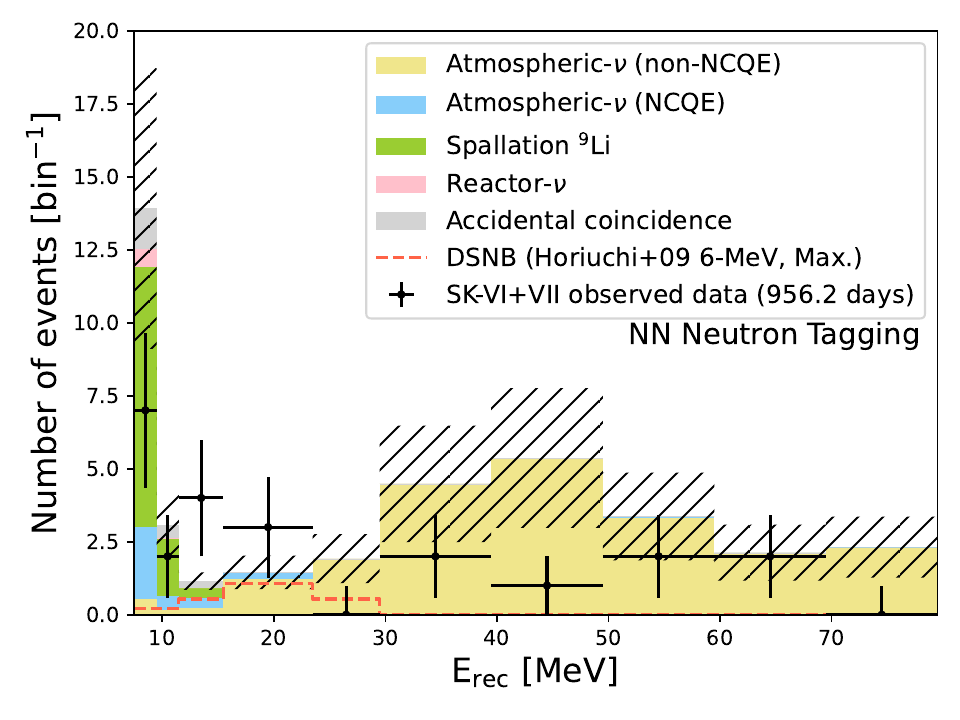}
\plotone{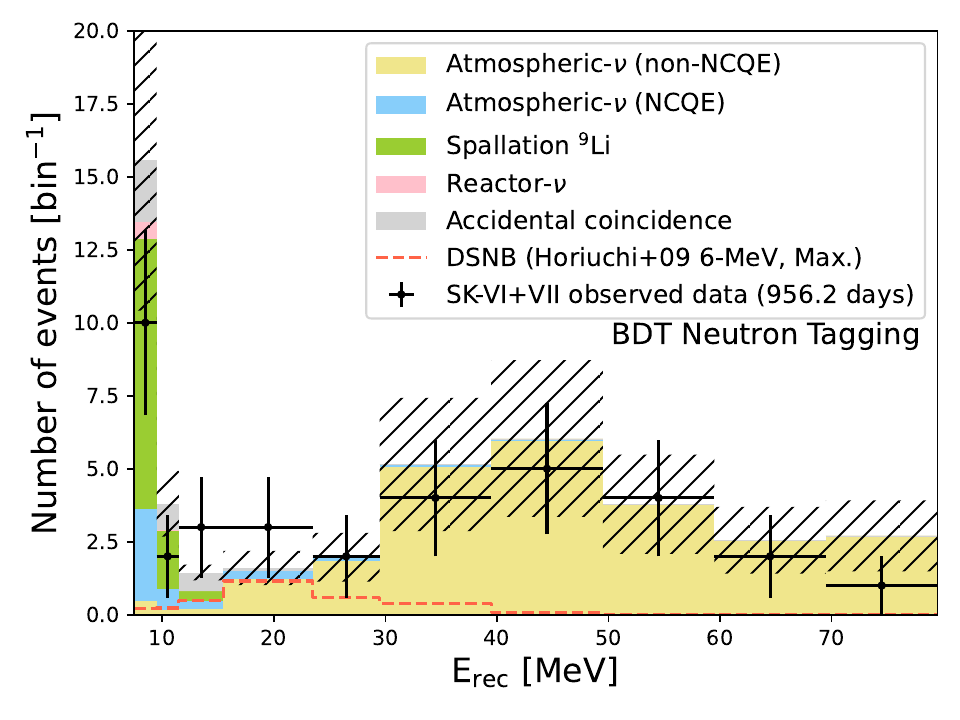}
\caption{
\label{fig:binned_spectrum}
Reconstructed positron equivalent kinetic energy spectra of the data and the expected background after all reductions, including the requirement of $N_{\rm n}=1$, for NN (top) and BDT (bottom) neutron tagging. 
The error bars in the data points represent the statistical error value estimated by taking the square root of the number of observed
events. 
Color-filled histograms show the expected backgrounds, which are stacked on top of each other.
The hatched areas filled onto the stacked background histograms represent the total absolute background systematic uncertainty for each bin.
The signal search window is $E_{\rm rec} \in [7.5,~29.5]$~MeV.
For illustrative purposes, a typical DSNB expectation from optimistic values of \citet{Horiuchi2009} is drawn separately from the background histograms as a red-dashed (bottom) line, superimposed.
}
\end{figure}

Data are divided into pre-determined energy bins as done by~\citet{Harada2023}, for which the first two ($E_{\rm rec} =7.5-9.5$, $9.5-11.5$~MeV) are spallation ${}^9$Li-dominated, and the next three ($E_{\rm rec} =11.5-15.5$, $15.5-23.5$, $23.5-29.5$~MeV) contain the lowest background levels. %mainly atmospheric non-NCQE backgrounds. %lowest background levels. 
In particular, the third bin contains the high-energy tail of $\rm ^9 Li$ events, while non-NCQE backgrounds start to dominate in the fourth bin. 
Finally, in the fifth bin, almost all events are non-NCQE events. 
The remaining bins ($E_{\rm rec} =29.5-79.5$~MeV divided into 10~MeV intervals) form the high-energy sideband. 
The final energy spectra after all reduction criteria are applied using either NN or BDT for neutron tagging are shown in Figure~\ref{fig:binned_spectrum}.
Again, the only difference in the two samples shown is the neutron tagging algorithm applied.

In each energy bin, we generate a probability distribution for the total event count under a background-only hypothesis. 
This is achieved by performing pseudo-experiments based on the expected value of each background category, 
varied according to its associated systematic uncertainty, assuming Gaussian distributions.
%varied assuming Gaussian distributions according to its associated systematic uncertainty. 
From these distributions, a background-only p-value, $p_b$, is calculated using the observed number of events from the data. 
For both NN and BDT final samples, we conclude that no significant excess is observed over the background, while the smallest $p$-value is 0.08.

\subsubsection{Astrophysical Electron Antineutrino Flux Upper Limit}

With no significant excess, we then place upper limits on the astrophysical $\bar{\nu}_e$ flux per energy bin. 
We adopt the CL$_s$ approach \citep{Read_2002}, for which a background-plus-signal p-value $p_{s+b}$ is modified by the rejection coming from the background-only hypothesis with $p$-value $p_{b}$ giving: 
\begin{equation}
    \text{CL}_s \equiv \frac{p_{s+b}}{1-p_b} \leq \alpha.
\end{equation}
This method is well suited when we expect an observation to be statistically consistent with both background-only and signal-plus-background hypotheses---especially when the signal is unknown---since $p_{s+b}$ is increased when $p_{b}$ is also large.

Both expected and observed upper limits are calculated per energy bin at 90\% CL$_s$, for which $\alpha = 0.1$. 
For the expected limit, $p_b$ is determined by the background-only expectation value of the number of events in that bin. 
In the case of the observed limit, $p_b$ is determined by the observed number of events per bin. 
For both scenarios, the amount of signal events is varied, changing the underlying signal-plus-background distribution, 
until the appropriate $p_{s+b}$ value meets the 90\% CL$_s$ criterion. 
%until the 90\% CL$_s$ criterion is met by the appropriate $p_{s+b}$ value. 
This defines an upper limit on the number of signal events per bin after all reduction steps, $N_{90}^{\text{limit}}$.

Using the $N_{90}^{\text{limit}}$ value in each bin, we can convert these quantities into limits on the $\bar{\nu}_e$ flux using
\begin{comment}
\begin{eqnarray} \label{eq:limit}
    \phi_{90}^{\text{limit}} = \frac{N_{90}^{\text{limit}}}{\bar{\epsilon}_{\rm IBD}\, \bar{\sigma}_{\rm IBD}\, N_p\, T\, \Delta E},
\end{eqnarray}
\end{comment}
\begin{eqnarray} \label{eq:limit}
    \phi_{90}^{\text{limit}} = \frac{N_{90}^{\text{limit}}}{\langle\bar{\epsilon}_{\rm IBD}\, \bar{\sigma}_{\rm IBD}\rangle\, N_p\, T\, \Delta E},
\end{eqnarray}
where $\bar{\epsilon}_{\rm IBD}$ is the 2~MeV-bin average IBD efficiency (shown in Figure~\ref{fig:signaleff}), $\bar{\sigma}_{\rm IBD}$ is the IBD cross-section~\citep{Strumia2003} at a mean neutrino energy in each 2~MeV-bin, 
$\langle\bar{\epsilon}_{\rm IBD}\, \bar{\sigma}_{\rm IBD}\rangle$ is the averaged value of $\bar{\epsilon}_{\rm IBD}\, \bar{\sigma}_{\rm IBD}$ in each energy bin width,
$N_p$ is the number of free protons in the fiducial volume, $T$ is the livetime, 
and $\Delta E$ is the energy bin width. 

\begin{figure*}[htb!]
\centering
%\epsscale{1.1}
\epsscale{0.73}
%\plotone{figure/srn_limit_comparison_integ2_sk67_ntagcomp.pdf}
%\plotone{figure/srn_limit_comparison_integ2_sk67_ntagcomp_CLs.pdf}
%\plotone{figure/srn_limit_comparison_integ2_sk67_conservative_wsk123.pdf}
\plotone{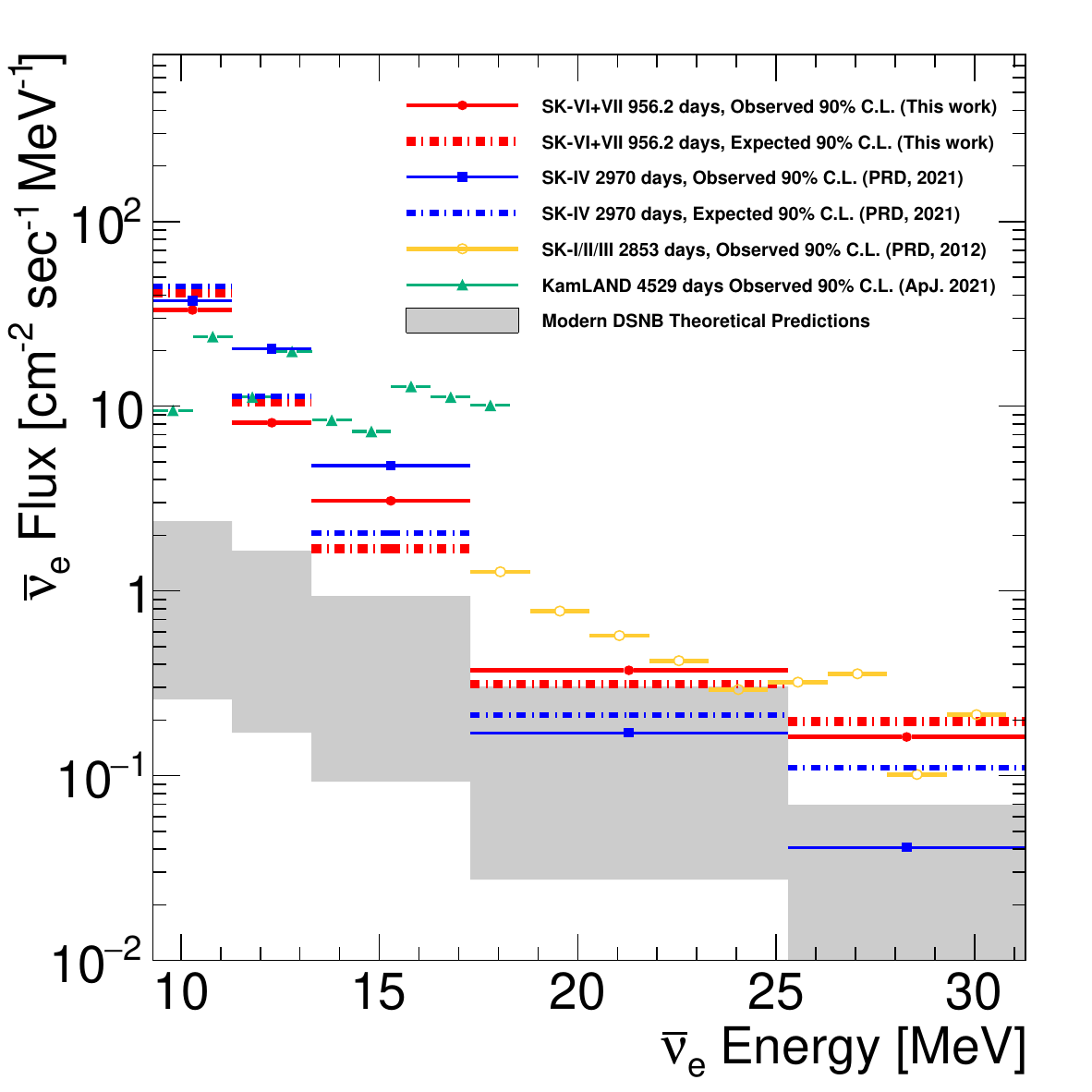}
\caption{\label{fig:binned_limits}
The 90\% C.L. upper limits on the astrophysical $\bar{\nu}_e$ flux, calculated by Equation~\ref{eq:limit}, per energy bin.
The red lines show the observed upper limit (solid, circle point) and expected sensitivity (dot-dashed) for SK-VI+VII. 
To be conservative, the lines represent the larger values of the two neutron tagging approaches listed in Table~\ref{tab:sumintepfluxlimit}.
%Conservatively, the lines represents the larger values for two neutron tagging approach listed in Table~\ref{tab:sumintepfluxlimit}.
%er limit type are shown per bin for the separate neutron identification approaches. 
The blue lines show the observed upper limit (solid, square point) and expected sensitivity (dot-dashed) for SK-IV~\cite{Abe2021}. 
The orange line displays the 90\% C.L. observed upper limit for SK-I/II/III without neutron tagging~\citep{Kirk2012}. 
The green line with triangle points represents the 90\% C.L. observed upper limit placed by KamLAND~\citep{KamLAND2022}. 
The gray-shaded regions represent the range of the theoretical flux expectation for the DSNB signal \citep{Malaney1997, Hartmann1997, Kaplinghat2000, Lunardini2009, Horiuchi2009, Galais2010, Priya2017, Horiuchi2018, Barranco2018, DeGouvea2020, Kresse2021, Horiuchi2021, Tabrizi2021, Nick2022, Ashida2023, Ivanez2023, Nakazato2024, Martinez2024}.
The theoretical prediction level is averaged within the bin width. 
}
\end{figure*}

Results are shown in Figure~\ref{fig:binned_limits}, and values are summarized in Table~\ref{tab:sumintepfluxlimit}.
Note that $E_\nu$ is calculated by $E_\nu=E_{\rm rec}+1.8~\rm MeV$.
The sensitivity above 17.3~MeV becomes comparable to some of the theoretical predictions, and the sensitivity in $E_{\nu}$ of $13.3-17.3$~MeV approaches models with large predicted fluxes to within a factor of two.
Compared to the previous SK pure-water search~\citep{Abe2021}, the new sensitivity from SK-Gd is better below 15.5~MeV, owing to the significantly higher neutron identification efficiency and lower levels of accidental coincidences. 
On the other hand, in the higher energy region, pure-water results still have the world's best sensitivity due to the smaller systematic uncertainty on non-NCQE events and larger dataset. %based on the larger statistics of the side-band energy region. 

%\begin{figure}[htb!]

\begin{deluxetable*}{c c c c c c c c}[htb!]
\centering
\tablecaption{ Summary of $\bar\nu_e$ flux upper limit and sensitivity for each period and neutron tagging algorithm, and optimistic and pessimistic DSNB expectation from \citet{Kaplinghat2000}, and \citet{Nakazato2015}, respectively. For SK-IV, only BDT-based neutron tagging is applied. } \label{tab:sumintepfluxlimit}
\tablecolumns{8}
\tablewidth{0pt}
\tablehead{
     \colhead{Neutrino Energy} 
    & \multicolumn{3}{c}{Observed Upper Limit} &\multicolumn{3}{c}{Expected Sensitivity} 
    & \colhead{DSNB Theoretical Expectation} \\
     \colhead{$\rm [MeV]$} 
    & \multicolumn{3}{c}{$\rm [cm^{-2}\, s^{-1}\, MeV^{-1}]$} & \multicolumn{3}{c}{$\rm [cm^{-2}\, s^{-1}\, MeV^{-1}]$} & \colhead{$\rm [cm^{-2}\, s^{-1}\, MeV^{-1}]$} \\
%    & \colhead{SK-IV} & \colhead{SK-VI} & \colhead{SK-IV} & \colhead{SK-VI} & \colhead{\citet{Kaplinghat2000}}
    & \colhead{SK-IV} 
    & \multicolumn{2}{c}{SK-VI+VII}
    & \colhead{SK-IV} 
    & \multicolumn{2}{c}{SK-VI+VII}
    & \\
    & \colhead{BDT} 
    & \colhead{NN} & \colhead{BDT} 
    & \colhead{BDT}
    &\colhead{NN} & \colhead{BDT} & 
}
%\todo[inline]{change later}
\startdata 
    9.29--11.29  & 37.30 & 23.79 & 33.20 & 34.07 & 38.26 & 40.89
    & 0.20 -- 2.40 \\
    11.29--13.29 & 20.43 & 7.48 & 8.14  & 11.35 & 10.32 & 10.50
    & 0.13 -- 1.66 \\
    13.29--17.29 & 4.77  & 3.07 & 2.76  & 2.05  & 1.67 & 1.69 
    & 0.67 -- 0.94 \\
    17.29--25.29 & 0.17  & 0.37 & 0.37  & 0.21  & 0.31 & 0.29
    & 0.02 -- 0.30 \\
    25.29--31.29 & 0.04  & 0.13 & 0.16  & 0.11  & 0.20 & 0.18
    & $<0.01$ -- 0.07 \\
\enddata
\end{deluxetable*}

\section{DSNB spectral fitting analysis} \label{sec:spectralfitting} 

In the spectral fitting analysis, we extract the normalization of each component (DSNB signal and backgrounds) by fitting their reconstructed PDFs to the data using an extended energy-unbinned likelihood maximization framework. 
%In the spectral fitting analysis, the overall prediction (DSNB signal and backgrounds) is fitted to the data using an extended energy-unbinned likelihood maximization framework. 
%This is a shape-driven analysis where the underlying background and signal contents are extracted by leveraging the shape of the data distribution in the considered parameter space. 
Thus, this analysis leads to a best-fit signal normalization for each DSNB prediction.
The main difference here from the spectrum-independent $\bar{\nu}_e$ search (Section~\ref{sec:binnedanalysis}) is that this approach introduces undetermined parameters, namely the absolute event rate of the DSNB signal and backgrounds, as well as certain nuisance parameters for each reconstructed energy PDF shape.
%as well as additional nuisance parameters for the shape of the reconstructed energy PDF of each event category. 
In order to further constrain the fit, instead of removing events with background-like $\theta_c$ and $N_{\rm n}$, the parameter space is extended to six regions: 
Three $\theta_c$ divisions ($\theta_c \in [20^\circ,\, 38^\circ]$, $\theta_c \in [38^\circ,\, 53^\circ]$, $\theta_c \in [78^\circ,\, 90^\circ]$), and two $N_{\rm n}$ regions ($N_{\rm n}\neq1$, $N_{\rm n}=1$).
%Additionally, in order to, further constrain the fit, the parameter space is extended from one region ($\theta_c \in [38^{\circ}, 53^{\circ}] \otimes N_{\rm n}=1$) to six regions ($\{ \theta_c \in [20^{\circ}, 48^{\circ}], \theta_c \in [38^{\circ}, 53^{\circ}], \theta_c \in [78^{\circ},90^{\circ}] \}  \otimes \{ N_{\rm n}\neq1, N_{\rm n}=1 \}$).

Overall, the principle of the spectral analysis is the same as that detailed in~\citet{Abe2021}, with three notable differences.
%outlined below and further developed in~\citet{Beauchene2024}. 
% Significant differences are three-fold, listed below. 
First, for the detector, we benefit from enhanced neutron-tagging efficiency due to the Gd-loading in SK-VI and SK-VII, which enhances the DSNB signal detection in the IBD-like region ($\theta_c\in[38^\circ,53^\circ]$ and $N_{\rm n}=1$).
%($\theta_c \in [38^{\circ},\, 53^{\circ}]\, \otimes\, N_{\rm n}=1$).
Next, for the fit, we now profile over all nuisance parameters of the analysis (background rates and shape-only nuisance parameters).  
Finally, for the data and MC, we update the derivation of the spallation PDF (see Section~\ref{subsec:spall_modeling} below) and apply the new MSG cut. %for further atmospheric background mitigation. .% relative to the region ($\theta_c \in [38^{\circ},\, 53^{\circ}] \, \otimes\,  N_{\rm n}\neq1$).

\subsection{Samples}
In this analysis, samples are divided into six regions as described above. 
The upper bound of the signal energy region is extended to $E_{\rm rec} = 79.5$~MeV to take full advantage of the shape and normalization of the signal and backgrounds in the different regions of the parameter space. 
At the same time, the energy threshold of the analysis is set at $E_{\rm rec}=15.5~\rm MeV$ in all regions.
%The middle $\theta_c$ region with $N_{\rm n}=1$ is exactly the same as the final sample in the spectrum-independent search outlined in Section~\ref{sec:binnedanalysis}. 

The event selection criteria for the six analysis regions are  the same as in the spectrum-independent search from Section~\ref{sec:binnedanalysis}, with a few exceptions: 
For the $N_{\rm n}\neq 1$ region, solar neutrinos are removed based on the reconstructed direction of prompt events with the same criteria as in~\citet{Abe2021}. 
The new MSG cut is applied only to the middle $\theta_c$ region such that atmospheric neutrino events remain in the sideband. 
The $N_{\rm n}\neq1$ region contains more spallation events than for $N_{\rm n}=1$ due to the lack of a strict neutron tagging requirement.
%In general, the spallation cuts for the $N_{\rm n}\neq1$ region differ from the $N_{\rm n}=1$ region as the former contains more spallation events.
For this reason, tighter spallation likelihood cuts are applied to the $N_{\rm n}\neq1$ sample.
These are determined by first looking at data without atmospheric background reduction and neutron tagging.
Spallation cuts are then varied such that the predicted remaining spallation events in [15.5, 19.5] MeV are approximately at the same level as the predicted peak of decay electrons around 50 MeV.
The cut criteria are then progressively loosened until 23.5~MeV.
The IBD signal efficiencies of these unique cuts for solar and spallation events for $N_{\rm n}\neq1$ are given in Table~\ref{tab:eff_ibd_nontag}.

\begin{deluxetable*}{c c c c c c C}[htb!]
\tablecaption{ IBD signal efficiency of separate cuts for $N_{\rm n}\neq1$. The solar neutrino reduction is an additional step, while the spallation values can differ from those for $N_{\rm n}=1$. \label{tab:eff_ibd_nontag}} 
\tablecolumns{7}
\tablewidth{0pt}
\tablehead{
    Reduction & 
    \multicolumn{6}{c}{Energy Region [MeV]} \\
     & [15.5,~16.5] & [16.5,~17.5] & [17.5,~18.5] & [18.5,~19.5] & [19.5,~23.5] & [23.5,~79.5]
}
\startdata
Solar & 0.72 & 0.81 & 0.87 & 0.97 & 1.0 & 1.0 \\
SK-VI Spallation & 0.73 & 0.73 & 0.78 & 0.78 & 0.86 & 0.95 \\
SK-VII Spallation & 0.40 & 0.40 & 0.46 & 0.46 & 0.53 & 0.98 \\
\enddata
\end{deluxetable*}

%The energy threshold of the analysis is set at $E_{\rm rec}=15.5~\rm MeV$ in each region, as in~\citet{Abe2021},  
The background events are divided into five categories: one spallation, one NCQE, and three non-NCQE, contrary to the spectrum-independent analysis. % , the non-NCQE events of this analysis are divided into finer subcategories. 
The first of the non-NCQE subcategories is from events with a visible muon or pion in the final state ($\mu/\pi$ background), mainly appearing in the $\theta_c\in [20^\circ,38^\circ]$ region. 
The second comes from electrons stemming from the decay of invisible muons and pions (Decay-$e$ background), while the last is from the charged-current interactions of atmospheric electron neutrinos and antineutrinos with no visible muon or pion in the final state ($\nu_e$-CC background). These second and third components reconstruct to the $\theta_c\in [38^\circ,53^\circ]$ region.
In the $N_{\rm n}\neq1$ region, then, there is a mixture of NCQE and non-NCQE events with varied final state neutron multiplicities.
NCQE interactions produce final states with often multiple neutrons~\citep{Sakai2024}.
On the other hand, many CCQE events continue to higher energies with slowly increasing neutron multiplicity~\citep{Han2025}.

Each event category (backgrounds and DSNB signal) is associated with a PDF across the extended parameter space of the spectral analysis, whose overall event rate is to be fitted to the data samples.
% ... whose overall event rate across all samples/panels/etc. is to be fitted... ?

\subsection{Spallation Modeling}
\label{subsec:spall_modeling}
Spallation events above $E_{\rm rec}= 15.5~\rm MeV$ mainly reconstruct to the $N_{\rm n}\neq1$ and $\theta_c\in[38^\circ,53^\circ]$ region.
Some of these become accidental coincidences in the $N_{\rm n}=1$ and $\theta_c\in[38^\circ,53^\circ]$ region but are negligible due to the low misidentification rate of neutron tagging.
%We are sensitive to spallation events near the 15.5~MeV energy threshold. %due to the absence of requiring exactly one neutron.
%In the $N_{\rm n}\neq1$ sample in the middle $\theta_c$ region, we are sensitive to spallation events near the 15.5~MeV energy threshold. %due to the absence of requiring exactly one neutron.
%As a reminder, most spallation radioisotopes do not emit neutrons in coincidence with their $\beta$ decay. 
%To generate the baseline spallation PDF here, we start with the same shape as in~\citet{Abe2021}. 
Therefore, to generate the spallation PDF, we focus on three isotopes (${}^9$C, ${}^8$B, and ${}^8$Li shown in Figure~\ref{fig:spatimeyield}) that will remain in the $N_{\rm n}\neq1$ and $\theta_c\in[38^\circ,53^\circ]$ region after spallation reduction due to their large endpoint energies and high yields. 
We then combine the reconstructed energy spectra of each of the three spallation isotopes into one global spallation energy spectrum.
%This parametric approach is completed by fitting this spectral sum to an analytical function, as was done in~\citet{Abe2021}. 
As in~\citet{Abe2021}, the following analytical function is then fit to this spectral sum:

\begin{eqnarray}
    {\rm PDF_{spall}}(E_{\rm rec}) \propto \exp\left(-\frac{(E_{\rm rec}+0.511~{\rm MeV})^\alpha}{\beta} \right),
\end{eqnarray}
where $\alpha$ and $\beta$ are free parameters.
This is the baseline PDF shape before taking into account any energy-dependent effects from event selection steps.
%In SK-IV, the spallation cuts chosen were estimated to have no energy-dependent impact on the spallation spectral shape, i.e., we assumed that the spallation background efficiency (i.e., its acceptance rate) was constant across the energies covered by the spallation PDF. 

To proceed, we should incorporate the impact of applying an energy-dependent cut for reducing solar neutrino events, assuming the same efficiencies as IBD events, summarized in Table~\ref{tab:eff_ibd_nontag}.
These efficiencies rescale the baseline spallation PDF per energy bin.
Next, we consider any energy-dependent effects from spallation cuts applied to the spallation PDF.
In SK-IV, it was determined that there was no energy dependence on the spallation remaining rate due to the cuts chosen.
%, and the uncertainties on the relative abundances of the three isotopes included were assigned as a systematic uncertainty.
%Next, in SK-IV, the chosen spallation cuts were estimated to have no energy-dependent impact on the spallation spectral shape, i.e., we could safely assume that the spallation background remaining rate was constant across the energies covered by the spallation PDF. 
In contrast, in this analysis, we choose different spallation cut criteria for each energy bin, which induces an energy-dependent impact on the spallation remaining rates.
%In SK-Gd though, the chosen cuts do induce an energy-dependence of the background acceptance rates. 
The spallation PDF is therefore reshaped by these rates in each energy bin to obtain the final PDF shape. % to be included in the spectral analysis. 
%pattern A
At these energies, we estimate that our reduction steps after the spallation reduction, i.e., atmospheric neutrino reduction and neutron tagging, have a negligible impact on the spectral shape of the remaining spallation events. 
%This is because only neutron misidentification will pull these events from $N_{\rm n}\neq1$ to $N_{\rm n}=1$, and this is negligible impact to PDF since the misidentification rate is at most $\mathcal{O}(10^{-3})$ per event and is independent from the energy bin.
%pattern B

%original?
\begin{comment}
%Concerning the requirement $N_{\rm n} \neq 1$, there is negligible impact.
%This is because the spallation isotopes modeled by this PDF do no produce neutrons.
%Therefore, only misidentification of noise or radioactive backgrounds will pull these events from $N_{\rm n}\neq1$ to $N_{\rm n}=1$ and impact the PDF.
%Since the misidentification rate is at most $\mathcal{O}(10^{-3})$ per event, this is negligible.
\end{comment}

\subsection{Systematic Uncertainties}

The background-related systematic uncertainties encode the uncertainties in the overall shape of the background PDFs across the entire parameter space $\theta_c \otimes\ N_{\rm n} \otimes\ E_{\rm rec}$, while the uncertainty on the integrated signal efficiency is the only signal-related systematic uncertainty considered for the fit. 
In particular, the uncertainty on the energy scale is assumed to be negligible in this analysis, as it has an insignificant effect on the shape of the PDFs and a negligible impact on the fit results due to the large statistical uncertainties from the small size of our final data samples.% caused by the statistical limitations of the analysis. 

The systematic uncertainty estimates for backgrounds in this analysis remain unchanged from those in ~\citet{Abe2021}. There are four nuisance parameters to be fitted: $\eta_{\rm spall}$ for the uncertainty in the shape of the spallation PDF, $\eta_{\nu_e \rm CC}$ for the uncertainty on the predicted energy-dependent ascending slope of the $\nu_e \rm \ CC$ PDF, $\eta_{\rm NCQE}$ for the relative contribution of NCQE events in three $\theta_c$ regions, and $\eta_{\rm n}$ for the relative contribution of all event categories between two neutron-tagging regions. 
To these, we add the one nuisance parameter related to the signal efficiency $\eta_{s}$. 

As the background PDFs are area-normalized to one and should be positive across the energy range, these parameters have physical limits.
Taking these constraints into account, we assign each parameter a reduced and centered prior distribution, namely a normal distribution for $\eta_{s}$ and $\eta_{\rm spall}$, a folded normal distribution for $\eta_{\nu_e \rm CC}$ and $\eta_{\rm n}$, and a log-normal distribution for $\eta_{\rm NCQE}$ (for details, see Appendix~\ref{appendix:spectral_fit_results}).

\subsection{Extended Likelihood}

We denote $\vec{\eta}_{b}$ as the 4-vector of the background-related systematics nuisance parameters ($\eta_{\rm spall}$, $\eta_{\rm NCQE}$, $\eta_{\nu_e \rm CC}$, $\eta_{\rm n}$), $\vec{N}_b$ as the 5-vector of background event rates ($N_{\rm spall}$, $N_{\rm NCQE}$, $N_{\rm{Decay} \ e^-}$, $N_{\nu_e \rm CC}$, $N_{\mu/\pi}$), and $N_s$ as the number of DSNB events corrected from the signal efficiency $\varepsilon_{s}(\eta_{s})$, equal to the number of DSNB events with an energy $E_\nu > 17.3$ MeV that have occurred in the SK fiducial volume.
We note that $\varepsilon_{s}$ differs from $\bar{\epsilon}_{\rm IBD}$ used in Equation \ref{eq:limit} of the spectrum-independent search of Section~\ref{sec:binnedanalysis} in two aspects: $\varepsilon_{s}$ does not contain the neutron tagging efficiency, given all $N_{\rm n}$ outcomes are included in the two neutron-tagging regions of the spectral analysis; and $\varepsilon_{s}$ is the integrated efficiency over the entire energy range of the spectral analysis and is therefore dependent on the shape of the DSNB model considered. 
The extended likelihood \citep{Barlow1990} to be maximized per phase reads:
{\small
    \begin{eqnarray}
    \mathcal{L}\left( \mathrm{Data}\, \big|\, N_s,\, \vec{N}_b, \, \right.&\eta_{s}&, \, \vec{\eta}_{b} \Big) 
    = 
    \mathcal{L}\left(\vec{0}\, |\, \eta_{s}, \, \vec{\eta}_{b} \right)\, \label{eq:likelihood_spectral_fit}\\ 
    \nonumber
    &\times&
    e^{- \left( \varepsilon_{s}(\eta_{s})N_s \ + \ \sum_{j \in \vec{b}} N_j \right)} \\
    \nonumber
    &\times&
    \prod_{i = 1}^{N_\mathrm{data}} 
    \Bigl[ 
    \varepsilon_{s}(\eta_{s})N_s \cdot \mathrm{PDF}_{s} \left(E^i,\,  \theta_{C}^i,\, N_{\mathrm n}^i \right) \\ \nonumber 
    &&
    \ \ + \sum_{j \in \vec{b}} N_j \cdot \mathrm{PDF}_{j} \left(E^i,\, \theta_{C}^i,\, N_{\mathrm n}^i\,  |\, \vec{\eta}_b \right) 
    \Bigr] \nonumber,
    \end{eqnarray}
}
where $\mathcal{L}\left(\vec{0}\, |\, \eta_{s}, \, \vec{\eta}_{b} \right)$ is the penalty term coming from the product of prior distributions for the nuisance parameters, which have prior values considered to be 0. 
$\mathrm{PDF}_{s} \left(E^i,\, \theta_{C}^i,\, N_{\mathrm n}^i \right)$ is the signal-related PDF, and $\mathrm{PDF}_{j} \left(E^i,\, \theta_{C}^i,\, N_{\mathrm n}^i\,  |\, \vec{\eta_b} \right)$ are the background-related PDFs, whose shape may vary depending on the value of the nuisance parameters $\vec{\eta_b}$.
The exponential term and the $N_{s,\, j}$ parameters account for the Poissonian fluctuations of the rate for each category of event.

To derive the best-fit DSNB event rate across all SK-Gd data, we maximize the sum of the SK-VI and SK-VII log-likelihoods along the common signal efficiency-corrected $N_s$ parameter, which is thereafter converted to a DSNB flux value.
Confidence intervals for this parameter are then constructed by profiling the likelihood ratio (see Appendix~\ref{appendix:spectral_fit_results}).

\subsection{Results}

\begin{figure*}[htb!]
\centering
\epsscale{0.9}
%\plotone{figure/bestfit_sk6_nn.pdf}
%\plotone{figure/bestfit_sk7_nn.pdf}
\plotone{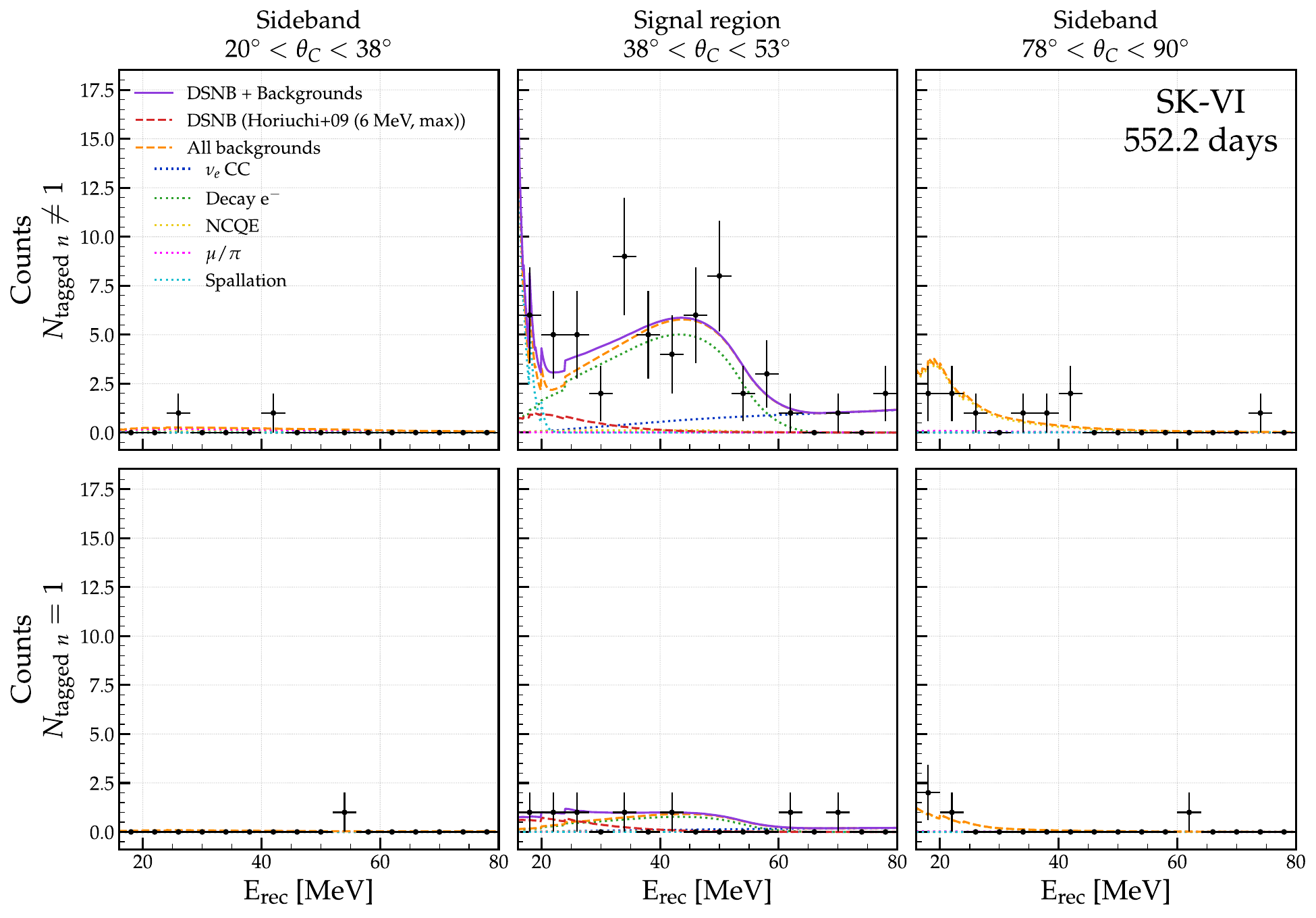}
\plotone{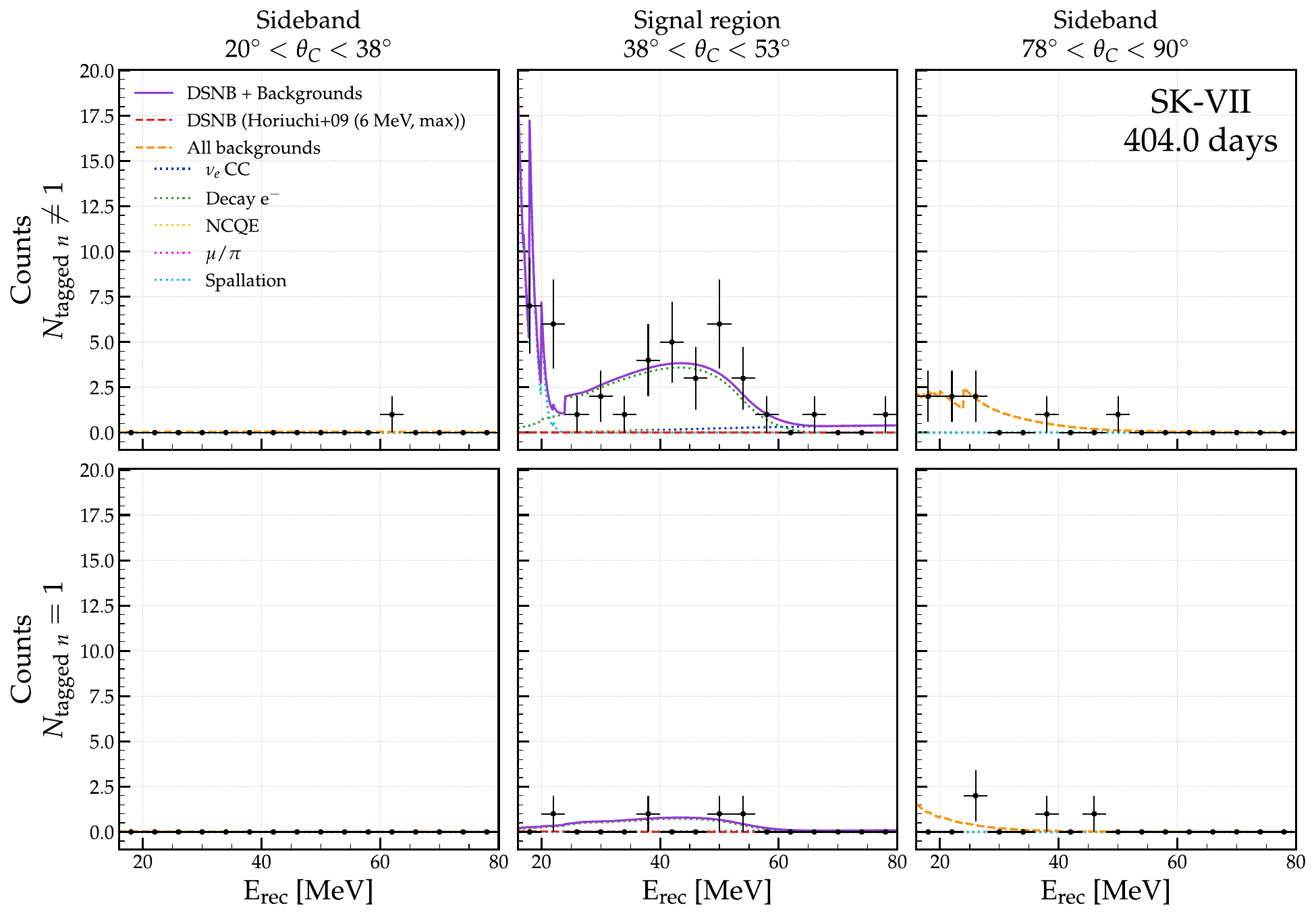}
\caption{\label{fig:bestfit_nn}
Best-fit results for SK-VI and SK-VII data samples, built out of the NN neutron-tagging algorithm. 
The top three panels show the events with zero or more than one neutron. 
The bottom three panels show events that found strictly one neutron, reducing accidental coincidence background rates by a factor of approximately $10^{4}$.
The input DSNB model used for the fit is \citet{Horiuchi2009}.}
\end{figure*}

\begin{figure*}[htb!]
\centering
\epsscale{0.8}
%\plotone{figure/Profile_likelihood_ratio.png}
\plotone{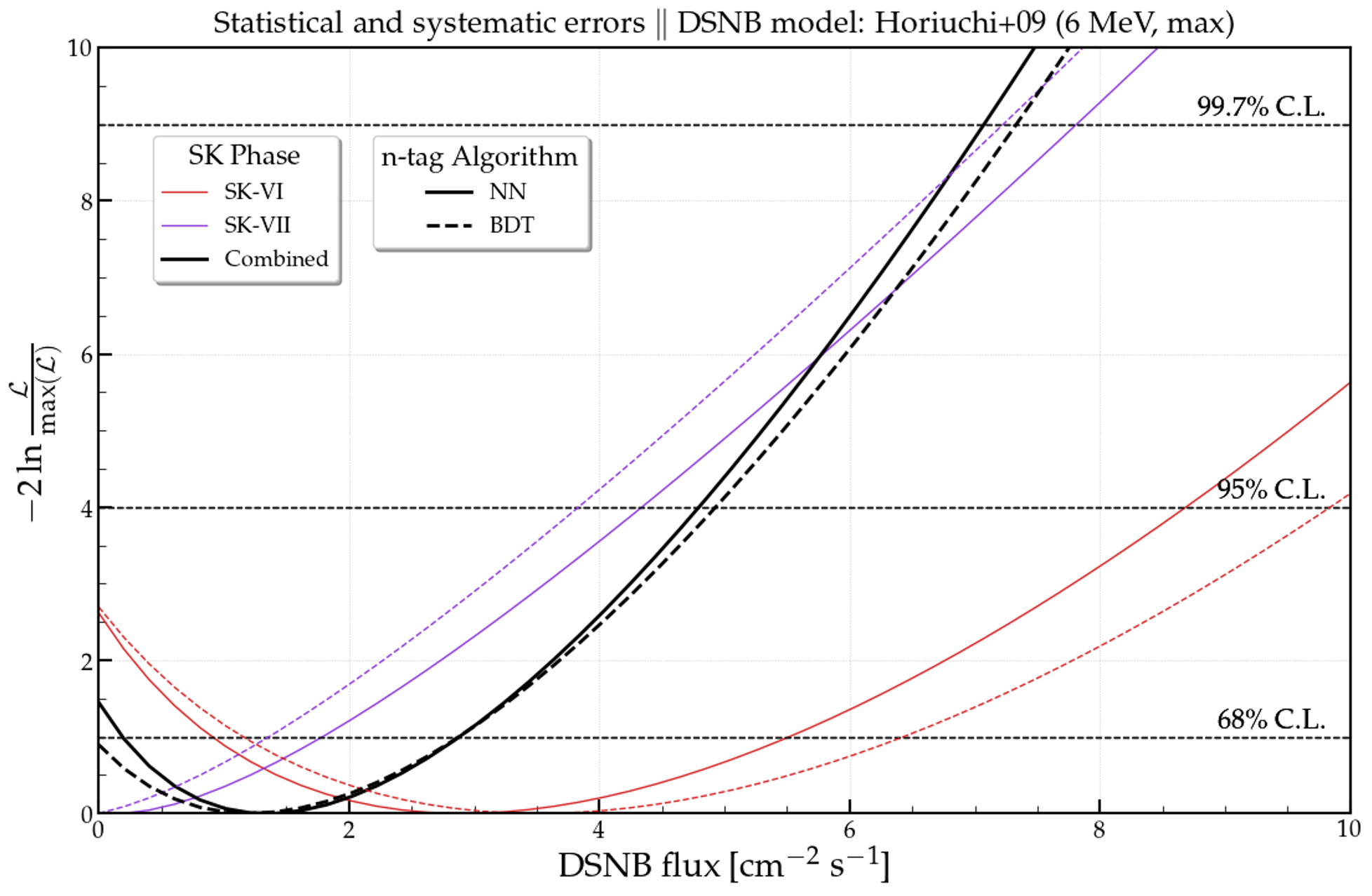}
\caption{\label{fig:profile_likelihood_ratio}
(Logarithm of) profile likelihood ratio as a function of the DSNB flux (for $E_{\nu} > 17.3$MeV), for both neutron-tagging algorithms and the DSNB model of \citet{Horiuchi2009}. 
%The combined fit of SK-Gd data (in black) rejects the background-only hypothesis at the 1.2$\sigma$ (resp. 0.9$\sigma$) level for the case using NN (resp. BDT).
The dotted line shows the significance corresponding to the confidence interval.
}
\end{figure*}

Using the model from \citet{Horiuchi2009} as a representative prediction of DSNB signal shape, we show in Figure~\ref{fig:bestfit_nn} the best-fit results for SK-VI and SK-VII using the NN neutron-tagging algorithm. 
The best-fit flux range of $2.9^{+2.6}_{-2.0}$ and $0.1^{+1.7}_{-0.1}~\rm cm^{-2}\, s^{-1}$ for SK-VI and SK-VII includes the predicted value of $1.9~\rm cm^{-2}\, s^{-1}$.
The best-fit results for SK-VI and -VII using the BDT neutron-tagging algorithm are reported in Appendix~\ref{appendix:spectral_fit_results}. 
Additionally, Figure~\ref{fig:profile_likelihood_ratio} displays the associated phase-combined profile likelihood ratio functions. 
We can see that samples built using the NN or BDT neutron-tagging algorithm yield statistically compatible results. 

The combined fit of SK-Gd data shown as a black line demonstrates a best-fit flux of $1.4^{+1.5}_{-1.2}$ ($1.2^{+1.7}_{-1.2})~\rm cm^{-2}\, s^{-1}$ for the NN (BDT) sample, and rejects the background-only hypothesis at the 1.2$\sigma$ (0.9$\sigma$) level for the case using NN (BDT) neutron tagging, a similar rejection level to the 1.5$\sigma$ result obtained using 5823 days of pure-water SK data~\citep{Abe2021}.

\begin{figure*}[htb!]
\epsscale{1.}
    \centering
    \plottwo{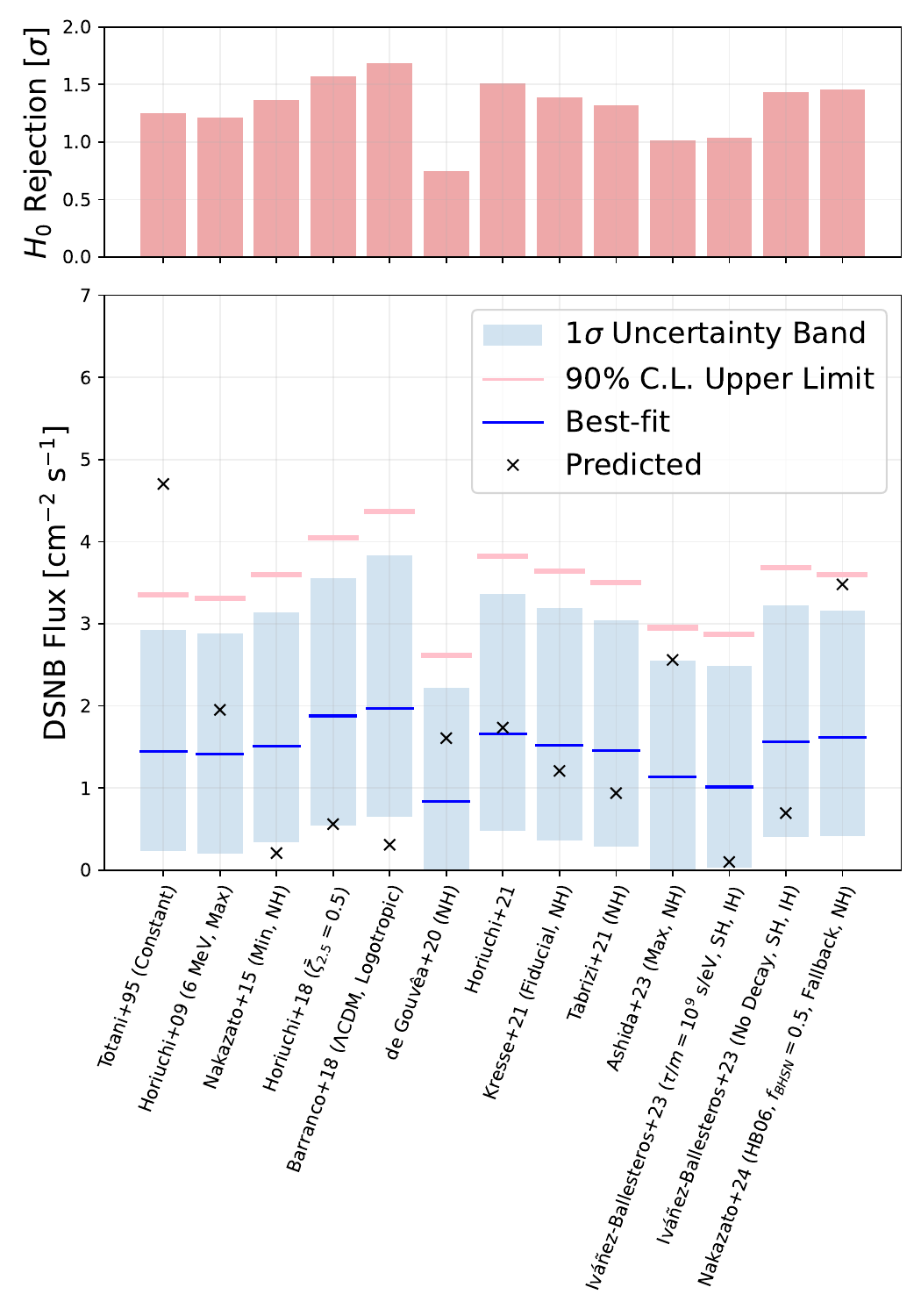}{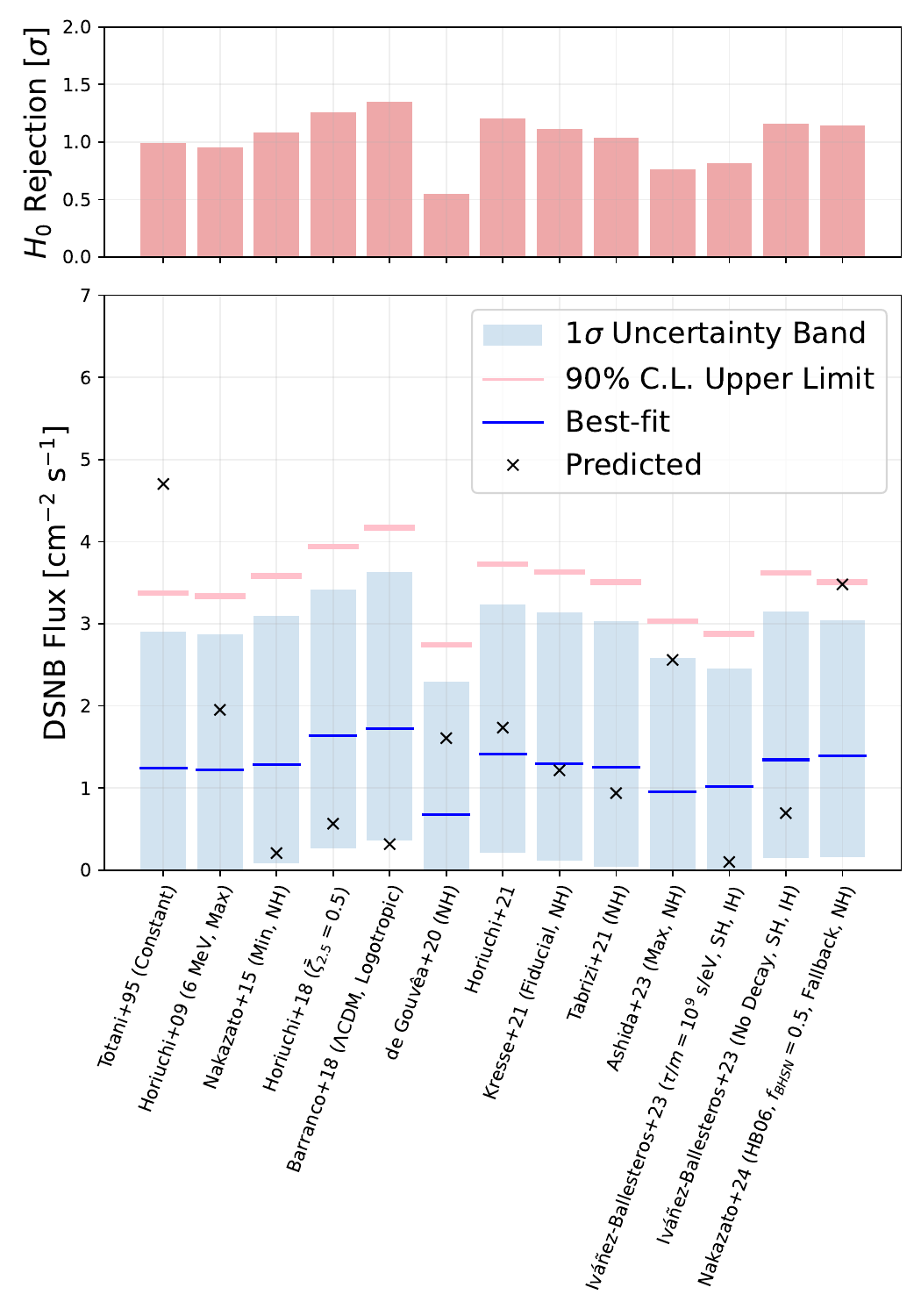}

\caption{\label{fig:spectral_analysis_model_dependent_results_MainBody} Model-dependent spectral analysis results for DSNB events with $E_\nu > 17.3$ MeV, obtained with (\textit{left}) NN and (\textit{right}) BDT neutron-tagging algorithms. (\textit{bottom}) $1\sigma$ uncertainty bands and 90\% C.L. upper limits on the DSNB flux, as well as (\textit{top}) rejection of the background-only hypothesis $H_0$ for a variety of input DSNB model shapes~\citep{Totani1995, Hartmann1997, Malaney1997, Kaplinghat2000, Ando2005, Fukugita_2003, Horiuchi2009, Lunardini2009, Galais2010, Nakazato2015, Priya2017, Barranco2018, Horiuchi2018, DeGouvea2020, Horiuchi2021, Kresse2021, Tabrizi2021, Ashida2023, Ivanez2023, Martinez2024, Nakazato2024}. For the sake of completeness, results with additional DSNB model shapes from \citep{ Kresse2021, Ivanez2023, Martinez2024, Nakazato2024} are shown in Appendix~\ref{appendix:spectral_fit_results}.  “NH” and “IH” refer to the normal and inverse neutrino mass hierarchies, respectively. “MD14” corresponds to the SFR calculations performed by \citet{Madau_2014}. For the models from \citet{Ivanez2023}, “SH” stands for “strongly hierarchical”. Finally, “CE”, “CGI” and “HMA” stand for “chemical evolution”, “cosmic gas infall” and “heavy metal abundance” respectively for models \citet{Hartmann1997}, \citet{Malaney1997} and \citet{Kaplinghat2000}. For more information on the models, please refer to the various publications.  }
\end{figure*}

Repeating the fitting procedure for different DSNB models~\citep{Totani1995, Hartmann1997, Malaney1997, Kaplinghat2000, Ando2005, Fukugita_2003, Horiuchi2009, Lunardini2009, Galais2010, Nakazato2015, Priya2017, Barranco2018, Horiuchi2018, DeGouvea2020, Horiuchi2021, Kresse2021, Tabrizi2021, Ashida2023, Ivanez2023, Martinez2024, Nakazato2024} yields similar confidence intervals, with an excess of $0.7-1.7\sigma$ with the NN-based approach and $0.5-1.3\sigma$ with the BDT-based approach. 
Frequentist upper limits on the DSNB flux at the 90\% C.L. are also derived as follows, in the frame of the Wald asymptotic approximation~\citep{Cowan}:
\begin{equation}
    \mu_{\rm upper, \, 90\% \, C.L.} = \hat{\mu} + \sigma \cdot \mathcal{N}^{-1}(90\%),
\end{equation}
where $\hat{\mu}$ is the best-fit DSNB flux, $\sigma$ is conservatively estimated as the upper uncertainty on the best-fit value, and $\mathcal{N}$ is the normal cumulative density function. 
We summarize the spectral fitting results (best-fit flux with $1\sigma$ fitting uncertainty, 90\% C.L. upper limit, and significance of excess over backgrounds) for these models in Tables~\ref{tab:specres_nn_appendix} and~\ref{tab:specres_bdt_appendix} in Appendix~\ref{appendix:spectral_fit_results}, and display some in Figure~\ref{fig:spectral_analysis_model_dependent_results_MainBody}.  Based on insufficient significance, we conclude that no excess beyond the background-only hypothesis is observed in the spectral analysis of the SK-Gd data. 
%As can be seen in the plot, the \citet{Totani1995} model is excluded at a high confidence level, similarly to the analysis of the SK-I to -IV data~\citep{Abe2021}.%, so that we ignore this model in the discussion below.

Yet, we should emphasize that the combined fit results in approximately $\sim 1.4 \ \rm{cm}^{-2} ~\rm{s}^{-1}$ uncertainty for the DSNB flux, for the Horiuchi+09 model. Noticeably, this is a considerable improvement with respect to the previous pure-water phases. Indeed, with only twice the size of the present SK-Gd dataset, the uncertainty should then become comparable to that of the $\sim$~6000 days of the SK pure-water phases ($\sim 0.9 \ \rm{cm}^{-2} \rm{s}^{-1}$), showing the enhanced sensitivity achieved in the Gd phase. 

%\begin{figure*}[htb!]
%    \centering
%    \gridline{ \fig{figure/spectral_test_v5.pdf}{0.8\textwidth}{}
%    }
%\caption{\label{fig:specfit_test} Test.}
%\end{figure*}

%\input{section/discussion}

\section{Discussion}
% --- result summary
%While the two statistical approaches we employ for the spectrum-independent and -dependent analyses are distinct, we can perform general comparisons of their results to check for consistency. 
%It is important to note that the spectrum fit has an energy threshold at $E_{\rm rec}=15.5$~MeV, which corresponds to the fourth lowest-energy bin in the spectrum-independent $\bar{\nu}_e$ search. 
%When comparing the p-values of the spectrum-independent search with the spectrum fit result above this energy, we find that both agree on the interpretation of around $1\sigma$~rejection of a background-only hypothesis. 
%->results?

% -- insight for best-fit
The spectral-fitting results indicate an excess over the background-only hypothesis at the $\sim$~$1\sigma$ level for many DSNB models. 
Despite the large variation of flux shapes from different modeling approaches, the best-fit values and $1\sigma$ intervals do not differ significantly. 
This suggests that changes in model parameters may not be distinguishable given the current statistical and systematic uncertainties.
%However, in some steeply decreasing flux models with extremely low event rates, such as certain parameter sets of \citet{Ivanez2023}, \citet{DeGouvea2020}, and \citet{Barranco2018}, the fitting band starts to follow their characteristic shape. 
However, in some steeply decreasing flux models with extremely low event rates, such as certain parameter sets of \citet{Ivanez2023}, \citet{DeGouvea2020}, and \citet{Barranco2018}, our fitting is already sensitive to their particular shape, which causes the best-fit DSNB flux and $1\sigma$ intervals to differ from the majority of models. 
%Next, we comment on the best-fit values compared to model predictions.
For example, the best-fit value of the minimum flux case of the \citet{Nakazato2015} model is comparable to other best-fit values, yet it is slightly above the theoretically predicted value. 
This suggests that the true flux level might be higher than conservative estimates indicate.
In contrast, models with a large black-hole-formation effect, such as \citet{Nakazato2024} with $f_{\rm BHSN}=0.5$, the maximum case of \citet{Kaplinghat2000}, and the maximum case of \citet{Ashida2023},
possibly overestimate their parameter assumption: These predicted flux values are above their $1\sigma$ best-fit ranges.
Finally, as another illustration, \citet{Ivanez2023} implements different neutrino decay scenarios which, depending on their lifetime and mass hierarchy, can modify the electron antineutrino flux to a greater or lesser extent. 

% --- update summary
Given the importance of neutron identification for the SK DSNB search, we employed two machine learning techniques --- the newly developed NN and the updated BDT. 
%Final samples were prepared such that the NN or BDT-based neutron tagging was the only reduction step that differed for both spectrum-dependent and -independent analyses. 
%Along the way, we demonstrated (e.g., Figures~\ref{fig:ntagroc_opt} and \ref{fig:signaleff}) that the performances of both algorithms are similar. 
Since they are constructed, trained, and tuned independently, this adds robustness to the results; indeed, the NN and BDT arrive at similar performance levels for distinct reasons. 
Moreover, the physics inferred from our data is consistent across neutron identification techniques for both statistical analysis approaches.

Enabled by the new MSG reduction targeting NCQE and other multi-cone events, we have demonstrated how these backgrounds become subdominant after cut optimization, as illustrated across all bins in Figure~\ref{fig:binned_spectrum}.
In the spectral fit example of Figure~\ref{fig:bestfit_nn}, we observe that a negligible amount of NCQE is fitted in the signal-rich $\theta_c$ region. 
While this NCQE reduction comes at a further cost to the IBD signal efficiency, the background removal is highly effective.
Further improvements may be achieved through machine learning approaches, such as that proposed by \citet{maksimovic_cnns_2021}.

%--- future discussion
Moving forward, improvement is still needed in the signal efficiency shown in Figure~\ref{fig:signaleff}, particularly at lower energies, although Gd-loading has already led to a significant increase. 
Furthermore, background events should be further reduced, for which there remain two dominant sources in the SK-Gd DSNB search.
The first are the decays of invisible muons and pions at higher energies. 
The second, in contrast, are spallation products that dominate at the lowest energies. 
%This underlines the importance of understanding and subsequently reducing these backgrounds even further. 
%Already, though, we expect that SK-Gd astrophysical $\bar{\nu}_e$ sensitivity will surpass SK-IV at higher energies, and it is already more sensitive than SK-IV at lower energies. 
The current data-driven method for extracting spallation event characteristics heavily relies on statistics.
In addition, evaluating the ability to remove spallation events relies on physics assumptions, which cause significant uncertainty above 50\% for the $^9\rm Li$ remaining rate. 
A better understanding of cosmic-ray muon interactions in water and the development of a reliable spallation simulation are crucial for improving background reduction, more accurate spallation event PDFs, and a reasonable estimation of the isotope remaining rate. This will lead to more strict constraints on the DSNB flux in the region where the flux is largest.

In this study, we presented a reliable data-driven method for estimating the non-NCQE normalization uncertainty.
However, further suppressing this uncertainty is limited by the statistics of the sideband samples.
The DSNB flux prediction in the higher energy region is largely affected by black hole formation: The longer accretion phase makes the neutron star hotter, such that the energy of emitted neutrinos is higher.
In the future, searches with larger SK datasets, including pure-water phase will enable access to the black-hole-formation history~\citep{Ashida2022, Ashida2023}, made possible by increased statistics in the sideband region. % can be used to constrain systematic uncertainties in the higher energy region more effectively, particularly those from invisible muons and pions. 
Additionally, new nuclear interaction models, once validated by the data from neutrino experiments, have the potential to reduce the uncertainties in the atmospheric neutrino background as studied by~\citet{zhou_first_2024}.  
%Additionally, new nuclear interaction models, once validated by the data from neutrino experiments, have the potential to reduce the uncertainties in the atmospheric neutrino background.  
Notably, better modeling of the neutron multiplicity in atmospheric neutrino interactions will be crucial for the discovery of DSNB, since we employ neutron tagging to enhance sensitivity.

\section{Conclusion}

We have analyzed 956.2~livedays of SK-Gd data with two parallel statistical approaches for the DSNB search. 
Given the importance of prompt and delayed signal coincidence, we used two machine learning algorithms to identify neutron captures for the first time in SK-Gd. 
In addition, we implemented a new background reduction technique targeting atmospheric neutrino interactions. 
In a DSNB spectrum-independent search, we searched for the DSNB signal in the 7.5 to 29.5~MeV energy range, and observed no significant excess over background predictions. 
Then, we set new upper limits on the astrophysical $\bar{\nu}_e$ flux. 
%For neutrino energies of 9.29--17.29~MeV, we updated the world's most stringent limit to be 33.20--2.76 $\rm cm^{-2}\, s^{-1}\, MeV^{-1}$.
In this time, we updated the world’s most stringent limits in the energy region 9.29-11.29~MeV to $33.2~\rm cm^{-2}\,s^{-1}\, MeV^{-1}$, in the energy region 11.29–13.29~MeV to $8.14~\rm cm^{-2}\,s^{-1}\, MeV^{-1}$, and 13.29-17.29~MeV to $2.76~\rm cm^{-2}\,s^{-1}\, MeV^{-1}$.
In a DSNB spectral fit, we observed an approximately 1.2$\sigma$ (0.9$\sigma$) rejection of a background-only hypothesis for the majority of DSNB models considered while using an NN-based (BDT-based) neutron capture identification algorithm.
%In a DSNB spectral fit, we observed a 1.2$\sigma$ (0.9$\sigma$) rejection of a background-only hypothesis using an NN (BDT) for neutron capture identification and a nominal DSNB spectral prediction.

%\input{section/acknowledegement}
%\begin{acknowledgements}    
\section*{Acknowledgments}
%\acknowledgement
%Paper Acknowledgements(as of May. 30, 2025)

%Short list (in case number of lines are limited in short letter papers)
%We gratefully acknowledge cooperation of the Kamioka Mining and Smelting Company. The Super-Kamiokande experiment was built and has been operated with funding from the Japanese Ministry of Education, Science, Sports and Culture, and the U.S. Department of Energy.

%Full list (for full papers and longer papers)
%\todo[inline]{ver. May 30, 2025}
We gratefully acknowledge the cooperation of the Kamioka Mining and Smelting Company. The Super-Kamiokande experiment has been built and operated from funding by the Japanese Ministry of Education, Culture, Sports, Science and Technology; the U.S. Department of Energy; and the U.S. National Science Foundation. Some of us have been supported by funds from the National Research Foundation of Korea (NRF-2009-0083526, NRF-2022R1A5A1030700, NRF-2202R1A3B1078756, RS-2025-00514948) funded by the Ministry of Science, Information and Communication Technology (ICT); the Institute for Basic Science (IBS-R016-Y2); and the Ministry of Education (2018R1D1A1B07049158, 2021R1I1A1A01042256, RS-2024-00442775); the Japan Society for the Promotion of Science; the National Natural Science Foundation of China under Grants No. 12375100; the Spanish Ministry of Science, Universities and Innovation (grant PID2021-124050NB-C31); the Natural Sciences and Engineering Research Council (NSERC) of Canada; the Scinet and Digital Research of Alliance Canada; the National Science Centre (UMO-2018/30/E/ST2/00441 and UMO-2022/46/E/ST2/00336) and the Ministry of  Science and Higher Education (2023/WK/04), Poland; the Science and Technology Facilities Council (STFC) and Grid for Particle Physics (GridPP), UK; the European Union’s Horizon 2020 Research and Innovation Programme H2020-MSCA-RISE-2018 JENNIFER2 grant agreement no.822070, H2020-MSCA-RISE-2019 SK2HK grant agreement no. 872549; and European Union's Next Generation EU/PRTR  grant CA3/RSUE2021-00559; the National Institute for Nuclear Physics (INFN), Italy.
%\end{acknowledgements}

\nopagebreak

\appendix
\restartappendixnumbering
\section{Neutron cloud cut with Gd} \label{appendix:ncloudcut}
%\section{Spallation cut update \label{appendix:spallation}}
\begin{comment}
At first, we eliminate events with a muon within 1~ms before the event time due to potential contamination from hadronic showers after muon spallation. The inefficiency of this reduction is negligible given our 2~Hz of cosmic-ray muon crossing rate.
Even after above ``1-ms cut'', multiple $\beta$-like events produced by hadronic showers from energetic muon spallation could form a cluster of MeV-scale events. 
To target these, we utilize spatial correlations between events in clusters instead of timing correlations wit muons, termed ``multiple-spallation cut.'' 
As such, we search and remove for other low-energy events with $E_{\rm rec}$ in 5.5--24.5~MeV, occurring within 60~s and 4~m from the DSNB candidate. 
This criterion was determined based on long-lived isotopes, such as $^{16}\rm N$. 

We evaluate the reduction efficiency of the multiple-spallation cut for the spallation event by requiring spatial correlation between DSNB candidates and low-energy events within a time range of 60~s, resulting in the remaining spallation event fraction of 57\%. 
The signal selection efficiency of this cut can be evaluated by calculating the efficiency for events without spatial correlation.
To do this, we randomly sample the vertices of low-energy events paired with DSNB candidates in the ID and apply the same cut, finding that the signal efficiency of this cut is 98\%.
\end{comment}

\begin{figure}[htb!]
\centering
\epsscale{0.6}
%\plotone{figure/illust_ncloud2.pdf}
\plotone{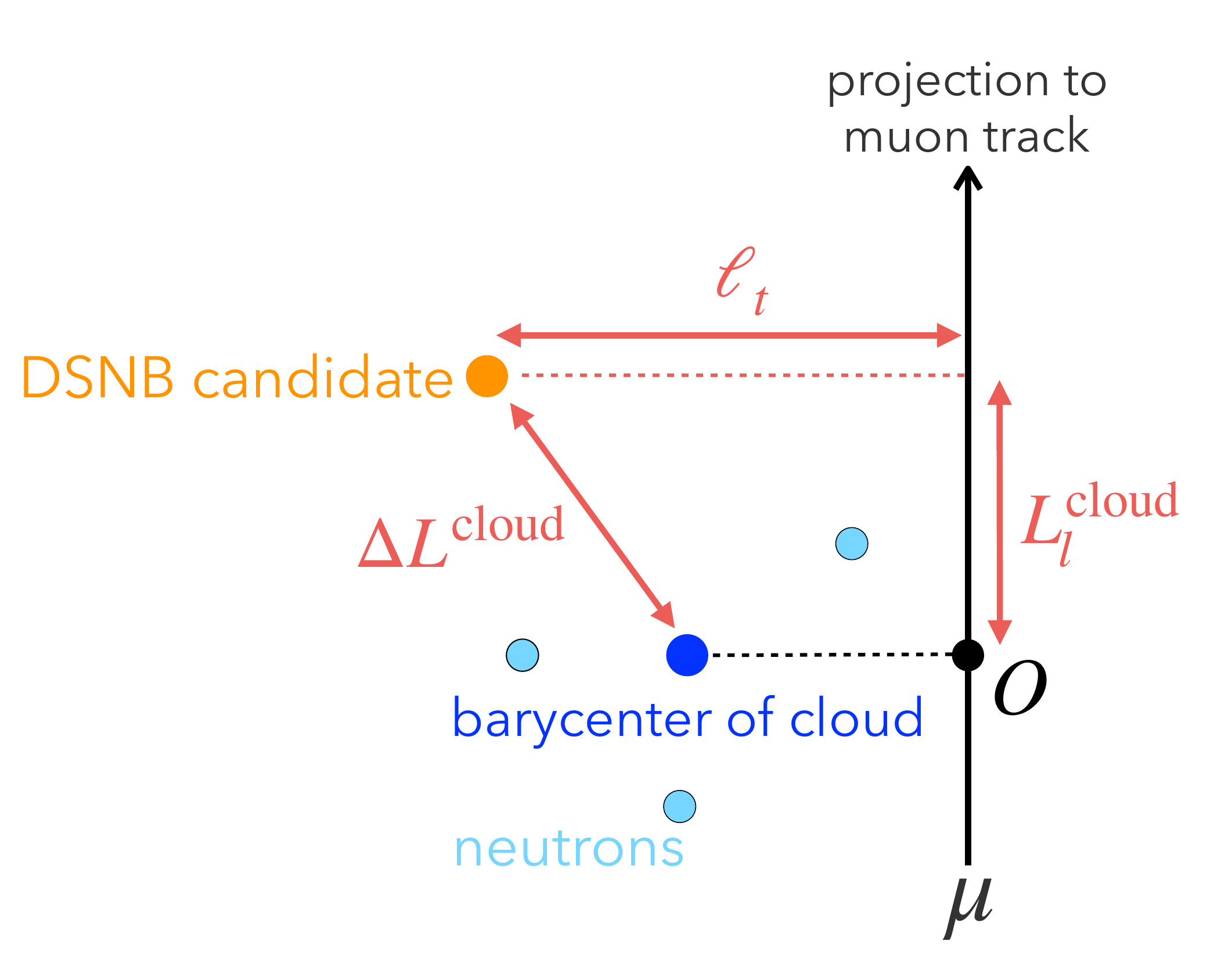}
\caption{
\label{fig:illustncloud}Illustration of neutron cloud variables.
}
\end{figure}

%\subsection{pre-treatment cut \label{appendix:spapre}}
%\subsection{Neutron cloud cut} \label{appendix:spapre}
Regarding the neutron cloud cut, the neutron detection method follows \citet{Shinoki2023} completely. 
To remove neutron-correlated spallation events, we use the timing difference of muons ($\Delta t$) and spatial correlation ($\Delta L^{\rm cloud}$) of the barycenter of reconstructed neutron cloud vertices from DSNB  candidates to reduce spallation events close to the hadronic shower. 
%To remove neutron-correlated spallation events, we use the timing difference of muons ($\Delta t$; shown in Figure~\ref{fig:ncloud_dt}) and spatial correlation ($\Delta L^{\rm cloud}$) of the barycenter of reconstructed neutron cloud vertices from DSNB  candidates to reduce spallation events close to the hadronic shower. 
In addition, a more sophisticated elliptical shape cut along with the reconstructed muon track is applied, using transverse distance from muon track($\ell_t$) and the position difference between neutron cloud and DSNB candidate along with the muon track ($L^{\rm cloud}_l$).
We utilize the same muon reconstruction algorithm as one used in \citet{Kitagawa2024}, detailed in \citet{conner1997} and  \citet{Desai2004}. 
%In addition, a more sophisticated elliptical shape cut along with the reconstructed muon track using distances along with the muon track ($dl^{\rm cloud}_l$) and transverse distance ($l_t$) express elliptical shape is applied. The muon reconstruction algorithm is detailed in \citet{conner1997, Desai2004}. 
The definitions of $\Delta L^{\rm cloud}$, $L^{\rm cloud}_l$, and $\ell_t$ are illustrated in Figure~\ref{fig:illustncloud}. 
Conservatively, we apply the same cut threshold as in the SK-IV analysis~\citep{Scott2024} since the timing difference between muon and spallation product should not have large differences. 
The vertex resolution improvement due to Gd has a minimal impact on the neutron cloud vertex. 

\begin{figure}[htb!]
\centering
\epsscale{0.8}
%\plotone{figure/illust_muon.pdf}
\plotone{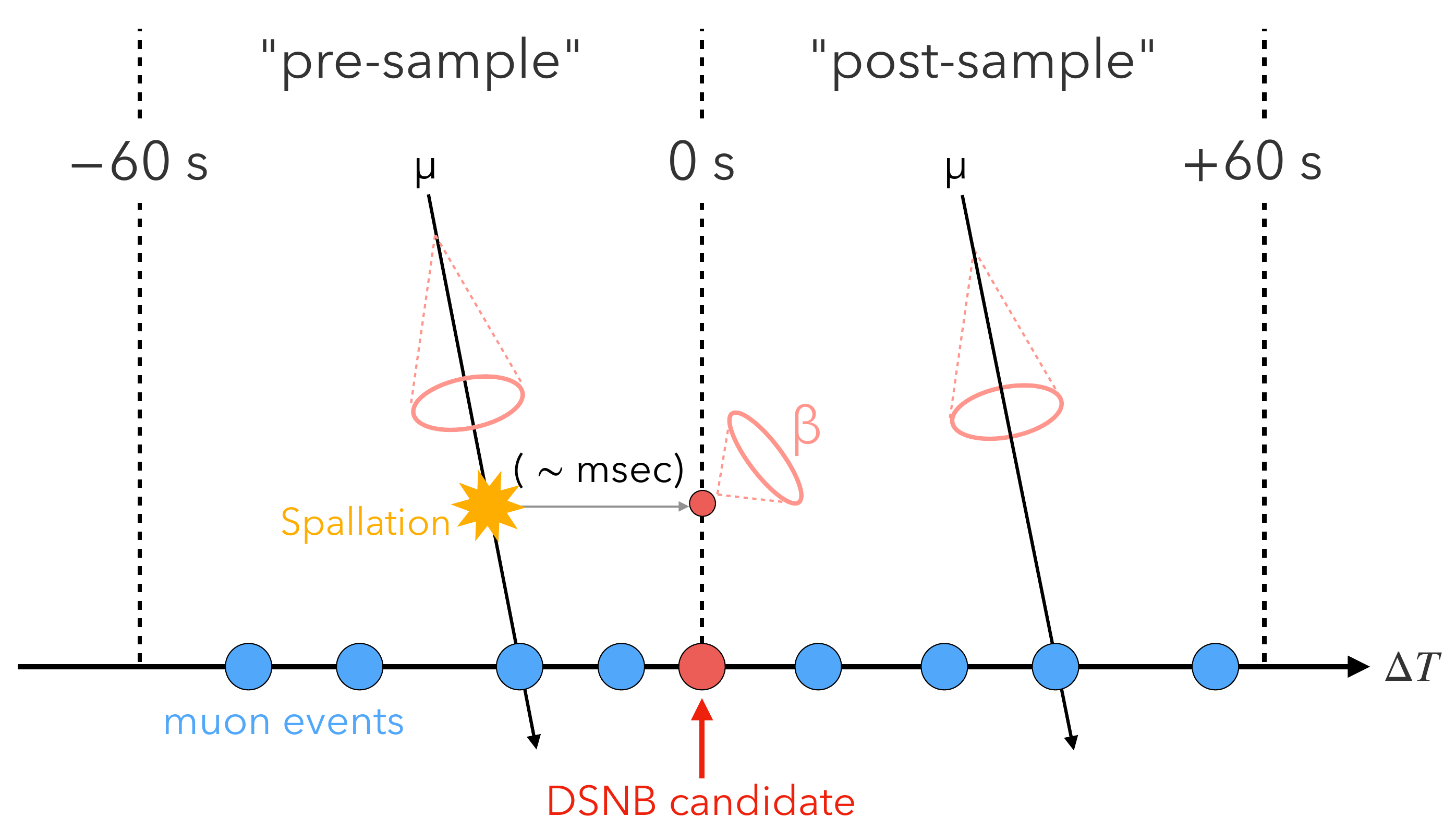}
\caption{
\label{fig:illustmuon} Illustration of the separation for pre- and post-sample regions.
}
\end{figure}

To extract the neutron cloud cut performance, at first, we separate muon samples within $\pm60~\rm s$ around DSNB candidates into pre-sample and post-sample, which are the prior and posterior timing muons, respectively, as illustrated in Figure~\ref{fig:illustmuon}. 
The muon responsible for causing spallation is included only in the pre-sample, and all other muons in the pre-sample and all post-sample muons should not be correlated with the DSNB candidates. 
This concept is used for the likelihood approach, as described below.

\begin{figure}[htb!]
\centering
%\plotone{figure/ncloud_dist_0.pdf}
\plotone{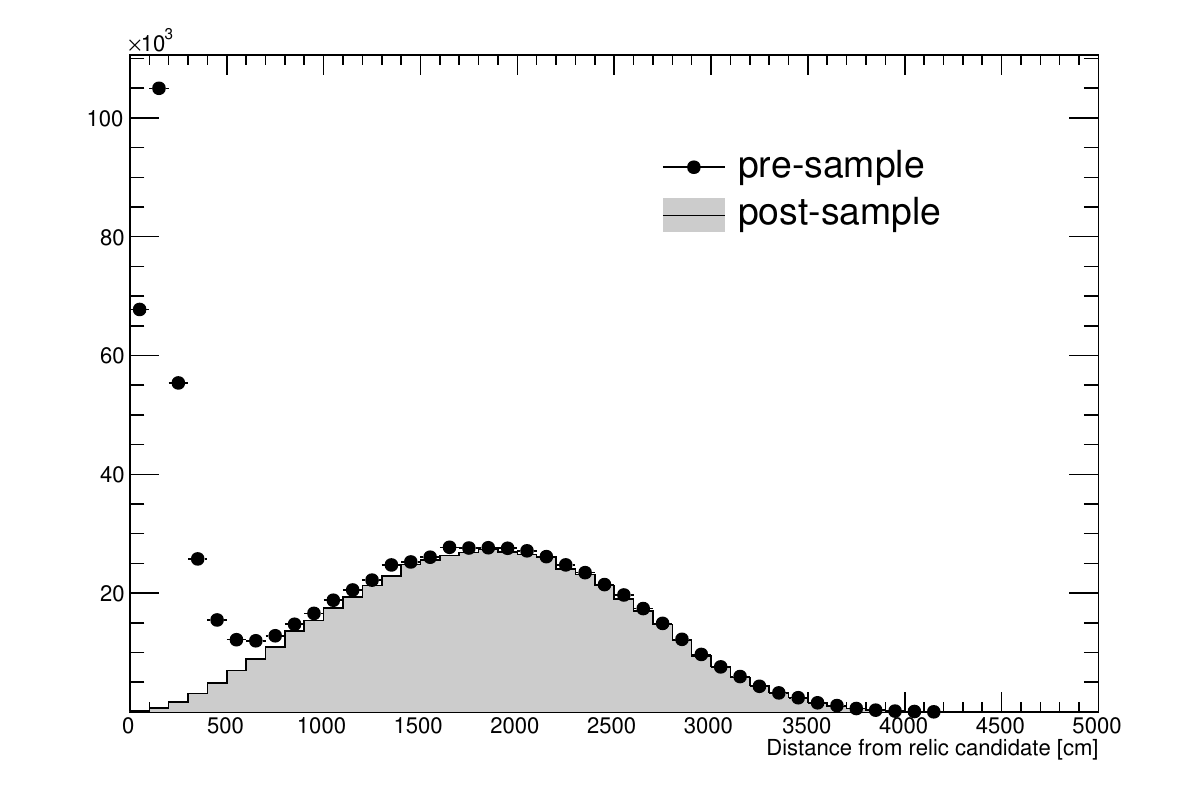}
%\PlaceHolder[0.3\textwidth]
\caption{
\label{fig:deltal_ncloud} $\Delta l$ distribution between a DSNB candidate and the muon events before and after the candidate.
}
\end{figure}

Figure~\ref{fig:deltal_ncloud} shows an example of the $\Delta L^{\rm cloud}$ for pre- and post-sample muons. 
A clear correlation is found in small $\Delta L^{\rm cloud}$ only in the pre-sample, and good consistency is seen in large distances exceeding 10~m.

The efficiency of the neutron cloud cut for both signal and background is calculated using pre- and post-sample data, following the same method as in previous works~\citep{Abe2021}. 
As a result, this cut removes 51\% of spallation events while keeping 98\% of the signal.
%The efficiency after both multiple spallation cut and neutron cloud cut, collectively called `spallation pre-cut,' is 97\% for the non-spallation events while eliminating 65\% of spallation events.

\section{Atmospheric neutrino background reduction \label{appendix:thirdred}}
\subsection{MSG Cut for NCQE Event Reduction}\label{appendix:msg}
\begin{figure}[ht!]
\centering
%\plotone{figure/srn_ncqe_eff.pdf}
\plotone{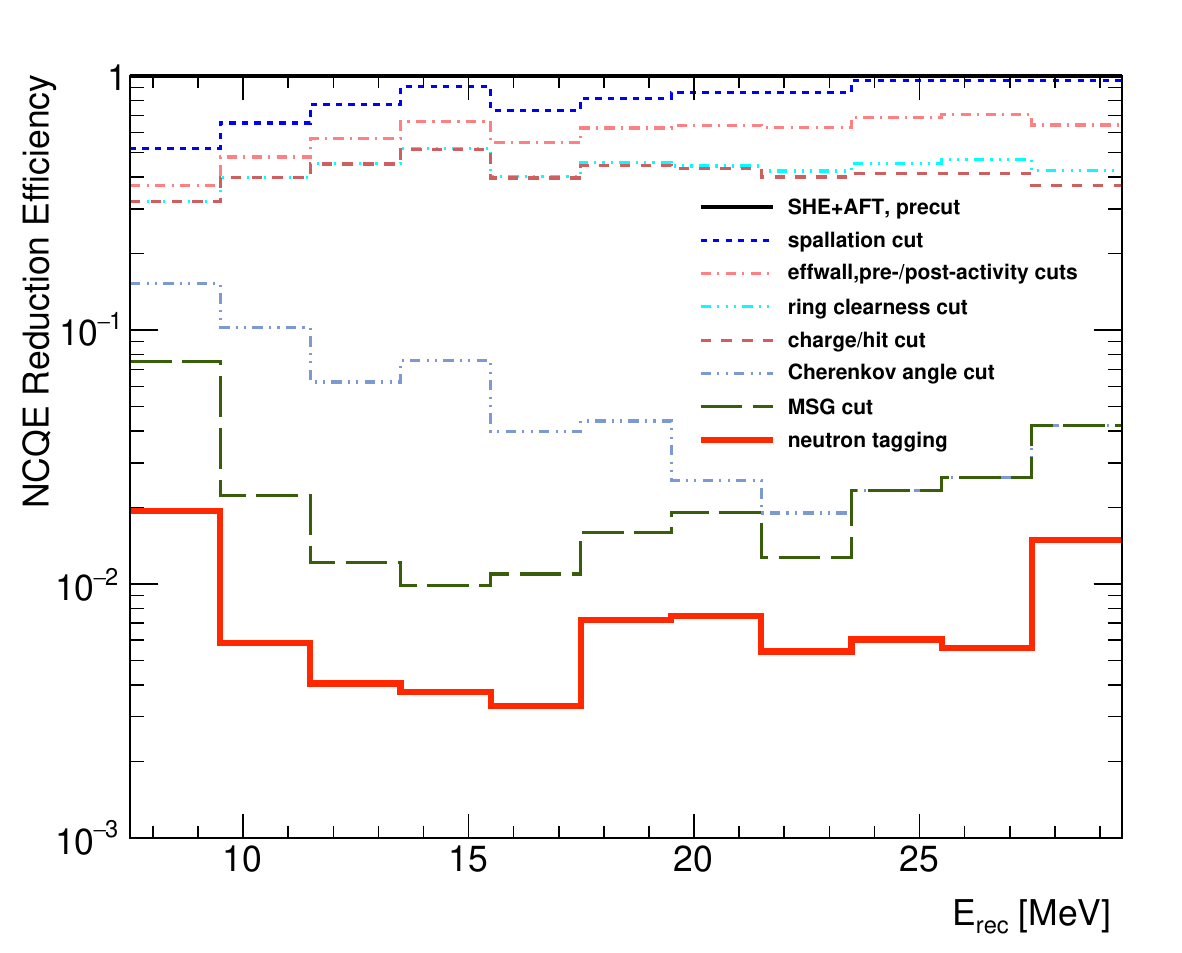}
\caption{
\label{fig:ncqe_eff}
NCQE event reduction efficiency for each cut step in SK-VI. These lines show accumulated efficiencies at each stage. 
}
\end{figure}

Figure~\ref{fig:ncqe_eff} shows the reduction efficiency of NCQE events. The Cherenkov angle cut and the MSG cut are the most effective in reducing NCQE events. %, and the other third reduction cuts are less effective in removing NCQE events. 
In higher energy regions, many NCQE events have multiple Cherenkov cones and can easily be reduced by the Cherenkov angle cut due to their topology. On the other hand, the lower energy events have a single-cone-like pattern or do not generate enough PMT hits to be identified as multiple cones, resulting in a worse reduction efficiency. The MSG cut is effective for these lower-energy events and complements the Cherenkov angle cut. This is because MSG exploits the finer structure of the PMT hit topology (again, originally to quantify the multiple scattering of electrons). Overall, the effect of MSG cut on NCQE is the strongest in regions for which NCQE events dominate the DSNB signal compared to other backgrounds, which is roughly in the energy range of $E_{\rm rec}\in$ [9.5,~19.5]~MeV.
The energy-dependent MSG threshold values in the analysis are given in Table~\ref{tab:msg_values}.

\begin{deluxetable}{l c}[htb!]
\tablecaption{MSG event selection threshold values as a function of energy $E_{\rm  rec}$.\label{tab:msg_values}} 
\tablecolumns{2}
\tablewidth{0pt}
\tablehead{
    Energy [MeV] & MSG Threshold Value
}
\startdata
$[7.5,~9.5]$ & 0.39 \\
$[9.5,~11.5]$ & 0.43 \\
$[11.5,~13.5]$ & 0.47 \\
$[13.5,~15.5]$ & 0.42 \\
$[15.5,~17.5]$ & 0.37 \\
$[17.5,~19.5]$ & 0.36 \\
$[19.5,~21.5]$ & 0.35 \\
$[21.5,~23.5]$ & 0.32 \\
\enddata
\end{deluxetable}

%\section{Atmospheric neutrino background reduction \label{appendix:thirdred}}
\subsection{Non-NCQE atmospheric neutrino background reduction}
\begin{figure}[ht!]
\centering
%\plotone{figure/NN_tagout}
\plotone{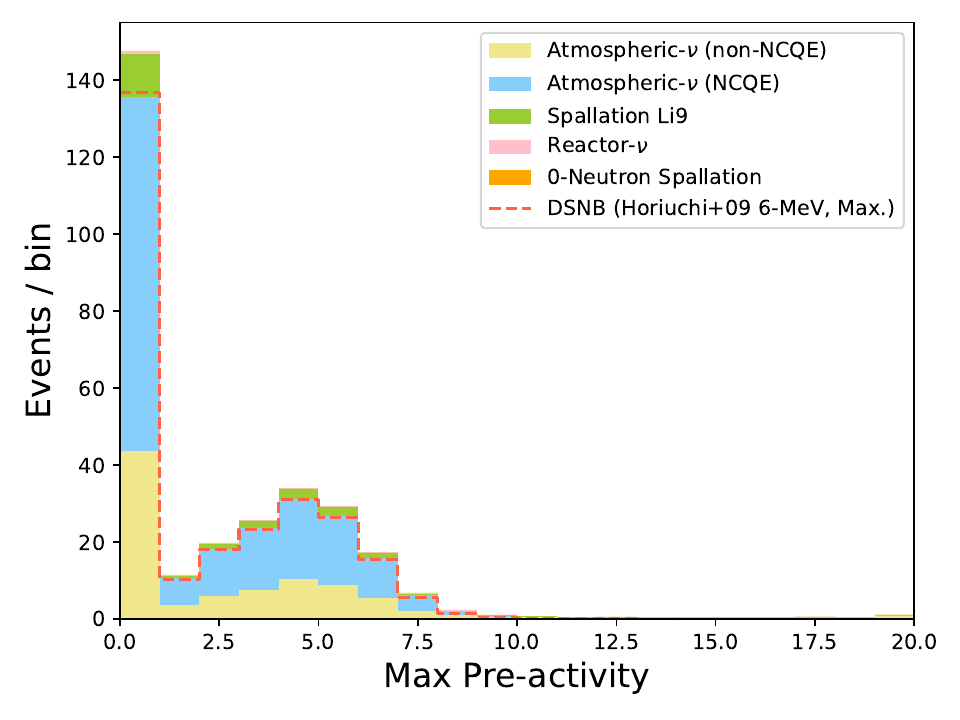}
%\PlaceHolder[0.3\textwidth]
\plotone{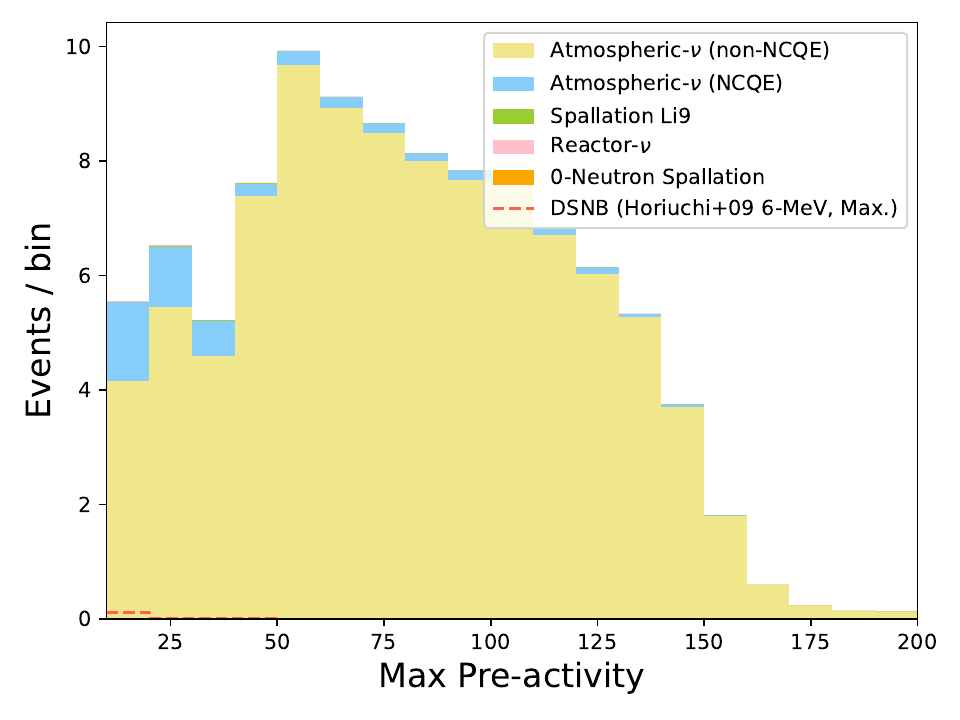}
\caption{%The maximum number of PMT hits in a 15-ns t-TOF sliding window between 5~$\mu$s before the main peak to 12~ns before it. 
Max pre-activity distribution for the small-value part (top) and the large-scale version (bottom). 
The DSNB signal model example is multiplied by a factor of 100 for demonstration. 
Event prompt energies are $E_{\text{rec}}\in[7.5,~29.5]$~MeV.
\label{fig:maxpre}
}
\end{figure}

%\textbf{
Since the major non-NCQE atmospheric background comes from Michel-electron events from invisible muon or pion decay, it is difficult to remove these events by hit-topology difference, such as MSG and Cherenkov angle.
One powerful technique for this type of event is examining the maximum PMT pre-activity before the prompt event peak.
%One powerful technique for rejecting atmospheric neutrino backgrounds---especially non-NCQE interactions---is examining the maximum PMT pre-activity before the prompt event peak.
This observable is calculated using the reconstructed prompt vertex to correct PMT hit timings by the time-of-flight.
Then, a 15~ns sliding window in time-of-flight-corrected timings searches for the maximum number of PMT hits within [$-$5~$\mu$s,~$-$12~ns] time window relative to the prompt peak timing.
Monte Carlo distributions for this maximum pre-activity for $E_{\text{rec}}\in[7.5,~29.5]$~MeV in prompt energy are shown in Figure~\ref{fig:maxpre}. A significant background peak appears in the large hits region, while the small peak around 5 is common to both the signal and the background.
%}

%\begin{figure}[ht!]
%\centering
%\plotone{figure/NN_tagout}
%\plotone{sk6_maxpre_nothirdred_wgt}
%\PlaceHolder[0.3\textwidth]
%\caption{The maximum number of PMT hits in a 15-ns t-TOF sliding window between 5~$\mu$s before the main peak to 12~ns before it. The DSNB signal model example is multiplied by a factor of 100 for demonstration. Event prompt energies are $E_{\text{rec}}\in[7.5,~29.5]$~MeV.
%\label{fig:maxpre}
%}
%\end{figure}

\begin{figure}[ht!]
\centering
%\plotone{figure/NN_tagout}
\plotone{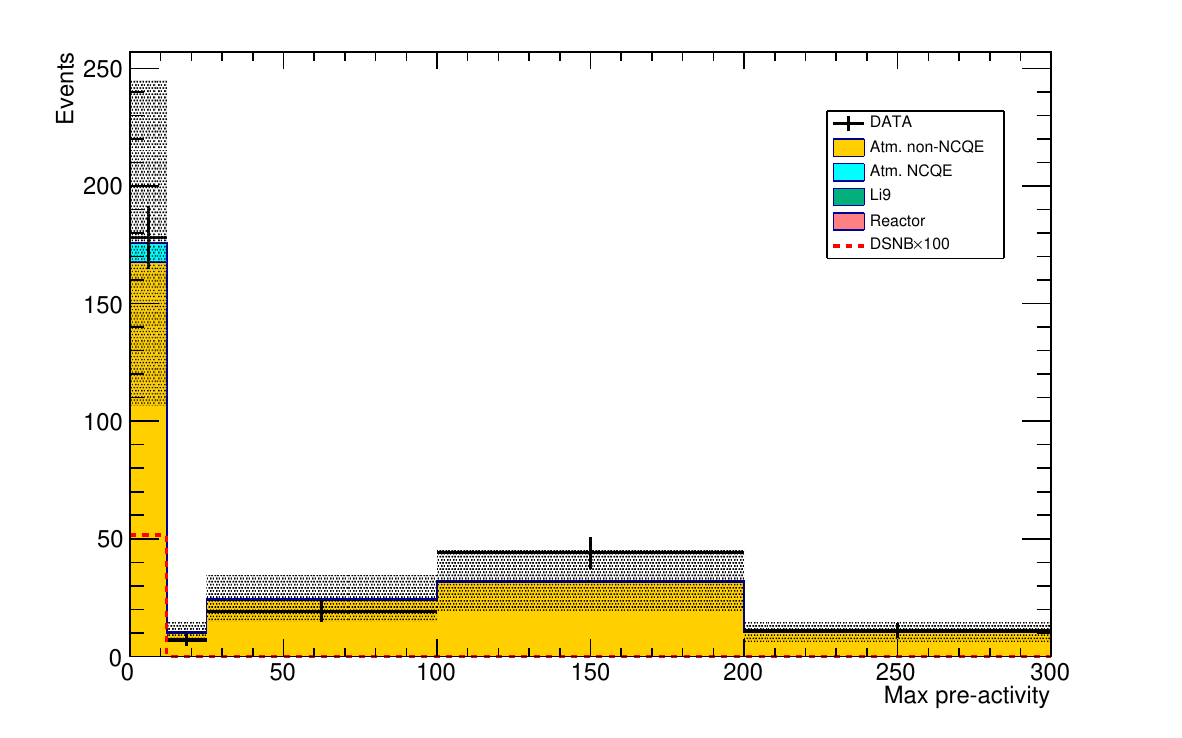}
%\PlaceHolder[0.3\textwidth]
\caption{Max pre-activity distribution of events with $E_{\rm rec} > 29.5~\rm MeV$. The hatched region is the final systematic uncertainty on MC prediction.
\label{fig:maxpre_comp}
}
\end{figure}

%\textbf{
%The MC predictions in Figure~\ref{fig:maxpre} are placed into two bins divided at the value 12, the same as applied in the SK DSNB search.
Approximately 38\% of non-NCQE events have gamma rays preceding the prompt event in our current modeling.
%Approximately 52\% of CCQE events have gamma rays preceding the prompt event in our current modeling and appear in the pre-timing peak, especially for $\nu_\mu$'s, and 48\% of these events can be rejected by this criterion.
In contrast, almost all NCQE and IBD interactions have low maximum pre-activity since they lack Cherenkov light emission before the main prompt peak.
%Essentially, all NCQE and IBD interactions have low maximum pre-activity from a lack of nuclear $\gamma$ emission before the main prompt peak or multiple charged leptons in the final state.
%As shown in Figure~\ref{fig:maxpre}, this selection removes around half of the non-NCQE events in an orthogonal way to the other prompt event selection steps.
We tested the pre-activity of non-NCQE background using the $E_{\rm rec} > 29.5~\rm MeV$ side-band dataset, which is dominated by non-NCQE background events. 
The pre-activity variable not only detects pre-gamma emission but also near-Cherenkov threshold particles with weak signatures that precede the prompt signal.
%Furthermore, the pre-activity values of these different events overlap.
The data are well explained by the MC predicted distribution as shown in Figure~\ref{fig:maxpre_comp}.
Applying the pre-activity selection, we find that 32\% of the data are rejected by this criterion, while 33\% of MC predictions are rejected. 
%Our model shows that this pre-activity explains the data well, as shown in Figure~\ref{}, and that 32\% of the data have a preceding event peak and are rejected by this criterion, while 33\% are rejected in the MC prediction.}
The main limitation of this approach is that the SK data structure limits scanning the pre-activity before 5~$\mu$s of the prompt peak.
Therefore, either longer timescale decays of Cherenkov-weak muon or preceding gamma-emission (pre-activity) to $e^\pm$ (prompt event) are not caught by the algorithm.
%Therefore, longer timescale decays of Cherenkov-weak $\mu^\pm$ (pre-activity) to $e^\pm$ (prompt event) are not caught by the algorithm.
%The same is true for Cherenkov-invisible $\mu^\pm$ going to visible $e^\pm$ (prompt event) preceded by $\gamma$-emission (pre-activity).
%The low and high pre-activity regimes are shown in Figures~\ref{fig:maxprelow} and \ref{fig:maxprehigh}.
%}

\section{NN Neutron Tagging \label{appendix:nn}}
The NN neutron tagging tool searches for peaks using a 14~ns sliding window with a 7-hit threshold to the time-of-flight corrected PMT-hit timing distribution.
For each cluster, we calculate feature variables; two types of the number of hits, such as 14~ns window ($N_{\rm hits}$) and $\pm$ 100~ns window ($N_{200}$), root-mean-square (RMS) of PMT hits from timing peak ($T_{\rm RMS}$), spherical harmonics parameters used in \citet{Bellerive2016} ($\beta_1$ and $\beta_5$), mean and RMS from the angle between each hit and averaged hit direction ($\theta_{\rm dir}^{\rm mean}$ and $\theta_{\rm dir}^{\rm RMS}$), mean, RMS, and skewness of the opening angle formed by three-hit combinations ($\theta_{\rm angle}^{\rm mean}$, $\theta_{\rm angle}^{\rm RMS}$, and $\theta_{\rm angle}^{\rm Skew}$), and the two kinds of distance of the prompt event from the ID wall ($d_{\rm wall}$) and $d_{\rm eff}$. 

For the classification algorithm, we adopt a feed-forward Multilayer-Perceptron (MLP) implemented using the TMVA library~\citep{Jan2010} as the NN algorithm. This NN is trained using $7\times 10^5$ events of IBD MC with an architecture of 0.02 as the learning rate, 14:15:13:1 as the layers, and using the sigmoid function for neuron activation.

To determine the selection criteria for the NN output score, we first explored the working neutron search window, which was 535~$\mu$s in SK-IV, because the increasing cross section of neutron captures in SK-Gd makes the neutron capture timing shorter. 
We applied an NN neutron selection to the signal MC with various combinations of NN score criteria and capture time criteria, ensuring $f_{\rm mis}$ remains about $\mathcal{O}(10^{-4})$ level. We optimized the time window to obtain the highest signal efficiency, as shown by the orange marker in Figure~\ref{fig:ntagroc_opt}. 
%We applied an NN cut to the signal MC with various combinations of NN cut and capture time criteria so that $f_{\rm mis}$ remains at $2.0\pm0.1\times10^{-4}$, and we optimized the time window to obtain the highest signal efficiency. 
%optimized to the most efficient signal efficiency is obtained.
%and optimized by comparing capture timing constant and misidentification rate. 
As a result, we successfully reduced the search time window range to $270~\mu$s and lowered the NN score threshold in SK-VII. % as shown in Figure~\ref{fig:nnwindow}.
The optimal criteria for the NN score are determined to be 0.99 for SK-VI and 0.98 for SK-VII.

\section{BDT Neutron Tagging \label{appendix:bdt}}

The BDT neutron search algorithm makes use of PMT hit clusters passing a pre-selection condition of at least 6~hits in a time-of-flight subtracted timing sliding window of 10~ns ($N_{10}$) starting 2~$\mu$s (in SK-Gd) after the prompt peak, going until 535~$\mu$s. This TOF is defined using the reconstructed prompt vertex with the understanding that IBD neutrons at these energies will not be captured far from the prompt vertex. After this, an attempt is made at reconstructing the neutron capture vertex to build other relevant observables. The BDT was trained on IBD MC, such that the relationship between the neutron capture vertex and the prompt vertex can be exploited with realistic neutron energies. Of the 22 total input variables, there are 9 (including $N_{10}$) that focus on the neutron capture vertex, another 7 focused on the Cherenkov light specifically, and a final 6 targeting noise hits. After $N_{10}$, the 5 most important input variables in SK-Gd (quantified by BDT F-score) are the number of PMT hits in clusters of at least 3 within 14.1$^\circ$ ($N_c$), the number of PMT hits with a low probability of originating from the reconstructed neutron capture vertex ($N_{\text{low}}$), the mean opening angle of the PMT hits ($\theta_m$), the distance of the neutron capture vertex from the detector walls (f$_{\text{wall}}$), and the distance between the reconstructed prompt vertex and neutron capture vertex (fp$_{\text{dist}}$). A list of all BDT input variables and their definitions is shown in Table~\ref{tab:bdt_inputs}.

For this study, it was important to ensure a low misidentification rate where no-neutron spallation dominates for $E_{\rm rec} < 29.5$~MeV. 
To this end, we added a pre-selection step for samples for BDT Neutron tagging, using $N_{10}$, $N_c$, and the number of reconstructed Cherenkov photons from the delayed vertex (\texttt{bse}) per $N_{10}$ window. 
Misidentification is more likely when there are low hits for $N_{10}$, few hits in clusters with $N_c$, and few Cherenkov photons reconstructed for \texttt{bse}. 
We reject neutron candidates for which $N_{10}<13$ and $N_c=0$, as well as those for which $N_{10}<13$ and \texttt{bse}~$< 20$. For SK-VII, this results in a misidentification rate of around 0.05\% at 64\% signal efficiency, which is about half the misidentification rate of the BDT's nominal performance at this signal efficiency. This pre-selection is taken into account for the performance shown in Figure~\ref{fig:ntagroc_opt}.

\begin{deluxetable*}{c | l}[htb!]
\centering
\tablecaption{ BDT input variable names and definitions. } \label{tab:bdt_inputs}
\tablecolumns{2}
\tablewidth{0pt}
\tablehead{
     \colhead{Variable} & \colhead{Definition} 
}
\startdata 
    $N_{10}$ & Number of PMT hits in a 10~ns TOF-corrected window from reconstructed prompt vertex. \\
    $\Delta N_{10}$ & Change in $N_{10}$ after using TOF correction from reconstructed delayed vertex instead. \\
    $t_{\rm RMS}$ & RMS of PMT hit times. \\
    $\Delta t_{\rm RMS}$ & Change in $t_{\rm RMS}$ after using TOF correction from reconstructed delayed vertex instead. \\
    $\rm fp_{dist}$ & Distance between reconstructed prompt vertex and delayed vertex. \\
    $\rm bp_{dist}$ & Difference between delayed vertex reconstructed position from two different approaches. \\
    \texttt{bse} & Reconstructed number of emitted Cherenkov photons emitted from the delayed vertex. \\
    $\rm f_{wall}$ & Distance of reconstructed delayed vertex (from minimizing TOF residuals within 2~m of prompt vertex) from detector wall. \\
    $\rm b_{wall}$ & Distance of BONSAI-reconstructed delayed vertex from detector wall. \\
    $\theta_{\rm mean}$ & Mean PMT hit opening angle calculated from the average PMT hit direction from the prompt vertex. \\
    $\theta_{\rm RMS}$ & RMS of the PMT hit opening angles. \\
    $\phi_{\rm RMS}$ & RMS of azimuthal separation between PMT hits. \\
    $N_{\rm back}$ & Number of PMT hits in the backward hemisphere from the average PMT hit direction. \\
    $N_{\rm low\theta}$ & Number of hits within opening angle of $20^\circ$. \\
    $N_{\rm clus}$ & Number of PMT hits inside a cluster of 3 separated by $14.1^\circ$. \\
    $N_{\rm low}$ & Number of PMT hits who are unlikely given the TOF information from the reconstructed delayed vertex. \\
    $Q_{\rm mean}$ & Mean charge deposited per PMT hit. \\
    $Q_{\rm RMS}$ & RMS of PMT hit charge deposited. \\
    $N_{\rm highQ}$ & Number of PMT hits with high charge deposited. \\
    $N_{300}$ & Number of PMT hits in 300~ns window around original $N_{10}$ window. \\
    $t_{\rm RMS}^{(3)}$ & Minimum RMS of hit times in subset of 3 hits. \\
    $t_{\rm RMS}^{(6)}$ & Minimum RMS of hit times in subset of 6 hits. \\
\enddata
\end{deluxetable*}

\section{Comparison of Americium-Beryllium calibration \label{appendix:ambe}}
\begin{figure}[ht!]
\centering
%\plotone{figure/NN_tagout}
\plotone{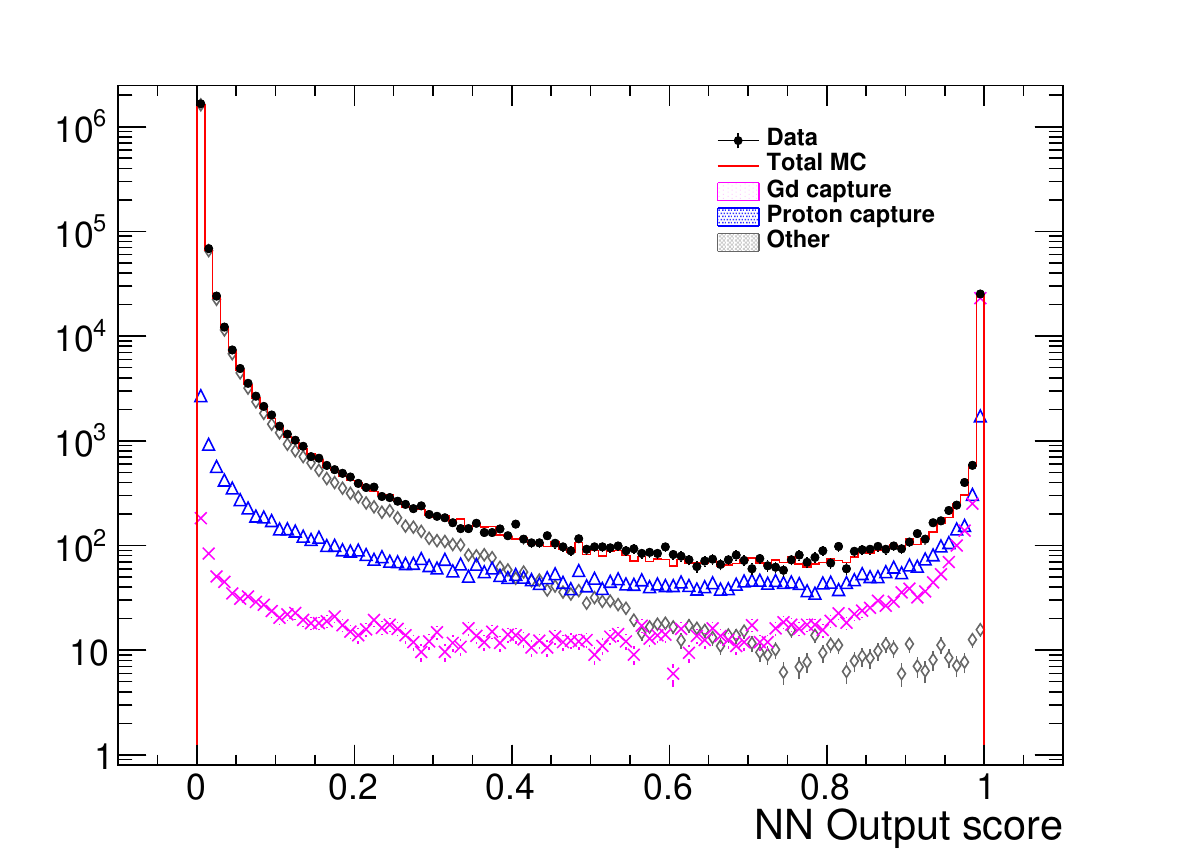}
%\PlaceHolder[0.3\textwidth]
\caption{
\label{fig:NNAmBescore}
NN score distribution of Am/Be data sample and MC in SK-VII. The red line shows the accumulated histogram among the Gd (Magenta cross) and proton (Blue triangle), and noise (Gray diamond). Each plot is normalized by the number of prompt events.
}
\end{figure}

\begin{figure}[ht!]
\centering
%\plotone{figure/ambe_sk6_log}
\plotone{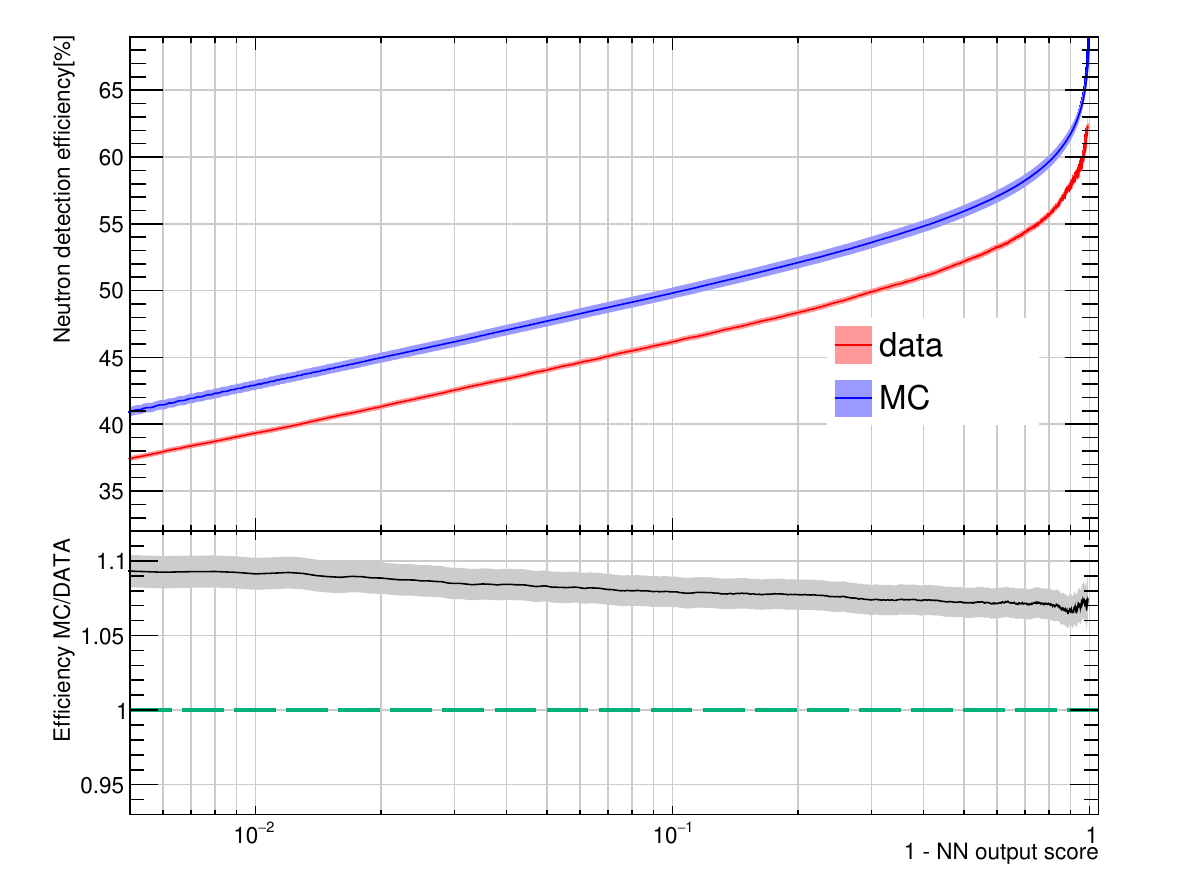}
%\plotone{figure/SK7_datamc_comparison_tageff_8BGO_weighted.retune_Logscale.pdf}
%\PlaceHolder[0.3\textwidth]
\caption{
\label{fig:NNAmBedatamc}
Neutron detection efficiency evaluated by Am/Be data and MC sample using NN algorithm (top) and its ratio between data/MC, as a function of cut criteria of NN score in SK-VI. The error band shown with the data represents statistical uncertainty, and with MC, it means the combination of statistical and systematic uncertainties.
}
\end{figure}

\begin{figure}[htb!]
\centering
\epsscale{1.0}
%\PlaceHolder[0.45\textwidth]
%\plotone{figure/sk7_ambe_eff_bdt_run91680}
\plotone{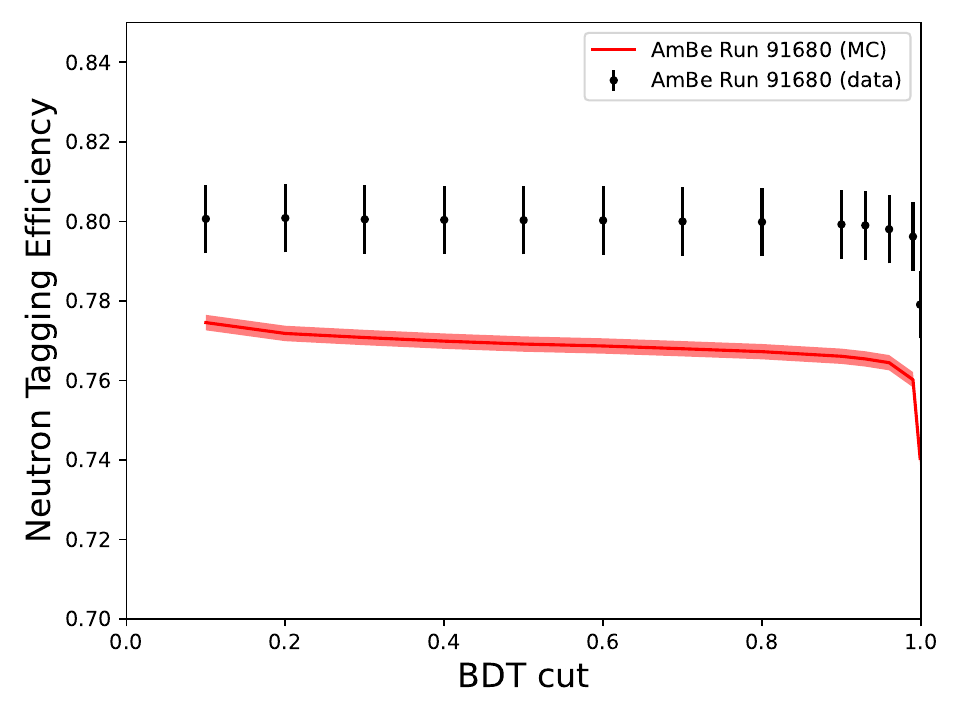}
\caption{Neutron selection efficiency for BDT as a function of BDT score threshold in one Am/Be run. Statistical error bars are shown on the data with a statistical error band for the MC prediction. \label{fig:eff_bdt_ambe}
}
\end{figure}

As mentioned in Section~\ref{subsec:validation}, we validate our neutron detection techniques using an Am/Be neutron source. 
The measurement configuration is detailed in~\citet{Harada2022}. %~\citet{Harada2023b}. 
We took data at nine source points in the SK tank equally $-12$, $0$, and $12~\rm m$ for the $x$ and $z$-axes, and monthly at the center point. 
Then, we evaluated the position dependence and time variation of neutron detection performance. 
The main purpose of this measurement is to check the consistency between the data and the MC sample for our neutron detection technique. 
Figure~\ref{fig:NNAmBescore} shows the distribution of the NN score of neutron candidates in the Am/Be measurement. 
The MC contains the true neutron capture events on Gd, protons, and noise candidates. The overall shape shows good agreement between the data and the MC samples.
We evaluated the neutron detection efficiency for both the data and MC samples, as well as their ratio as a function of the NN score. 
The results for SK-VII is shown in Figure~\ref{fig:NNAmBedatamc}. 
There are discrepancies between the data and the MC sample around the NN working point, with 8.4\% and 3.4\%, in SK-VI and SK-VII, respectively. 
The position dependence and time variation of the neutron detection efficiency are negligibly small and stable compared to the size of the discrepancy. 
Thus, we assigned this difference to the systematic uncertainty of neutron detection.
We also use this Am/Be data to estimate the systematic uncertainty associated with BDT neutron detection.
Following the same procedure of comparing data to MC for detection efficiency, we find a maximum discrepancy for the BDT of 5\% and 6\% for SK-VI and SK-VII, respectively.
Figure~\ref{fig:eff_bdt_ambe} shows the BDT comparison for one Am/Be run in SK-VII as an example.

\section{Spectral fit details \label{appendix:spectral_fit_results}}

\subsection{Nuisance parameters \label{appendix:nuisance_parameters}}

We provide the details below about the five nuisance parameters considered in the spectral analysis, as detailed in~\citet{Beauchene2024}. All parameters are expressed in units of standard deviation and assigned a centered and reduced probability density function, i.e., whose mean and standard deviation are, respectively, 0 and 1.

\subsubsection{\texorpdfstring{Signal efficiency: $\eta_{s}$}{Signal efficiency}}

The number of DSNB signal events to be fitted in the likelihood of Equation (\ref{eq:likelihood_spectral_fit}) is $\varepsilon_{s}(\eta_{s})N_s$. More specifically, we have: $\varepsilon_{s}(\eta_{s}) = \varepsilon_{s}^{0} \times (1 + \sigma_{s} \eta_{s})$, with $\varepsilon_{s}^{0}$ the nominal signal efficiency and $\sigma_{s}$ the total uncertainty on this value. This uncertainty includes the uncertainties related to the event selection criteria, the IBD cross section, and the phase livetime, which are summed in quadrature, amounting to $\sim 3\%$ total uncertainty for both SK-VI and -VII. $\eta_{s}$ is associated with a standard normal probability density function. 

\subsubsection{\texorpdfstring{Atmospheric $\nu_e$ CC interactions: $\eta_{\nu_e \rm CC}$}{Atmospheric nue CC interactions}}

As reported in the SK-I to -IV analysis \citep{Abe2021}, Monte-Carlo simulations predict a linear increase in the reconstructed energy of the spectrum for this event category. A 50\% systematic uncertainty on the slope of the spectrum is assigned by the formula to the right of the right arrow, fully correlated between the two neutron-detected regions:
%as following right-side formula by right-arrow, fully correlated between the two neutron-detected regions:
\begin{eqnarray}
    &\mathrm{PDF}&_{\nu_e \rm CC}(E_{\rm rec}) \to \mathrm{PDF}_{\nu_e \rm CC}(E_{\rm rec}) \nonumber
    \\ &\times&\left( 
    1 + \sigma_{\nu_e \rm CC}\, \eta_{\nu_e \rm CC}\, \frac{E_{\rm rec} - 15.5 \ [\mathrm{MeV}]}{79.5 - 15.5 \ [\mathrm{MeV}]}
    \right)
\end{eqnarray}
%\begin{equation}
%    \begin{aligned}
%        \mathrm{PDF}_{\mathrm{NCQE}}^{\mathrm{High}\,\theta_c} \to      &\mathrm{PDF}_{\mathrm{NCQE}}^{\mathrm{High}\,\theta_c}
%        \\ &\times (1 - \sigma_{\mathrm{NCQE}} A^{\mathrm{High}\,\theta_c}_{\mathrm{NCQE}} \eta_{\mathrm{NCQE}}) \\
%        \mathrm{PDF}_{\mathrm{NCQE}}^{\mathrm{Medium}\,\theta_c} \to      &\mathrm{PDF}_{\mathrm{NCQE}}^{\mathrm{Medium}\,\theta_c}
%        \\ &\times (1 + \sigma_{\mathrm{NCQE}} \eta_{\mathrm{NCQE}}) \\
%        \mathrm{PDF}_{\mathrm{NCQE}}^{\mathrm{Low}\,\theta_c} \to    &\mathrm{PDF}_{\mathrm{NCQE}}^{\mathrm{Low}\,\theta_c}
%        \\ &\times (1 - \sigma_{\mathrm{NCQE}} A^{\mathrm{Low}\,\theta_c}_{\mathrm{NCQE}} \eta_{\mathrm{NCQE}})
%    \end{aligned}
%\end{equation}
where $15.5$ (resp. $79.5$) MeV is the lower (resp. upper) boundary of the spectral analysis in the reconstructed energy window, and $\sigma_{\nu_e \rm CC}= 50\%$. Therefore, $\eta_{\nu_e \rm CC}$ is bound to be greater than $-2$ and is assigned a folded normal probability density function to preserve this feature.

\subsubsection{\texorpdfstring{NCQE interactions: $\eta_{\rm NCQE}$}{NCQE interactions}}

NCQE-type events exhibit a distinct angular distribution in the large Cherenkov angle region, associated with a significant uncertainty due to the challenging modeling of multiple secondary gammas emitted in these interactions. A significant portion of these events may therefore be misclassified in the signal Cherenkov angle region. Following the approach of \citet{Abe2021}, we parameterize this effect as a 100\% systematic uncertainty on the number of NCQE events in the signal (medium) Cherenkov angle region, fully correlated between neutron-tagged regions:

%\begin{eqnarray}
%        \mathrm{PDF}_{\mathrm{NCQE}}^{\mathrm{High}\, \theta_c} &\to \mathrm{PDF}_{\mathrm{NCQE}}^{\mathrm{High}\, \theta_c} \\ 
%        &\times (1 - \sigma_{\mathrm{NCQE}}\, A^{\mathrm{High}\, \theta_c}_{\mathrm{NCQE}} \, \eta_{\mathrm{NCQE}}) \nonumber\\
%        \mathrm{PDF}_{\mathrm{NCQE}}^{\mathrm{Mid}\, \theta_c} &\to \mathrm{PDF}_{\mathrm{NCQE}}^{\mathrm{Mid}\, \theta_c}  \times (1 + \sigma_{\mathrm{NCQE}}\, \eta_{\mathrm{NCQE}}) \\
%        \mathrm{PDF}_{\mathrm{NCQE}}^{\mathrm{Low}\, \theta_c} &\to \mathrm{PDF}_{\mathrm{NCQE}}^{\mathrm{Low}\, \theta_c}  \times (1 - \sigma_{\mathrm{NCQE}}\, A^{\mathrm{Low}\, theta_c}_{\mathrm{NCQE}}) \nonumber
%\end{eqnarray}

\begin{equation}
    \begin{array}{l} 
        \mathrm{PDF}_{\mathrm{NCQE}}^{\mathrm{High}\, \theta_c} \to \mathrm{PDF}_{\mathrm{NCQE}}^{\mathrm{High}\, \theta_c} \times (1 - \sigma_{\mathrm{NCQE}}\, A^{\mathrm{High}\, \theta_c}_{\mathrm{NCQE}} \, \eta_{\mathrm{NCQE}}), \\
        \mathrm{PDF}_{\mathrm{NCQE}}^{\mathrm{Medium}\, \theta_c} \to \mathrm{PDF}_{\mathrm{NCQE}}^{\mathrm{Medium}\, \theta_c} \times (1 + \sigma_{\mathrm{NCQE}}\, \eta_{\mathrm{NCQE}}), \\
        \mathrm{PDF}_{\mathrm{NCQE}}^{\mathrm{Low}\, \theta_c} \to \mathrm{PDF}_{\mathrm{NCQE}}^{\mathrm{Low}\, \theta_c} \times (1 - \sigma_{\mathrm{NCQE}}\, A^{\mathrm{Low}\, \theta_c}_{\mathrm{NCQE}} \, \eta_{\mathrm{NCQE}}),
    \end{array}
\end{equation}    
where $\sigma_{\mathrm{NCQE}} = 100\%$, $A^{\mathrm{High}\, \theta_c}_{\mathrm{NCQE}}$ and $A^{\mathrm{Low}\, \theta_c}_{\mathrm{NCQE}}$ are normalization factors introduced to preserve the overall normalization of the PDF across the six regions. $\eta_{\rm NCQE}$ is bound to be greater than $-1$. To fulfill this condition while maintaining a zero mean and unit standard deviation, this parameter is assigned a log-normal probability density function.

\subsubsection{\texorpdfstring{Spallation events: $\eta_{\rm spall}$}{Spallation events}}

Spallation events, mainly those from $^8$B, $^8$Li, and $^9$C isotopes, contribute to the background in the signal region of the spectral analysis. To model the shape uncertainty associated with the spallation spectrum, we follow the procedure in \citet{Abe2021} and introduce a nuisance parameter $\eta_{\rm spall}$, modifying the spallation event PDF as:

\begin{eqnarray}
    \mathrm{PDF}_{\rm spall}&(E_{\rm rec})& \\ &\rightarrow& \mathrm{PDF}_{\rm spall}(E_{\rm rec}) \times \left( 
    1 + \eta_{\rm spall}\, \mathcal{P}_{3}(E_{\rm rec})
    \right), \nonumber
\end{eqnarray}
where $\mathcal{P}_{3}(E_{\rm rec})$ is a third-order polynomial denoting the energy-dependent 1$\sigma$ uncertainty on the spallation PDF shape, stemming from the uncertainty on the relative contribution of each isotope to the spallation background event rate. $\eta_{\rm spall}$ is assigned a standard normal probability density function.

\subsubsection{\texorpdfstring{Neutron tagging: $\eta_{\rm n}$}{Neutron tagging}} 
    
As in \citet{Abe2021}, the uncertainty related to neutron tagging efficiency is dominated by the uncertainty in the neutron multiplicity and found to be approximately $40\%$. The effect is parametrized by a dedicated nuisance parameter as follows:

\begin{equation}
    \begin{array}{rl}
        \mathrm{PDF}_{j}^{N_{\mathrm{n}} = 1} &\to \mathrm{PDF}_{j}^{N_{\mathrm{n}} = 1} \times (1 + \sigma_{N_{\mathrm{n}}} \, \eta_{N_{\mathrm{n}}}), \\
        \mathrm{PDF}_{j}^{N_{\mathrm{n}} \neq 1} &\to \mathrm{PDF}_{j}^{N_{\mathrm{n}} \neq 1} \times (1 - \sigma_{N_{\mathrm{n}}} \, A_{N_{\mathrm{n}}} \, \eta_{N_{\mathrm{n}}}),
    \end{array}
\end{equation}
where $\sigma_{N_{\mathrm{n}}} = 40\%$, and $A_{N_{\mathrm{n}}}$ is a relative normalization parameter. $\eta_{\rm n}$ is therefore bound to be greater than $-2.5$ and is assigned a folded normal probability density function. 

\begin{figure}[ht!]
\centering
%\epsscale{0.8}
%\plotone{figure/systematic_prior_PDFs.pdf}
\plotone{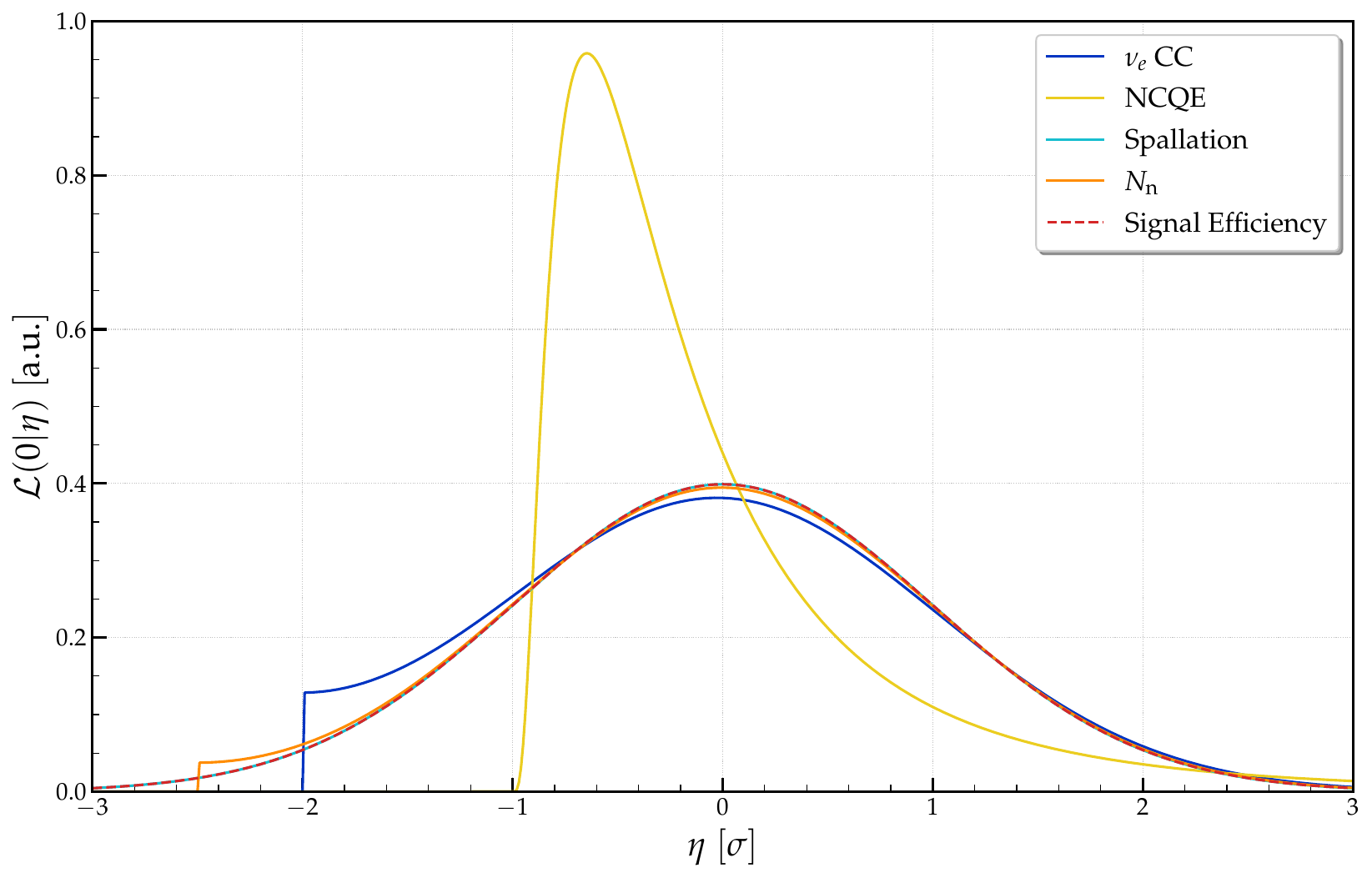}
\caption{
\label{fig:systematic_priors}
Prior probability density functions for the nuisance parameters of the spectral analysis.}
\end{figure}

We show in Figure~\ref{fig:systematic_priors} the probability density functions associated with the aforementioned nuisance parameters $\mathcal{L}(0\, |\, \eta)$. The overall penalty term in Equation (\ref{eq:likelihood_spectral_fit}) reads $\mathcal{L}\left(\vec{0}\, |\, \vec{\eta} \right) = \prod_{\eta} \mathcal{L}(0\, |\, \eta)$.

\subsection{Likelihood ratio \label{appendix:likelihood_ratio}}

In the fit of the spectral analysis, the chosen statistical approach is to profile the likelihood ratio over all the nuisance parameters of the analysis, namely the background rates $\vec{N}_b$, the PDFs shape-only nuisance parameters and signal efficiency nuisance parameter $\vec{\eta}_{b},\, \eta_{s} := \vec{\eta}$. This means that the  likelihood ratio function $\mathcal{L_R}$, whose logarithm is displayed in Figure~\ref{fig:profile_likelihood_ratio}, is obtained as follows: 
\begin{equation}
\mathcal{L_R}(N_s) = 
\frac
{\mathcal{L}\left( \mathrm{Data}\, |\, N_s,\, \hat{\hat{\vec{N}}}_b, \, \hat{\hat{\vec{\eta}}} \right)}
{\mathcal{L}\left( \mathrm{Data}\, |\, \hat{N_s},\, \hat{\vec{N}}_b, \, \hat{\vec{\eta}} \right)},
\end{equation}
where $\mathcal{L}\left( \mathrm{Data}\, |\, \hat{N_s},\, \hat{\vec{N}}_b, \, \hat{\vec{\eta}} \right)$ is the likelihood maximized over the entire parameter space $\{N_s,\, \vec{N}_b, \, \vec{\eta} \}$, whereas $\mathcal{L}\left( \mathrm{Data}\, |\, N_s,\, \hat{\hat{\vec{N}}}_b, \, \hat{\hat{\vec{\eta}}} \right)$ is the likelihood maximized over the restricted parameter space $\{\vec{N}_b, \, \vec{\eta}\}$ at fixed $N_s$.

This derivation of the likelihood ratio function slightly differs from the one adopted in the SK-I to -IV analysis of~\citet{Abe2021}, where the profiling was done only over background rate, whereas the PDFs shape-only and signal efficiency nuisance parameters were marginalized, i.e., integrated over in the computation of the likelihoods.

\subsection{Other results \label{appendix:other_results}}

\begin{figure*}[ht!]
\centering
\epsscale{0.8}
%\plotone{figure/bestfit_sk6_bdt.pdf}
%\plotone{figure/bestfit_sk7_bdt.pdf}
\plotone{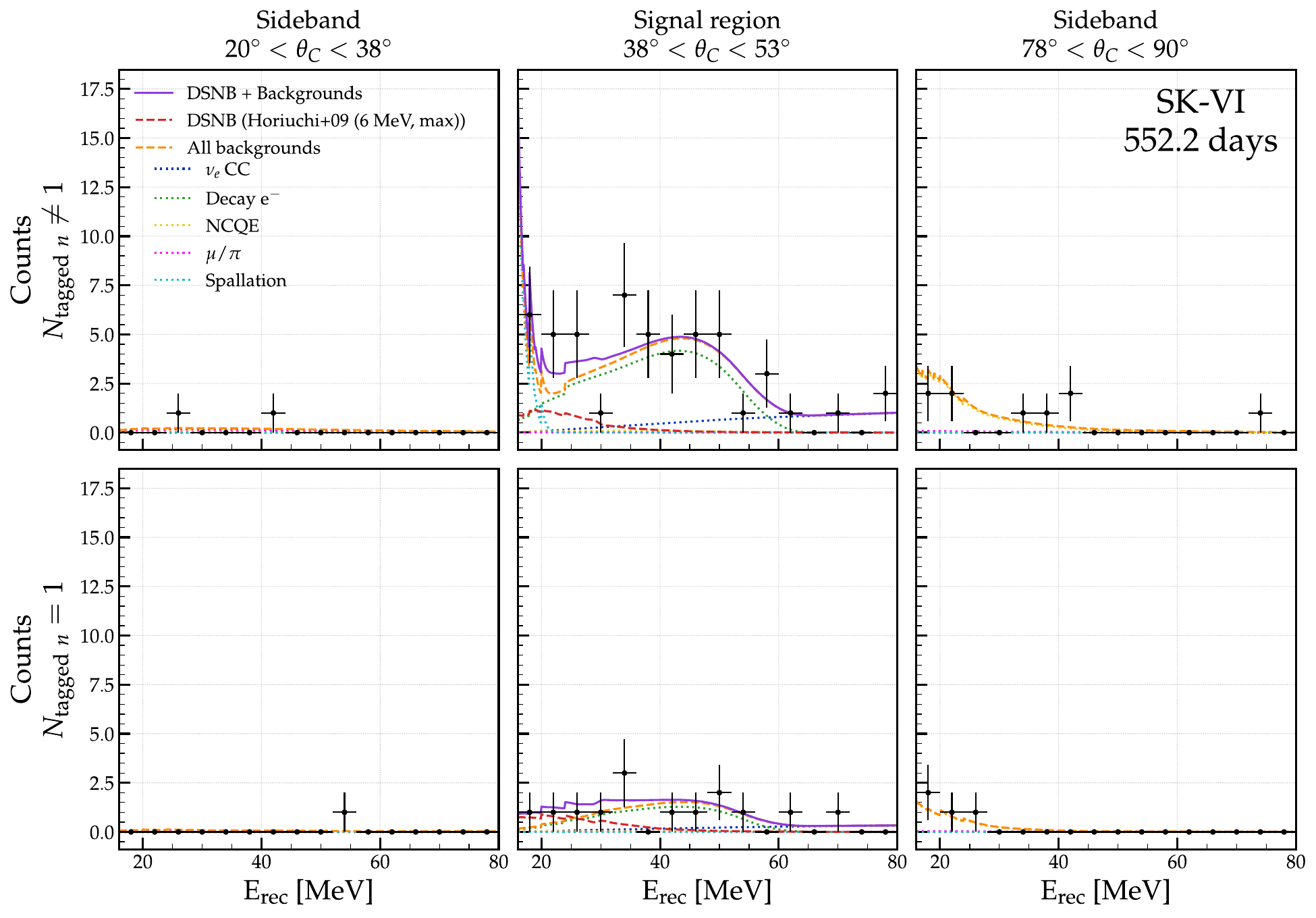}
\plotone{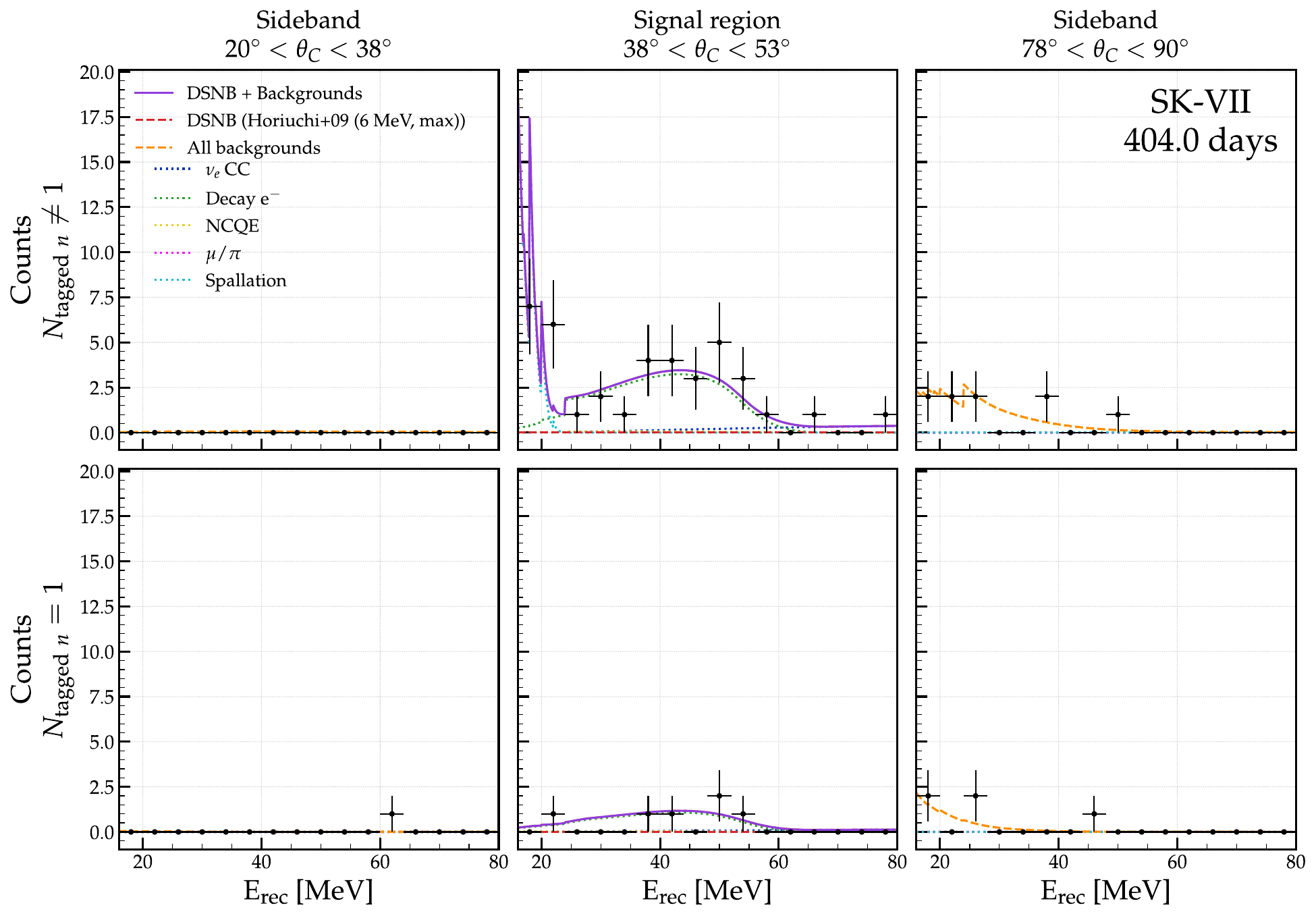}
\caption{\label{fig:bestfit_sk6_sk7_bdt}
Best-fit results for SK-VI and -VII data samples, built out of BDT neutron-tagging algorithm. The input DSNB model used for the fit is \citet{Horiuchi2009}.}
\end{figure*}

In Figure~\ref{fig:bestfit_sk6_sk7_bdt}, we display the result of the best-fit for the SK-VI and -VII BDT-samples, obtained with the~\citet{Horiuchi2009} model as an input DSNB signal shape. 
%In Figure~\ref{fig:bestfit_sk7_nn}, we display the result of the best-fit for the SK-VII NN-sample, as well as the result for the SK-VI and -VII BDT-samples in Figure~\ref{fig:bestfit_sk6_sk7_bdt}, obtained with the~\citet{Horiuchi2009} model as an input DSNB signal shape. 
Finally, Tables~\ref{tab:specres_nn_appendix} and~\ref{tab:specres_bdt_appendix} list the results of the spectral analysis for various DSNB model shapes, including the ones displayed in Figure~\ref{fig:spectral_analysis_model_dependent_results_MainBody}.
%Finally, Figures~\ref{fig:spectral_analysis_model_dependent_results_nn_Appendix} and~\ref{fig:spectral_analysis_model_dependent_results_bdt_Appendix} depict results of the spectral analysis for a variety of recent DSNB model shapes, as complements to Figure~\ref{fig:spectral_analysis_model_dependent_results_MainBody}.

\newpage

\begin{deluxetable*}{lcccc}
\tablecaption{\label{tab:specres_nn_appendix} Model-dependent spectral analysis results for DSNB events with $E_\nu > 17.3$ MeV, obtained with the NN-based neutron-tagging algorithm. “NH” and “IH” refer to the normal and inverted neutrino mass hierarchies, respectively. “HB06” and “MD14” correspond to the SFR calculations performed by \citet{Hopkins_2006} and \citet{Madau_2014} respectively. For the models from \citet{Ivanez2023}, “SH” stands for “strongly hierarchical”.}
\tablehead{
\colhead{Model} & 
\colhead{\parbox[c]{2cm}{Best-fit \\ $[$cm$^{-2}$ s$^{-1}]$}} & 
\colhead{\parbox[c]{2.5cm}{90\% Upper Limit \\ $[$cm$^{-2}$ s$^{-1}]$}} & 
\colhead{\parbox[c]{2cm}{Predicted \\ $[$cm$^{-2}$ s$^{-1}]$}} & 
\colhead{\parbox[c]{2cm}{$H_0$ Rejection \\ $[\sigma]$}} 
}
\startdata
Totani+95 (Constant) & $1.4^{+1.5}_{-1.2}$ & 3.4 & 4.7 & 1.2 \\
Hartmann+97 (CE) & $1.5^{+1.6}_{-1.2}$ & 3.6 & 0.6 & 1.3 \\
Malaney+97 (CGI) & $1.6^{+1.7}_{-1.1}$ & 3.8 & 0.3 & 1.4 \\
Kaplinghat+00 (HMA, Max) & $1.5^{+1.7}_{-1.2}$ & 3.7 & 3.0 & 1.4 \\
Kawasaki+03 & $1.5^{+1.7}_{-1.2}$ & 3.6 & 0.7 & 1.3 \\
Ando+03 (Updated 05) & $1.5^{+1.6}_{-1.2}$ & 3.5 & 0.7 & 1.3 \\
Horiuchi+09 (6 MeV, Max) & $1.4^{+1.5}_{-1.2}$ & 3.3 & 2.0 & 1.2 \\
Lunardini+09 (Failed SN) & $1.4^{+1.6}_{-1.2}$ & 3.4 & 0.7 & 1.2 \\
Galais+10 (NH) & $1.5^{+1.5}_{-1.2}$ & 3.4 & 1.6 & 1.3 \\
Galais+10 (IH) & $1.5^{+1.5}_{-1.2}$ & 3.5 & 1.5 & 1.3 \\
Nakazato+15 (Min, NH) & $1.5^{+1.6}_{-1.2}$ & 3.6 & 0.2 & 1.3 \\
Nakazato+15 (Max, IH) & $1.4^{+1.4}_{-1.2}$ & 3.3 & 0.5 & 1.2 \\
Priya+17 (NH) & $1.5^{+1.7}_{-1.2}$ & 3.7 & 0.4 & 1.4 \\
Horiuchi+18 ($\bar{\zeta}_{2.5} = 0.1$) & $1.4^{+1.4}_{-1.2}$ & 3.3 & 1.2 & 1.2 \\
Horiuchi+18 ($\bar{\zeta}_{2.5} = 0.5$) & $1.9^{+1.7}_{-1.3}$ & 4.1 & 0.6 & 1.6 \\
Barranco+18 ($\Lambda$CDM, Logotropic) & $2.0^{+1.9}_{-1.3}$ & 4.4 & 0.3 & 1.7 \\
De Gouvêa+20 (NH) & $0.8^{+1.4}_{-0.8}$ & 2.6 & 1.6 & 0.7 \\
Horiuchi+21 & $1.6^{+1.7}_{-1.2}$ & 3.8 & 1.7 & 1.5 \\
Kresse+21 (Fiducial, NH) & $1.5^{+1.7}_{-1.2}$ & 3.7 & 1.2 & 1.4 \\
Kresse+21 (Fiducial, IH) & $1.5^{+1.7}_{-1.2}$ & 3.6 & 1.0 & 1.4 \\
Kresse+21 (Low, NH) & $1.5^{+1.7}_{-1.2}$ & 3.7 & 1.0 & 1.4 \\
Kresse+21 (High, NH) & $1.5^{+1.6}_{-1.2}$ & 3.5 & 1.6 & 1.3 \\
Kresse+21 (Low, IH) & $1.5^{+1.7}_{-1.2}$ & 3.7 & 0.8 & 1.4 \\
Kresse+21 (High, IH) & $1.5^{+1.6}_{-1.2}$ & 3.5 & 1.2 & 1.3 \\
Tabrizi+21 (NH) & $1.5^{+1.6}_{-1.2}$ & 3.5 & 0.9 & 1.3 \\
Ashida+23 (Min, NH) & $1.4^{+1.5}_{-1.2}$ & 3.4 & 0.9 & 1.2 \\
Ashida+23 (Max, NH) & $1.1^{+1.4}_{-1.1}$ & 3.0 & 2.6 & 1.0 \\
Iváñez-Ballesteros+23 (No Decay, SH, NH) & $1.6^{+1.7}_{-1.2}$ & 3.7 & 0.8 & 1.4 \\
Iváñez-Ballesteros+23 (No decay, SH, IH) & $1.6^{+1.7}_{-1.2}$ & 3.7 & 0.7 & 1.4 \\ 
Iváñez-Ballesteros+23 ($\tau/m=10^{11}$ s/eV, SH, NH) & $1.6^{+1.7}_{-1.2}$ & 3.7 & 0.8 & 1.4 \\
Iváñez-Ballesteros+23 ($\tau/m=10^{11}$ s/eV, SH, IH) & $1.6^{+1.6}_{-1.2}$ & 3.7 & 0.7 & 1.4 \\
Iváñez-Ballesteros+23 ($\tau/m=10^{10}$ s/eV, SH, NH) & $1.6^{+1.7}_{-1.2}$ & 3.7 & 0.8 & 1.4 \\
Iváñez-Ballesteros+23 ($\tau/m=10^{10}$ s/eV, SH, IH) & $1.5^{+1.5}_{-1.2}$ & 3.5 & 0.5 & 1.3 \\
Iváñez-Ballesteros+23 ($\tau/m=10^{9}$ s/eV, SH, NH) & $1.6^{+1.7}_{-1.2}$ & 3.7 & 0.7 & 1.4 \\
Iváñez-Ballesteros+23 ($\tau/m=10^{9}$ s/eV, SH, IH) & $1.0^{+1.5}_{-1.0}$ & 2.9 & 0.1 & 1.0 \\
Martínez-Miravé+24 ($f_{\rm MR}=0.03$) & $1.5^{+1.6}_{-1.2}$ & 3.6 & 0.6 & 1.4 \\
Martínez-Miravé+24 ($f_{\rm MR}=0.1$) & $1.5^{+1.6}_{-1.2}$ & 3.5 & 0.8 & 1.3 \\
Martínez-Miravé+24 ($f_{\rm MR}=0.2$) & $1.5^{+1.5}_{-1.2}$ & 3.5 & 0.9 & 1.3 \\
Nakazato+24 (HB06, $f_{\rm BHSN}=0.1$, Fallback, NH) & $1.6^{+1.5}_{-1.2}$ & 3.6 & 2.2 & 1.4 \\
Nakazato+24 (HB06, $f_{\rm BHSN}=0.1$, Fallback, IH) & $1.6^{+1.5}_{-1.2}$ & 3.5 & 1.9 & 1.4 \\
Nakazato+24 (HB06, $f_{\rm BHSN}=0.5$, Fallback, NH) & $1.6^{+1.5}_{-1.2}$ & 3.6 & 3.5 & 1.4 \\
Nakazato+24 (HB06, $f_{\rm BHSN}=0.5$, Fallback, IH) & $1.6^{+1.4}_{-1.2}$ & 3.4 & 2.4 & 1.4 \\
Nakazato+24 (MD14, $f_{\rm BHSN}=0.1$, Fallback, NH) & $1.6^{+1.5}_{-1.2}$ & 3.6 & 1.8 & 1.4 \\
Nakazato+24 (MD14, $f_{\rm BHSN}=0.1$, Fallback, IH) & $1.6^{+1.5}_{-1.2}$ & 3.5 & 1.6 & 1.4 \\
Nakazato+24 (MD14, $f_{\rm BHSN}=0.5$, Fallback, NH) & $1.6^{+1.5}_{-1.2}$ & 3.5 & 2.8 & 1.4 \\
Nakazato+24 (MD14, $f_{\rm BHSN}=0.5$, Fallback, IH) & $1.6^{+1.5}_{-1.2}$ & 3.6 & 1.9 & 1.4 \\
\enddata
\end{deluxetable*}

\newpage
\begin{deluxetable*}{lcccc}
\tablecaption{\label{tab:specres_bdt_appendix} Model-dependent spectral analysis results for DSNB events with $E_\nu > 17.3$ MeV, obtained with the BDT-based neutron-tagging algorithm. “NH” and “IH” refer to the normal and inverted neutrino mass hierarchies, respectively. “HB06” and “MD14” correspond to the SFR calculations performed by \citet{Hopkins_2006} and \citet{Madau_2014} respectively. For the models from \citet{Ivanez2023}, “SH” stands for “strongly hierarchical”.}
\tablehead{
\colhead{Model} & 
\colhead{\parbox[c]{2cm}{Best-fit \\ $[$cm$^{-2}$ s$^{-1}]$}} & 
\colhead{\parbox[c]{2.5cm}{90\% Upper Limit \\ $[$cm$^{-2}$ s$^{-1}]$}} & 
\colhead{\parbox[c]{2cm}{Predicted \\ $[$cm$^{-2}$ s$^{-1}]$}} & 
\colhead{\parbox[c]{2cm}{$H_0$ Rejection \\ $[\sigma]$}} 
}
\startdata
Totani+95 (Constant) & $1.2^{+1.7}_{-1.2}$ & 3.4 & 4.7 & 1.0 \\
Hartmann+97 (CE) & $1.3^{+1.8}_{-1.2}$ & 3.5 & 0.6 & 1.1 \\
Malaney+97 (CGI) & $1.6^{+1.7}_{-1.4}$ & 3.7 & 0.3 & 1.2 \\
Kaplinghat+00 (HMA, Max) & $1.3^{+1.8}_{-1.2}$ & 3.6 & 3.0 & 1.1 \\
Kawasaki+03 & $1.3^{+1.9}_{-1.2}$ & 3.6 & 0.7 & 1.1 \\
Ando+03 (Updated 05) & $1.2^{+1.8}_{-1.2}$ & 3.5 & 0.7 & 1.0 \\
Horiuchi+09 (6 MeV, Max) & $1.2^{+1.7}_{-1.2}$ & 3.3 & 2.0 & 0.9 \\
Lunardini+09 (Failed SN) & $1.2^{+1.8}_{-1.2}$ & 3.5 & 0.7 & 1.0 \\
Galais+10 (NH) & $1.3^{+1.7}_{-1.3}$ & 3.4 & 1.6 & 1.0 \\
Galais+10 (IH) & $1.3^{+1.7}_{-1.2}$ & 3.4 & 1.5 & 1.0 \\
Nakazato+15 (Min, NH) & $1.3^{+1.8}_{-1.2}$ & 3.6 & 0.2 & 1.1 \\
Nakazato+15 (Max, IH) & $1.2^{+1.7}_{-1.2}$ & 3.3 & 0.5 & 0.9 \\
Priya+17 (NH) & $1.3^{+1.8}_{-1.2}$ & 3.6 & 0.4 & 1.1 \\
Horiuchi+18 ($\bar{\zeta}_{2.5} = 0.1$) & $1.2^{+1.7}_{-1.2}$ & 3.3 & 1.2 & 0.9 \\
Horiuchi+18 ($\bar{\zeta}_{2.5} = 0.5$) & $1.6^{+1.8}_{-1.4}$ & 3.9 & 0.6 & 1.3 \\
Barranco+18 ($\Lambda$CDM, Logotropic) & $1.7^{+1.9}_{-1.4}$ & 4.2 & 0.3 & 1.3 \\
De Gouvêa+20 (NH) & $0.7^{+1.6}_{-0.7}$ & 2.7 & 1.6 & 0.5 \\
Horiuchi+21 & $1.4^{+1.8}_{-1.2}$ & 3.7 & 1.7 & 1.2 \\
Kresse+21 (Fiducial, NH) & $1.3^{+1.9}_{-1.2}$ & 3.6 & 1.2 & 1.1 \\
Kresse+21 (Fiducial, IH) & $1.3^{+1.9}_{-1.2}$ & 3.6 & 1.0 & 1.1 \\
Kresse+21 (Low, NH) & $1.5^{+1.6}_{-1.4}$ & 3.7 & 1.0 & 1.1 \\
Kresse+21 (High, NH) & $1.3^{+1.8}_{-1.2}$ & 3.5 & 1.6 & 1.0 \\
Kresse+21 (Low, IH) & $1.5^{+1.6}_{-1.4}$ & 3.7 & 0.8 & 1.1 \\
Kresse+21 (High, IH) & $1.3^{+1.7}_{-1.2}$ & 3.5 & 1.2 & 1.0 \\
Tabrizi+21 (NH) & $1.2^{+1.8}_{-1.2}$ & 3.5 & 0.9 & 1.0 \\
Ashida+23 (Min, NH) & $1.2^{+1.7}_{-1.2}$ & 3.4 & 0.9 & 1.0 \\
Ashida+23 (Max, NH) & $0.9^{+1.6}_{-0.9}$ & 3.0 & 2.6 & 0.8 \\
Iváñez-Ballesteros+23 (No Decay, SH, NH) & $1.3^{+1.8}_{-1.2}$ & 3.6 & 0.8 & 1.1 \\
Iváñez-Ballesteros+23 (No decay, SH, IH) & $1.3^{+1.8}_{-1.2}$ & 3.6 & 0.7 & 1.1 \\ 
Iváñez-Ballesteros+23 ($\tau/m=10^{11}$ s/eV, SH, NH) & $1.3^{+1.8}_{-1.2}$ & 3.6 & 0.8 & 1.1 \\
Iváñez-Ballesteros+23 ($\tau/m=10^{11}$ s/eV, SH, IH) & $1.3^{+1.8}_{-1.2}$ & 3.6 & 0.7 & 1.1 \\
Iváñez-Ballesteros+23 ($\tau/m=10^{10}$ s/eV, SH, NH) & $1.3^{+1.8}_{-1.2}$ & 3.6 & 0.8 & 1.1 \\
Iváñez-Ballesteros+23 ($\tau/m=10^{10}$ s/eV, SH, IH) & $1.3^{+1.7}_{-1.2}$ & 3.5 & 0.5 & 1.0 \\
Iváñez-Ballesteros+23 ($\tau/m=10^{9}$ s/eV, SH, NH) & $1.4^{+1.8}_{-1.2}$ & 3.7 & 0.7 & 1.1 \\
Iváñez-Ballesteros+23 ($\tau/m=10^{9}$ s/eV, SH, IH) & $1.0^{+1.4}_{-1.0}$ & 2.9 & 0.1 & 0.8 \\
Martínez-Miravé+24 ($f_{\rm MR}=0.03$) & $1.3^{+1.8}_{-1.2}$ & 3.6 & 0.6 & 1.1 \\
Martínez-Miravé+24 ($f_{\rm MR}=0.1$) & $1.3^{+1.7}_{-1.2}$ & 3.5 & 0.8 & 1.1 \\
Martínez-Miravé+24 ($f_{\rm MR}=0.2$) & $1.3^{+1.7}_{-1.3}$ & 3.4 & 0.9 & 1.0 \\
Nakazato+24 (HB06, $f_{\rm BHSN}=0.1$, Fallback, NH) & $1.4^{+1.7}_{-1.3}$ & 3.5 & 2.2 & 1.1 \\
Nakazato+24 (HB06, $f_{\rm BHSN}=0.1$, Fallback, IH) & $1.4^{+1.6}_{-1.3}$ & 3.4 & 1.9 & 1.1 \\
Nakazato+24 (HB06, $f_{\rm BHSN}=0.5$, Fallback, NH) & $1.4^{+1.7}_{-1.2}$ & 3.5 & 3.5 & 1.1 \\
Nakazato+24 (HB06, $f_{\rm BHSN}=0.5$, Fallback, IH) & $1.3^{+1.5}_{-1.3}$ & 3.3 & 2.4 & 1.1 \\
Nakazato+24 (MD14, $f_{\rm BHSN}=0.1$, Fallback, NH) & $1.4^{+1.6}_{-1.3}$ & 3.5 & 1.8 & 1.1 \\
Nakazato+24 (MD14, $f_{\rm BHSN}=0.1$, Fallback, IH) & $1.4^{+1.6}_{-1.3}$ & 3.4 & 1.6 & 1.1 \\
Nakazato+24 (MD14, $f_{\rm BHSN}=0.5$, Fallback, NH) & $1.4^{+1.6}_{-1.3}$ & 3.4 & 2.8 & 1.1 \\
Nakazato+24 (MD14, $f_{\rm BHSN}=0.5$, Fallback, IH) & $1.4^{+1.6}_{-1.3}$ & 3.5 & 1.9 & 1.1 \\
\enddata
\end{deluxetable*}

\newpage
\bibliography{sample631}{}
\bibliographystyle{aasjournal}



\end{document}